\documentclass{ar-astro2e}
\usepackage{ARAstroBib}
\usepackage{amsfonts}
\usepackage{amssymb}
\def\pasp{Pub. Astron. Soc. Pac.}
\def\apj{Ap. J.}
\def\apjl{Ap. J. Let.}
\def\aj{Astron. J.}
\def\apss{Astrophy. Space Sci}
\def\apjs{Ap. J. Suppl.}
\def\mnras{MNRAS}

\def\aap{Astron. Astrophys.}
\def\aaps{Astron. Astrophys. Suppl.}
\def\araa{Annu. Rev. Astron. Astrophy.}
\def\nat{Nature}
\def\nar{New Astron. Rev.}

\def\Msun{M_{\odot}}
\def\teff{T_{\rm eff}}
\def\lbol{L_{\rm bol}}
\def\logg{{\rm log}\,g}
\def\kms{\ {\rm km\, s}^{-1}}
\newcommand{\be}{\begin{equation}}
\newcommand{\ee}{\end{equation}}
\newcommand\ionn[2]{#1$\;${\scshape{#2}}}

\begin{document}

\input psfig.sty

\jname{Annu. Rev. Astron. Astrophys.}
\jyear{2013}
\jvol{}
\ARinfo{1056-8700/97/0610-00}

\title{Modeling the Panchromatic Spectral Energy Distributions of Galaxies}

\author{Charlie Conroy
\affiliation{Department of Astronomy and Astrophysics, University of
  California, \\Santa Cruz, California, 95064; email: cconroy@ucolick.org}}


\begin{abstract}

  The spectral energy distributions (SEDs) of galaxies are shaped by
  nearly every physical property of the system, including the star
  formation history, metal content, abundance pattern, dust mass,
  grain size distribution, star-dust geometry, and interstellar
  radiation field.  The principal goal of stellar population synthesis
  (SPS) is to extract these variables from observed SEDs.  In this
  review I provide an overview of the SPS technique and discuss what
  can be reliably measured from galaxy SEDs.  Topics include stellar
  masses, star formation rates and histories, metallicities and
  abundance patterns, dust properties, and the stellar initial mass
  function.

\end{abstract}

\maketitle

\section{INTRODUCTION}

Many of the fundamental properties of unresolved stellar populations
are encoded in their SEDs.  These properties include the star
formation history (SFH), stellar metallicity and abundance pattern,
stellar initial mass function (IMF), total mass in stars, and the
physical state and quantity of dust and gas.  Some of these properties
are easier to measure than others, and each provides important clues
regarding the formation and evolution of galaxies.  It is precisely
these quantities, measured from the SEDs of galaxies, that have
provided the foundation for our modern understanding of galaxy
formation and evolution.

Over the past several decades considerable effort has been devoted to
extracting information from the SEDs of galaxies, exploiting
information from the FUV to the FIR.  Early attempts at understanding
the visible and NIR spectral windows approached the problem by
combining mixtures of stars in ad hoc ways until a match was achieved
with observations \citep[e.g.,][]{Spinrad71}.  More sophisticated
versions of this technique were developed that incorporated physical
constraints and automated fitting techniques \citep{Faber72}.  At
about the same time, synthesis models were being developed that relied
on stellar evolution theory to constrain the range of possible stellar
types at a given age and metallicity \citep[e.g.,][]{Tinsley68,
  Searle73, Tinsley76, Bruzual83}.  The substantial progress made in
stellar evolution theory in the 1980s and 1990s paved the way for the
latter approach to become the de facto standard in modeling the SEDs
of galaxies \citep[e.g.,][]{Charlot91, Bruzual93, Bressan94,
  Worthey94, Fioc97, Leitherer99, Vazdekis99}.  This modeling
technique, which will be described in detail in the next section, is
sometimes referred to as `evolutionary population synthesis'
\citep[e.g.,][]{Maraston98}, although the term `stellar population
synthesis' (SPS) has garnered wider use.  The latter term will be used
throughout this review.

The UV and IR spectral windows are rather more difficult to probe
owing to the obscuring effects of the atmosphere.  Nonetheless,
numerous balloon and space-based observatories have opened up the
ultraviolet and infrared to detailed investigations.  In these
spectral regions dust plays a major role; it absorbs and scatters much
of the UV light emitted by stars and re-radiates that energy in the
IR.  In young stellar populations the UV is dominated by hot massive
stars, while in old stellar populations the UV can be influenced by
hot evolved stellar types such as post-AGB and extreme HB stars
\citep[see the review by][for details]{Oconnell99}.

The development of models for the IR SEDs of galaxies
\citep[e.g.,][]{Draine84, Zubko04} has proceeded in parallel with the
development of models for UV, optical, and NIR SEDs, and it is only
recently that models have been developed to simultaneously and
self-consistently predict the FUV through FIR SEDs
\citep[e.g.,][]{Devriendt99, Silva98, daCunha08, Groves08, Noll09b}.

Several broad questions will serve to focus this review.  They are:
\begin{itemize}
\item What have we learned about the physical properties of galaxies
  from their observed SEDs?
\item How reliable are the quantities thus derived?
\item What can be learned, in principle, from the SEDs of galaxies?
\end{itemize}
The first question is relatively straightforward to address, while the
second and third are necessarily more complicated.  Attention will be
given to cases where answers to these questions are presently unknown,
but knowable (the `unknown unknowns', in Rumsfeld's sense, are of
course the most interesting, but most difficult to discuss).

There are surprisingly few thorough reviews of stellar population
synthesis and its application to modeling SEDs.  The foundational
review by \citet{Tinsley80} is highly recommended to anyone seeking a
broad yet intuitive understanding of stellar populations and galaxy
formation.  The reviews by \citet{Faber77} and \citet{Frogel88} are
more narrowly focused on old stellar populations but they are
informative because many of the issues raised therein are still
relevant today.  Recently, \citet{Walcher11} has presented an
excellent review of many aspects related to the modeling of galaxy
SEDs.

The topic of modeling galaxy SEDs is vast, and it would be impossible
to provide a thorough review of the entire field.  Hard decisions
therefore had to be made.  With regard to wavelength, energies higher
than the FUV (roughly the Lyman limit), and lower than the
submillimeter (roughly $1 mm$) will not be discussed.  Nebular
emission lines will also be neglected, except in a few cases.  The
measurement of and uncertainty induced by photometric redshifts will
not be discussed and active galactic nuclei (AGN) will be ignored.
The technique of fitting models to data was reviewed recently by
\citet{Walcher11} and so will not be discussed here.  A whole review
could (and should) be written on the evaluation and comparison of
existing SPS models.  This will not be undertaken herein, except in
cases where different models, when applied to data, produce starkly
different results.  The reader is referred to \citet{Conroy10c} for a
recent comparison of several popular models.  Finally, it is worth
emphasizing what this review is not: it is not a summary of science
results derived with SPS models.  There is an extensive literature
devoted to applying SPS models to data in order to derive insights
into the formation and evolution of galaxies.  Results of this nature,
though fascinating in their own right, will not feature prominently in
this review.  Readers are referred to \citet{Renzini06} and
\citet{Blanton09} for recent reviews along these lines.

This review is organized as follows.  An overview of SPS model
construction and application is provided in Section 2.  We then turn
to the topics of mass-to-light ratios and stellar masses (Section 3),
star formation rates, histories, and stellar ages (Section 4), stellar
metallicities and abundances patterns (Section 5), dust (Section 6),
and the initial mass function (Section 7).  Several concluding remarks
are provided in Section 8.


\section{OVERVIEW OF STELLAR POPULATION SYNTHESIS}

\begin{figure}
\centerline{\psfig{figure=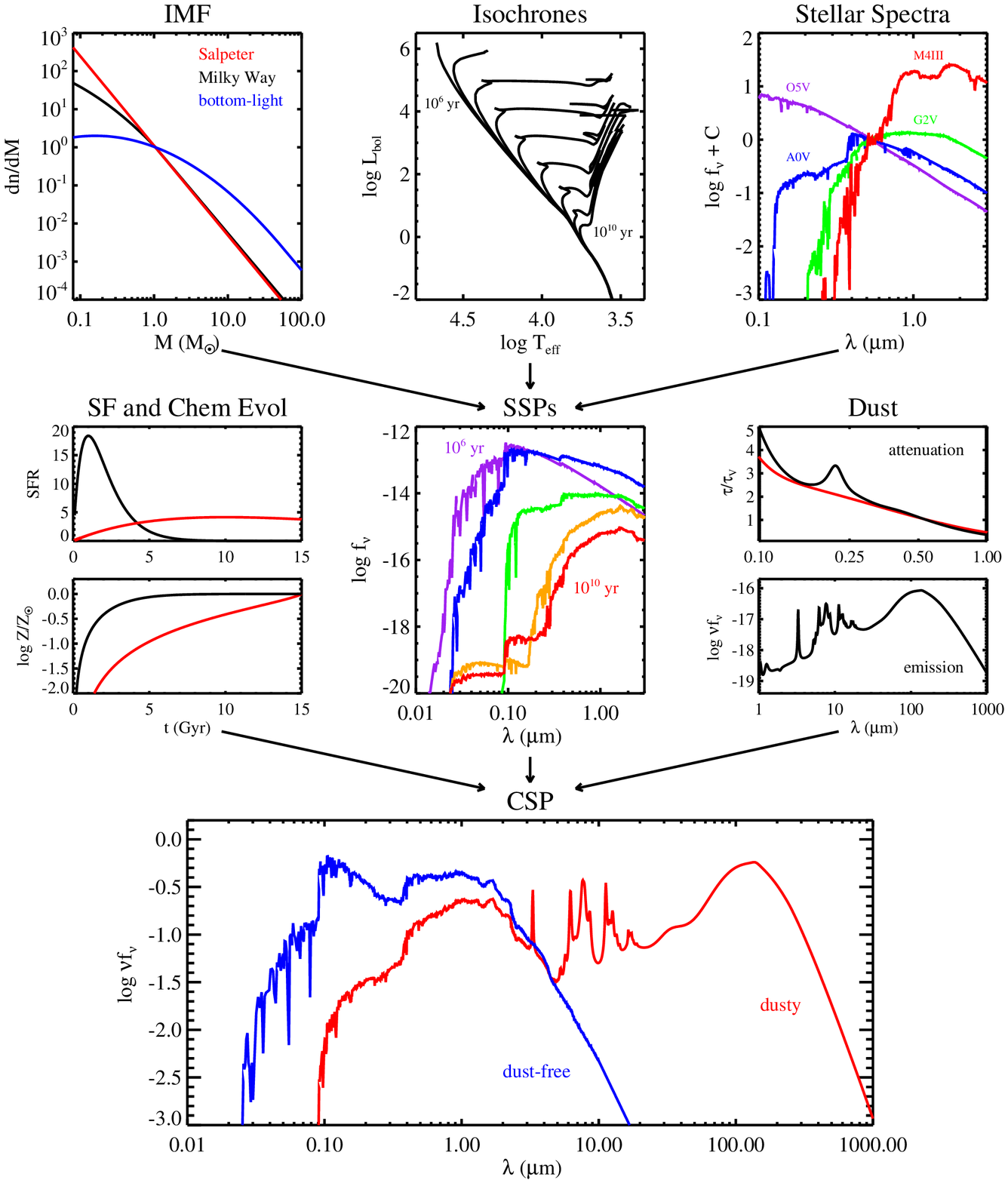,height=7.3in}}
\caption{Overview of the stellar population synthesis technique.  The
  upper panels highlight the ingredients necessary for constructing
  simple stellar populations (SSPs): an IMF, isochrones for a range of
  ages and metallicities, and stellar spectra spanning a range of
  $\teff$, $\lbol$, and metallicity.  The middle panels highlight the
  ingredients necessary for constructing composite stellar populations
  (CSPs): star formation histories and chemical evolution, SSPs, and a
  model for dust attenuation and emission.  The bottom row shows the
  final CSPs both before and after a dust model is applied.}
\label{f:sps}
\end{figure}

The construction of models for simple and composite stellar
populations is conceptually straightforward.  There are however
certain constraints that make the creation of such models rather
difficult in practice (incomplete isochrone tables, incomplete empirical
stellar libraries, poorly calibrated physics, etc.).  In this section
the ingredients necessary for constructing model SEDs will be
discussed.  Areas of particularly large uncertainties will be
highlighted, but it is not the goal of this review to provide an
exhaustive inter-comparison of various possible model choices.  An
overview of the entire process of constructing composite stellar
populations is given in Figure \ref{f:sps}.

\subsection{The Simple Stellar Population}
\label{s:ssp}

The starting point of any SPS model is the simple stellar population
(SSP), which describes the evolution in time of the SED of a single,
coeval stellar population at a single metallicity and abundance
pattern.  An SSP therefore requires three basic inputs: stellar
evolution theory in the form of isochrones, stellar spectral
libraries, and an IMF, each of which may in principle be a function of
metallicity and/or elemental abundance pattern.  These components are
typically combined in the following way:
\noindent
\be
f_{\rm SSP}(t,Z) = \int_{m_{\rm lo}}^{m_{\rm up}(t)} f_{\rm
  star}[\teff(M),\logg(M)|t,Z]\,\Phi(M)\,{\rm d}M,
\ee
\noindent
where $M$ is the initial (zero-age main sequence) stellar mass,
$\Phi(M)$ is the initial mass function, $f_{\rm star}$ is a stellar
spectrum, and $f_{\rm SSP}$ is the resulting time and
metallicity-dependent SSP spectrum.  The lower limit of integration,
$m_{\rm lo}$, is typically taken to by the hydrogen burning limit
(either 0.08 or $0.1\Msun$ depending on the SPS code), and the upper
limit is dictated by stellar evolution.  The isochrones determine the
relation between $\teff$, $\logg$, and $M$ for a given $t$ and $Z$.
This approach to constructing SSPs is common but not universal.
Alternatives include the fuel consumption theorem \citep{Renzini86,
  Maraston98}, or the use of empirical spectra of star clusters as
templates for SSPs \citep{Bica86}.

Nearly all SPS models provide SSPs as a `black-box'
\citep[e.g.,][]{Leitherer99, Bruzual03, Maraston05, Vazdekis10}.  The
user therefore has little working knowledge of how SSPs are built and
where major issues lie.  The following discussion is therefore largely
pedagogical and geared toward users of SPS models, rather than model
builders.

\subsubsection{Stellar Evolution \& Isochrones}

An isochrone specifies the location in the Hertzsprung-Russell (HR)
diagram of stars with a common age and metallicity.  Isochrones are
constructed from stellar evolution calculations for stars from the
hydrogen burning limit ($\approx0.1\,\Msun$) to the maximum stellar
mass ($\approx100\,\Msun$).  The construction of isochrones is
straightforward for stellar evolution tracks that are infinitely
well-sampled in mass and time.  In practice, evolutionary tracks are
discretely sampled and this can lead to issues in isochrone
construction for fast evolutionary phases.  Modern sets of isochrones
have constructed specifically to ensure that the models are relatively
immune to these effects \citep{Charlot91}.

A number of widely-used isochrone tables exist in the literature.  The
most popular models span a wide range in ages (masses), chemical
compositions, and cover most relevant evolutionary phases.  Models in
this category include the Padova models \citep{Bertelli94, Girardi00,
  Marigo08}, and the BaSTI models \citep{Pietrinferni04, Cordier07}.
The Geneva models \citep{Schaller92, Meynet00} are tailored to follow
high-mass stars through advanced evolutionary phases, including the
Wolf-Rayet (WR) phase, but they do not model low mass stars.  Other
models have focused on the main sequence, red giant branch (RGB), and
horizontal branch (HB) evolution of low-mass stars ($M<3\,\Msun$),
including the $Y^2$ models \citep{Yi01, Yi03}, the Dartmouth models
\citep{Dotter08b}, and the Victoria-Regina models \citep{Vandenberg85,
  Vandenberg06}.  Finally, there are isochrones tailored to very
low-mass stars and brown dwarfs.  The Lyon models are the most widely
used in this regime \citep{Chabrier97, Baraffe98}.  It is noteworthy
that none of the isochrones listed here cover the post-AGB
evolutionary phase.  Remarkably, the post-AGB isochrones computed by
\citet{Schoenberner83}, \citet{Vassiliadis94}, and \citet{Blocker95b}
are still widely used in modern SPS codes.

Implementing isochrones in an SPS model is challenging because no
single set spans the necessary range of ages, metallicities and
evolutionary phases.  It is common to use the Padova isochrones for
the bulk of the age and metallicity range and to supplement with the
Geneva models at young ages.  Little attention is normally paid to the
lowest mass portion of the isochrones, since low mass stars contribute
only $\sim1\%$ of the light of an old stellar population.  An
exception in this regard is the model of \citet{Conroy12a}, who paid
special care to the modeling of low-mass stars for SPS. The
splicing together of various isochrone sets can be difficult.  For
example, different codes make different assumptions regarding
convection, rotation, etc., and so the age at which certain stars
evolve off the main sequence varies between codes.

Modules for Experiments in Stellar Astrophysics (MESA) is a new,
highly modular and sophisticated stellar evolution code that includes
the latest stellar interior ingredients including opacity tables,
equations of state, nuclear reaction networks, and surface boundary
conditions \citep{Paxton11}.  There is great hope that MESA will be
employed to produce high-quality isochrones over the full age and
metallicity range and for all evolutionary phases.

In addition to the practical difficulty of implementing current
isochrone libraries, stellar evolution calculations contain a number
of uncertainties relevant to SPS.  For example, all of the stellar
models mentioned above are based on one dimensional codes.  As such,
they require approximations to inherently three dimensional processes
such as convection, rotation, mass-loss, close binary interactions,
and thermal pulses during AGB evolution.  These processes lead to
major uncertainties in the isochrones that impact the resulting SPS
model predictions \citep[e.g.,][]{Charlot96a, Charlot96b, Yi03a,
  LeeHC07, Conroy09a}.

The cores of stars more massive than $M\sim1.1\,\Msun$ are convective,
as are the envelopes of evolved giants and low mass stars.
Classically, the boundary between convective and radiative regions is
specified by the Schwarzschild criterion.  However, this criterion is
effectively a requirement that the acceleration of a convective fluid
element be zero.  The fluid element will likely have a non-zero
velocity as it crosses this boundary, and so some amount of
`overshooting' is expected.  This will result in a larger convective
region than would be expected from the Schwarzschild criterion alone.
With regards to the convective core, a wide body of observational
evidence favors the existence of a moderate amount of overshooting in
the mass range $1.1<M/\Msun<7$ \citep{Stothers91, Nordstrom97,
  VandenBerg04, Keller06}.  The amount of core overshooting required
to fit the data results in a $\sim25\%$ increase in the main sequence
lifetime compared to models that do not include overshooting.  Nearly
all SPS models adopt isochrones with a modest amount of overshooting
in the convective core, as supported by observations.  The exception
to this is the SPS model of \citet{Maraston98, Maraston05}, which uses
isochrones without core convective overshooting.  The treatment of
core overshooting has a noticeable effect on the color evolution of
SSPs in the age range of $\sim0.1-1$ Gyr, by as much as $\approx0.1$
mag \citep{Yi03a, Conroy10c} and can therefore be an important source
of systematic uncertainty when modeling SEDs.

As discussed in \citet{Cassisi04}, the amount of overshooting in the
convective envelopes of evolved giants is less constrained.  This is
unfortunate because the treatment of envelope convection affects,
among other observables, the ratio of red to blue helium-burning
giants \citep{Renzini92}.  These stars can contribute several tens of
percent to the integrated light, depending on the star formation
history \citep{Melbourne12a}, and so uncertainty in the amount of
envelope overshooting should affect the integrated light predictions
at a significant level.  Overshooting in the convective envelope also
affects the location of the luminosity bump in the RGB luminosity
function.  This fact was used by \citet{Alongi91} and more recently by
\citet{Cassisi11} to argue for a modest amount of overshooting in the
convective envelope.

The importance of stellar rotation has been investigated by the Geneva
group, amongst others, over the past two decades \citep[see][for a
review]{Maeder00, Maeder12}.  Perhaps most importantly, rotation
increases the main sequence lifetimes (by $\sim25$\%) due to
rotation-induced mixing bringing fresh fuel to the convective core.
Rotation also lowers the effective surface gravity, lowers the opacity
in the radiative envelope, increases the luminosity, and changes the
ratio of red-to-blue supergiants.  The mixing to the surface of
H-burning products caused by rotation will also affect the number and
type of Wolf-Rayet stars.  \citet{Vazquez07} and \citet{Levesque12}
investigated the impact of rotating massive star models on integrated
light properties by comparing to non-rotating models and found that
the number of ionizing photons increased by as much as an order of
magnitude and colors became bluer by as much as $0.1-1$ mag, depending
on wavelength and age.  The signatures of WR stars as a function of
time and metallicity in integrated spectra are also different between
rotating and non-rotating models.

Most massive stars are in binary systems, and there is some evidence
that massive star binaries preferentially have comparable masses
\citep{Kobulnicky07, Sana11}.  The interaction of close binary stars
via mass-transfer and common envelope evolution will affect the
evolution of such stars and bring about further changes to their
observable properties.  \citet{Eldridge08} demonstrated that in many
respects the effects of binary star models without rotation are similar
to single star models with rotation.  \citet{Eldridge12} argued that
SPS models including binary star evolution produced a better fit to UV
spectra of star-forming galaxies.  Regardless of the details, it is
clear that where massive stars matter in galaxy SEDs, the effects of
both rotation and binary evolution will play an important role.  In
fact, binary star evolution may also create blue straggler stars and
extreme HB stars, which would suggest that binary evolution can affect
older stellar populations as well \citep{Han02, Han03, Zhang05}.  No
popular SPS model includes the effects of binary star evolution.

The potential importance of thermally-pulsating AGB (TP-AGB) stars in
the context of SPS models was emphasized by \citet{Maraston06}, and
has since become a controversial topic.  This phase occurs for stars
in the mass range $\approx1<M/\Msun<8$ (depending on metallicity) and
is difficult to model for several reasons, including the fact that
nuclear burning occurs in alternate hydrogen-rich and helium-rich
shells.  When the helium shell burns it does so explosively because of
the thin shell instability \citep{Schwarzschild68}, which gives rise
to the thermal pulses.  Mass-loss becomes catastrophic during this
phase, thereby terminating the life of the star.  Recently the Padova
group has developed a new suite of isochrones with updated TP-AGB
models calibrated against observations in the Large Magellanic Cloud
\citep[LMC][]{Marigo07a, Marigo08}.  However, \citet{Conroy10c} found
that the updated Padova models failed to reproduce the colors of
intermediate-age Magellanic Cloud star clusters, by as much as 0.5 mag
in some cases.  These authors provided recalibrated SPS models in
which the weight given to TP-AGB stars was reduced in order to match
the LMC data.  More recently, \citet{Melbourne12a} analyzed resolved
color-magnitude diagrams in nearby dwarf galaxies and concluded that
these updated models produce twice as much luminosity in the TP-AGB
phase as observed, roughly independent of the inferred mass fraction
in young stars.  In addition, \citet{Melbourne12a} found that the
latest Padova model predictions for the luminosity contributed by red
core helium burning stars is a factor of two higher than observed in
dwarf galaxies.  This problem in the models may be related to the
treatment of convection in the envelopes of evolved giants.

Mass-loss is another critical parameter in stellar evolution models.
At $M\lesssim8\,\Msun$, it determines when a star will end its life as
a white dwarf, and how massive the white dwarf will be.  At higher
masses mass-loss can significantly alter the course of advanced
evolutionary phases, especially at $M>40\,\Msun$.  In high mass stars
the mass loss mechanism is thought to be via line-driven winds, while
in lower mass stars ($\lesssim8\,\Msun$) it is believed to be due to
pulsation-induced dust-driven winds \citep{Willson00}.  In any event,
the mass-loss prescription is another free parameter in these models,
and it is a critical one because it strongly affects the lifetimes of
advanced (and luminous) evolutionary phases.  For example, the
lifetime of TP-AGB stars and the number of thermal pulses they undergo
depend strongly on the mass-loss prescription \citep{Ventura10b}.

In summary, the computation of isochrones for use in SPS depends on
many uncertain aspects of stellar evolution including the treatment
of convection, close binary evolution, rotation effects, and
mass-loss, amongst many others.  The importance of these uncertainties
on derived SPS results will be highlighted throughout this review.

\subsubsection{Stellar Remnants} 

Stars eventually die, usually leaving behind stellar remnants in the
form of white dwarfs, neutron stars, or black holes \citep[theory
predicts that a certain class of very massive metal-poor stars
undergoes pair-instability supernovae that leave behind no stellar
remnant;][]{Heger03}.  The relation between the initial, zero-age main
sequence stellar mass and the final remnant mass is not
well-constrained observationally, especially for massive stars.  The
initial-final mass relation for white dwarfs can be reasonably well
constrained by measuring white dwarf masses in open clusters with
known ages \citep[e.g.,][]{Kalirai08}.  The initial-final mass
relation is predicted to be a function of metallicity
\citep{Marigo01b, Heger03}, further complicating the situation.

In SPS models stellar remnants are usually included in the total
stellar mass budget, and their contribution can be significant.  For
example, if a certain amount of mass is formed into stars instantly,
then after 13 Gyr only 60\% of the mass remains in stars or stellar
remnants; the other 40\% has returned to the interstellar medium.  Of
the remaining stellar mass, 25\% is locked in stellar remnants.  Of
the total remnant mass, 73\% is comprised of white dwarfs, 7\% is in
neutron stars, and 20\% is in black holes.  These numbers were
computed assuming solar metallicity and a \citet{Kroupa01} IMF, with
the initial-final mass relations from \citet{Renzini93}.  The
\citet{Maraston05} SPS model predicts similar numbers (largely because
the same initial-final mass relation was used).  The point is that
stellar remnants can make an important contribution to the total
stellar mass of a system.

\subsubsection{Stellar Spectral Libraries}

Stellar spectral libraries are required to convert the outputs of
stellar evolution calculations --- surface gravities, $g$, and
effective temperatures, $T_{\rm eff}$ --- as a function of
metallicity, $Z$, into observable SEDs.  There is however no single
spectral library, whether theoretical or empirical, that covers the
entire range of parameter space necessary for constructing SPS models.
Stitching together various libraries, often of widely varying quality,
is therefore necessary.  Some modelers take the approach of
constructing spectral libraries entirely from theoretical
calculations, while others use purely empirical libraries.  The
benefits and drawbacks of these two approaches will be discussed in
this section.

\paragraph{Theoretical Libraries}

Theoretical libraries offer the great advantage of densely covering
parameter space, including spectral resolution, and of producing
spectra that are not subject to observational issues such as flux
calibration and atmospheric absorption.  The clear disadvantage is
that the libraries are only as good as the input atomic and molecular
parameters and the approximations made in the computation of the
models, as discussed below.

There are a number of decisions that must be made when computing
synthetic spectral libraries including how to treat convection and the
microturbulent velocity profile, and whether or not to model
departures from LTE and plane-parallel geometries.  Additional
important limitations include the incomplete and often inaccurate
atomic and molecular line lists.  As emphasized by \citet{Kurucz11},
models still fail to reproduce all the observed features in ultra
high-resolution spectra of the Sun due to incomplete atomic line
lists.  The situation is even more serious for cooler stars because
the molecular line lists can carry fairly large uncertainties, in
particular TiO \citep{Allard11}.  A related problem is that a
significant fraction of the atomic and molecular lines are derived
from theoretical calculations, rather than measured in the laboratory,
and so they have uncertain strengths and central wavelengths.
Unfortunately, these `predicted lines' can be important for
determining the overall SED shape of stars and hence of galaxies.
Worse still, the partition function for many molecules is poorly
known, so even the total abundance of particular molecules is
uncertain.

The quality and state of the art of the theoretical models varies
considerably as a function of $\teff$.  At the high temperature end,
i.e., Wolf-Rayet and O-type main sequence stars, the state-of-the-art
libraries are from \citet{Smith02} and \citet{Lanz03}, while for
hotter compact objects (e.g., post-AGB stars) the most up-to-date
models are from \citet{Rauch03}.  By far the widest range in parameter
space is covered by the BaSeL atlas \citep{Lejeune97, Lejeune98,
  Westera02}.  This library is comprised of a variety of theoretical
models from Kurucz (1995), \citet{Fluks94}, \citet{Bessell89b,
  Bessell91}, and \citet{Allard95}.  The broadband SEDs of the stars
in this library have been modified to agree with observed $UBVRIJHKL$
color--temperature relations for individual stars.  The BaSeL atlas is
almost universally used in modern SPS codes, despite the fact that the
input theoretical spectra are now almost 20 years old.  At the very
coolest temperatures the MARCS and NEXTGEN/PHOENIX grids are the
state-of-the-art because of their comprehensive molecular line lists
and treatment of spherical effects on the atmospheres and spectra of
cool giants.  Finally, the \citet{Aringer09} models fill out the low
temperature end by providing carbon star spectra over a range of
parameter space.

There are several modern spectral libraries computed by
\citet{Munari05}, \citet{Martins05}, and \citet{Coelho05} that offer
fairly wide coverage in parameter space and are at high spectral
resolution.  The \citet{Martins05} library has even been incorporated
into a fully theoretical high resolution SPS model
\citep{GonzalezDelgado05}.  However, these libraries appear to be
geared toward high resolution spectroscopic analyses because they do
not include the predicted lines.  They therefore cannot be used to
model broadband SEDs, or any spectral region that contains a
significant contribution from predicted lines.  However, progress can
be made by making corrections to these high resolution libraries for
the effect of predicted lines on the broadband SED, as for example
done in \citet{Coelho07}.

\begin{figure}
\centerline{\psfig{figure=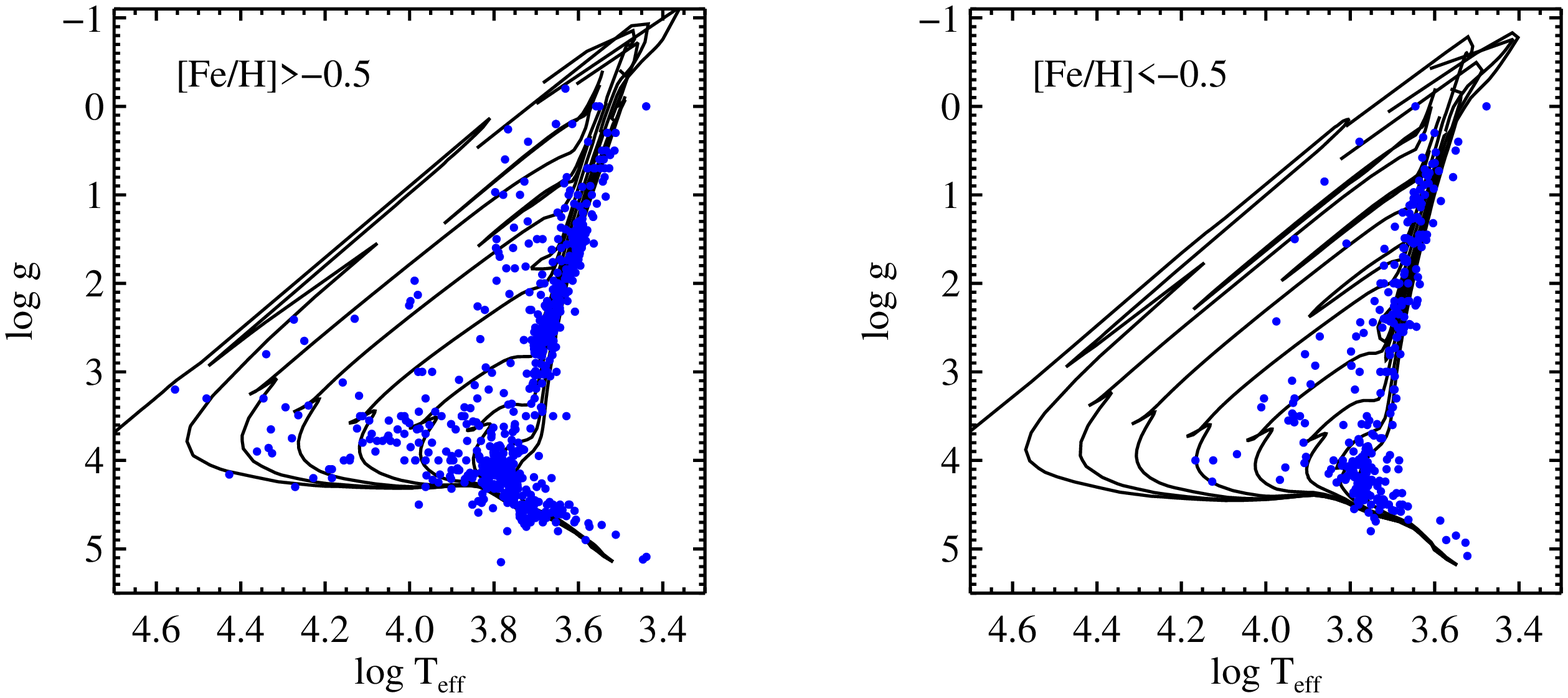,height=3.0in}}
\caption{Physical properties of the MILES empirical spectral library,
  separated by [Fe/H] (solid symbols).  Isochrones from the Padova
  group are overplotted for $t=3\times10^6-10^{10}$ yr.  Typical
  errors on $\teff$ and $\logg$ are $60-100$ K and $0.2$ dex,
  respectively \citep{Cenarro07}.  Notice that the MILES library
  covers the lower main sequence and RGB both at high and low
  metallicity, but it only sparsely covers the upper main sequence and
  supergiants.  This figure highlights the difficulty in constructing
  SPS models based on empirical stellar libraries, especially at low
  metallicity.}
\label{f:miles}
\end{figure}

\paragraph{How Accurate are the Theoretical Libraries?}

Normally, observed stars are assigned physical parameters based on
comparison to models.  However, if one wants to constrain the models
with observed stellar data then one requires an independent estimate
of the physical properties of stars.  This is a significant obstacle
to assessing the accuracy of the models.  One can obtain essentially
model-independent estimates of $\teff$ via angular diameter
measurements, but these are very difficult to obtain for giants, let
alone main sequence stars \citep{Perrin98, Boyajian12a}.  A more
widely used technique is the infrared flux method (IRFM), first
proposed by \citet{Blackwell77}.  The basic idea is that the flux of a
star depends only weakly on $\teff$ in the IR, and so one can deduce
angular diameters from fluxes alone with only a weak model dependence.
This technique underpins nearly all existing color-$\teff$ relations
\citep[e.g.,][]{Alonso96, Ramirez05, Casagrande10}.

The latest generation of model spectra succeed in reproducing the
broadband colors for FGK dwarfs and warm giants \citep{Bertone04,
  Kucinskas05, Martins07, Boyajian12b}.  Not surprisingly, the models
continue to have difficultly fitting the SEDs of the coolest stars and
the flux in the Wien tail of the flux distribution (typically
$\lesssim5000$\,\AA\, for types G and later).  The latter difficulty
arises because the Wien tail is extremely sensitive to $\teff$ and so
even minor changes to the model assumptions lead to large changes in
that spectral region.  In addition, metal line blanketing becomes very
strong in the blue/UV, and so the requirements on the line lists are
demanding.  \citet{Martins07} explored the ability of three modern
synthetic libraries to reproduce a variety of spectral features of
observed stars.  Overall the agreement was found to be satisfactory,
although there were notable deficiencies.  Some models were unable to
reproduce the observed CH features in cool stars, while other models
failed to match the observed Mg{\it b} and MgH features.  The majority
of the model shortcomings were restricted to stars with $\teff<4500$
K.  Modern models also have difficulty reproducing the strength of the
TiO bands in cool stars \citep{Kucinskas05, Allard11}.

The situation regarding the H$_2$O band strengths in the NIR should
serve as a cautionary tale: models have for decades predicted water
bands in cool stars that are stronger than observations.  Part of the
problem seems to have been the molecular line lists, which improved
steadily over the years, culminating in the \citet{Barber06} line
list, which contains 500 million transitions.  However,
\citet{Allard11} demonstrated that the recent revision of the solar
abundance scale by \citet{Asplund09}, which entails a factor of two
reduction in the oxygen abundance, has a much larger effect on the
predicted strength of the water bands.  Their new models with updated
solar abundances reproduce remarkably well the NIR colors of M dwarfs,
which have strong water band features.  Thus, not only the atomic and
molecular parameters but also the abundance patterns can have a very
significant effect on theoretical spectra.

\paragraph{Empirical Libraries}

The strengths and weaknesses of the theoretical libraries are the
weaknesses and strengths of the empirical libraries.  Empirical
spectra of course do not suffer from issues with line lists, treatment
of convection, etc., but they are plagued by standard observational
constraints such as correction for atmospheric absorption, flux
calibration, and limited wavelength coverage and spectral resolution.
Worse, the empirical libraries are woefully incomplete in their
coverage of parameter space.  This is a long-standing issue that is
difficult to address owing to the fact that empirical libraries are
drawn from samples of stars in the solar neighborhood.  For example,
hot main sequence stars at low metallicity are very rare, as are stars
in rapid phases of stellar evolution such as WR and TP-AGB stars.

One of the first comprehensive empirical stellar spectral libraries
was constructed by \citet{Gunn83}.  Later optical/NIR libraries
included \citet{Pickles98}, Jones (1999), ELODIE \citep{Prugniel01},
STELIB \citep{LeBorgne03}, Indo-US \citep{Valdes04}, NGSL
\citep{Gregg06, Heap11}, MILES \citep{Sanchez-Blazquez06}, IRTF
\citep{Rayner09}, and the X-shooter library \citep{XSL1}.  Other,
specialized libraries include an ultraviolet atlas compiled from IUE
data \citep{Fanelli92}, a TP-AGB library \citep{Lancon00}, and the
{\it Spitzer Space Telescope} IR stellar library \citep{Ardila10}.
The Sloan Digital Sky Survey (SDSS) has obtained spectra of several
hundred thousand stars and has devoted considerable effort to
measuring stellar parameters from the spectra \citep{LeeYS08}.  As yet
there is no official stellar spectral library based on the SDSS stars,
but it would clearly provide a great leap forward both in terms of
coverage in parameter space and uniformity in spectral quality.

An example of the difficulties posed by empirical libraries for SPS
model construction is shown in Figure \ref{f:miles}.  This figure
shows the location in the HR diagram of all stars from the MILES
spectral library, with stellar parameters determined by
\citet{Cenarro07}.  MILES is the empirical library today that provides
the greatest coverage in terms of $\logg$, $\teff$, and [Fe/H] with a
total of 985 stars.  In this figure the metal-rich and metal-poor
stars are split into two panels.  Overplotted are isochrones from the
Padova group for ages ranging from $10^{6.6}-10^{10.1}$ yr.  While the
lower main sequence and red giant branch are well covered by the
empirical library over a wide range in metallicity, hotter stars are
rare, especially at lower metallicity.  Because the HR diagram is
covered so sparsely in this regime, constructing SPS models based on
empirical libraries at young ages ($\lesssim1$ Gyr) is clearly
challenging.

A second, related problem with empirical libraries is irregular
coverage in the HR diagram.  Interpolation within the library,
necessary for SPS model construction, can be difficult.
\citet{Vazdekis10} has developed a sophisticated algorithm to deal
with this issue.  These authors assign quality numbers to the
resulting SPS models based on the number of empirical spectra
comprising each model.  This is a profitable approach that should be
standard in the field because it allows users to assess the
reliability of various regions of parameter space.

A third problem associated with empirical libraries is the assignment
of physical parameters to stars, including $\logg$, $\teff$, and
[Fe/H].  Of particular note is the $\teff$ determination, which
carries an uncertainty of order 100 K \citep{Alonso96, Ramirez05,
  Casagrande10}.  Using theoretical models, \citet{Percival09}
explored the impact of uncertainties in $\teff$, $\logg$, and [Fe/H]
for individual stars on SPS models.  They found that for models with
old ages ($>4$ Gyr), the effect of changing the temperature scale by
100 K had significant effects ($0.1-0.4$\,\AA in EW in the most
extreme cases) on a variety of spectral absorption features, including
the hydrogen Balmer lines and several commonly employed iron and
magnesium lines used to interpret the stellar populations of
early-type galaxies.  The work of Percival \& Salaris was mostly
exploratory, and it would be desirable to consider the full
propagation of these sorts of uncertainties into the derived
properties of galaxies.

Yet another issue with empirical libraries is the abundance patterns
of the stars.  It is well-known that low metallicity stars in the
Galaxy tend to be $\alpha-$enhanced, such that [Mg/Fe]$\approx0.0$ at
[Fe/H]$\approx0.0$ and [Mg/Fe]$\approx0.4$ at [Fe/H]$<-1.0$
\citep{Edvardsson93}.  Thus, any model based on empirical stars at low
metallicity must somehow correct for this [$\alpha$/Fe] `bias' in the
models \citep[see e.g.,][]{Thomas03, Schiavon07}.

\begin{figure}
\centerline{\psfig{figure=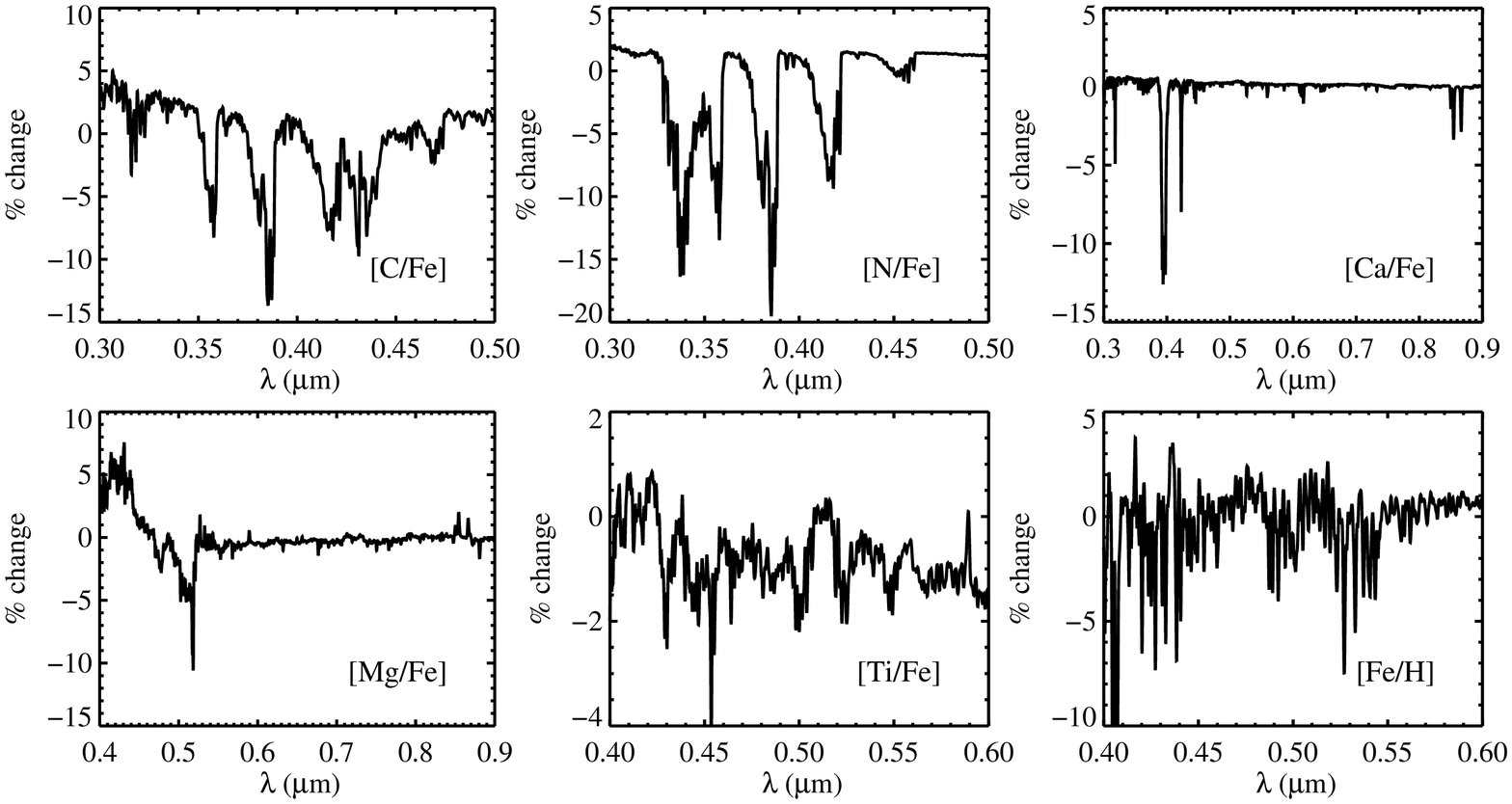,height=3.5in}}
\caption{Variation in the spectrum of a 13 Gyr SSP due to changes in
  individual elemental abundances.  All abundance changes are $+0.3$
  dex except for C and Ca which are varied by $+0.15$ dex.  Spectra
  have been broadened to a velocity dispersion of $150\kms$. Notice
  the different $x$ and $y$-axes in each panel.  Prominent spectral
  features include the CH, CN, NH, C$_2$, \ionn{Ca}{i}, \ionn{Ca}{ii}
  H\&K, \ionn{Ca}{ii} triplet, MgH, and Mg {\it b} (\ionn{Mg}{i})
  features, along with numerous atomic Fe and Ti features.  Computed
  from the models of \citet{Conroy12a}.}
\label{f:specabund}
\end{figure}

\paragraph{Variable Abundance Patterns}

Moderate resolution spectra (i.e., $R\sim1000-5000$) contain a wealth
of information on the detailed abundance patterns of the stellar
populations.  In the context of SPS models, abundance ratios have
historically been estimated through the analysis of spectral indices.
Ideally, an index is sensitive to a single feature (such as an atomic
absorption line or a molecular band head).  One usually defines a
feature bandpass and one or two pseudocontinua, and then an equivalent
width (EW), or in some cases a magnitude, can be measured.  The
Lick/IDS system is the most popular index system, defining 21 indices
in the wavelength range $4000-6500$\,\AA\, \citep{Burstein84,
  Worthey94b}.  Other index systems have been defined at other
wavelengths \citep[e.g.,][]{Fanelli92, Alvarez00, Cenarro01,
  Serven05}.  In practice, the use of indices is greatly complicated
by the fact that there are rarely if ever clean regions of the
spectrum from which one can estimate the continuum level.  The
strength of each index thus depends not only on the feature of
interest but also the features in the pseudocontinua.

\citet{Tripicco95} were the first to assess the sensitivity of the
Lick/IDS indices to variation in the abundances of 10 elements with
theoretical spectral models.  \citet{Korn05} provided an update to
these `response functions' with updated line lists and transition
probabilities, and for a range in metallicity.  \citet{Serven05}
considered variation in 23 elements, including several neutron-capture
elements.  These authors defined new indices sensitive to a host of
elements not considered in the earlier work.  The Serven models were
based on theoretical spectra of only two stars per abundance pattern,
a turnoff star and a luminous giant, and the model effectively had a
fixed age of $\sim5$ Gyr.  \citet{LeeHC09} provided response functions
for Lick indices by employing a larger number of synthetic spectra.
\citet{Conroy12a} produced theoretical stellar spectra with variation
in 11 elements.  These models contained theoretical spectra at 20
points along the isochrone, for ages from $3-13.5$ Gyr.  Rather than
focusing on spectral indices, these models made predictions for the
response of the full spectrum from $3500$\,\AA$-2.4\,\mu m$ to
abundance variations.  Sample response spectra from that work are
shown in Figure \ref{f:specabund}.  In this figure the relative
response of the spectrum to an increase in the abundance of a
particular element is shown.  The full spectrum is rich in diagnostic
features beyond what any index system can capture.

To construct models with arbitrary abundance patterns one would want
to create synthetic spectra for each possible abundance pattern.  When
considering large numbers of elements, such an approach becomes
computationally infeasible.  Instead, the standard technique is to
create arbitrary abundance patterns by treating the effect of each
element on the spectrum as being independent of the other elements
\citep[e.g.,][]{Thomas03, Schiavon07, LeeHC09, Conroy12a}.  This is a
reasonable assumption for trace elements, but it is less realistic for
elements such as C, N, O, Na, Ti, and Fe, which affect the opacity,
free electron density, and/or molecular equilibrium.  There has been
surprisingly little work in the literature testing this assumption.

\begin{figure}
\centerline{\psfig{figure=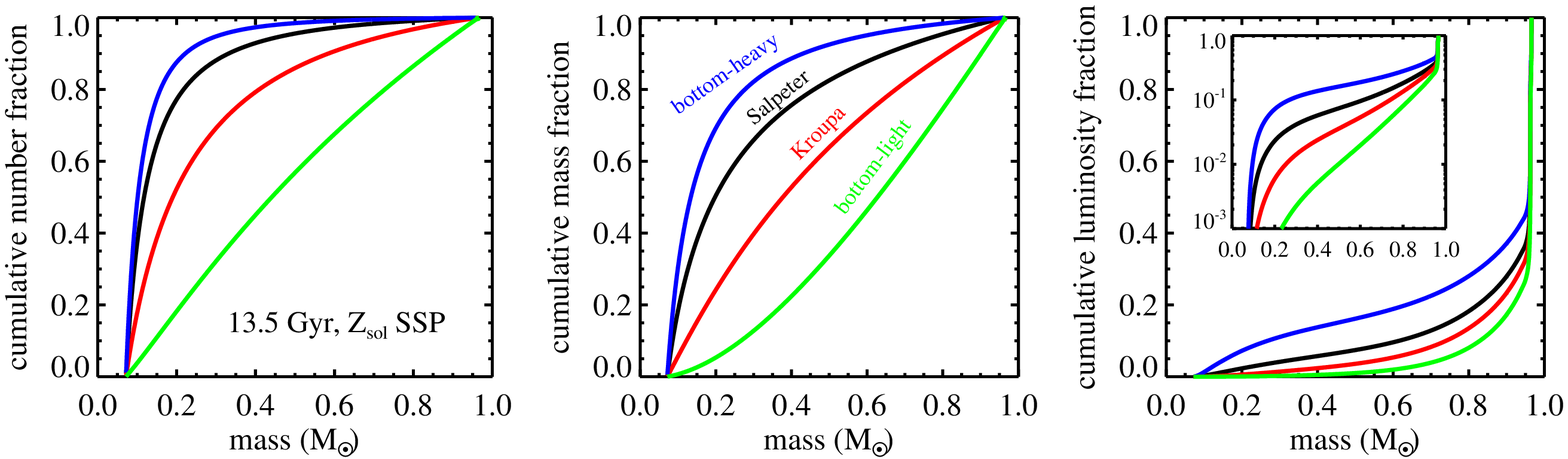,height=1.7in}}
\caption{Fractional contribution to the total number, mass, and
  bolometric luminosity as a function of stellar mass for a 13.5 Gyr
  solar metallicity model.  Lines correspond to different IMFs: a
  bottom-heavy with logarithmic slope $x=3.0$ (blue line); Salpeter
  ($x=2.35$; black line); MW IMF (specifically a Kroupa IMF; red
  line); a bottom-light IMF (specifically of the form advocated by van
  Dokkum (2008); green line).  The inset in the right panel shows the
  cumulative luminosity fraction in logarithmic units.  Low mass stars
  dominate the total number and mass in stars, but contribute a tiny
  fraction of the luminosity of old stellar populations.}
\label{f:fracflux}
\end{figure}

\subsubsection{The IMF}

The initial distribution of stellar masses along the main sequence,
known as the stellar initial mass function (IMF), has been studied
extensively for decades \citep[e.g.,][]{Salpeter55, Scalo86, Scalo98,
  Kroupa01, Chabrier03}.  As emphasized in a recent review by
\citet{Bastian10}, there is no compelling evidence for variation in
the IMF from direct probes, e.g., star counts.  The canonical Salpeter
IMF has the form $dN/dM\propto M^{-x}$ with $x=2.35$.  The IMF
measured in the solar neighborhood deviates from the Salpeter form
only at $M<1\,\Msun$, where $x$ becomes shallower \citep{Kroupa01,
  Chabrier03}.

From the perspective of SPS, the IMF (1) determines the overall
normalization of the stellar mass-to-light ratio, $M/L$; (2)
determines the rate of luminosity evolution for a passively evolving
population; (3) affects the SED of composite stellar populations; (4)
has a small effect on the shape of the SED of single stellar
populations.  The last point is due to the fact that the integrated
light of a coeval population is overwhelmingly dominated by stars at
approximately the same mass, i.e., the turnoff mass.  (4) will not
hold for IMFs that depart dramatically from the Salpeter IMF.  (3)
arises because composite populations have a range of turnoff masses
that contribute to the integrated light.  The fractional contribution
of stars of various masses to the total number of stars, stellar mass,
and bolometric luminosity is shown in Figure \ref{f:fracflux}.  This
figure demonstrates quantitatively that low mass stars dominate the
stellar mass and number of stars in a galaxy, but contribute only a
few percent to the bolometric light of an old stellar population.  At
younger ages the light contribution from low mass stars is even less.

\citet{Tinsley80} demonstrated that the evolution in $M/L$ for a
passively evolving stellar population is sensitive to the logarithmic
slope of the IMF, $x$, at the main sequence turnoff point.  The
logarithmic evolution of the luminosity per logarithmic time
(dln$L$/dln$t$) scales linearly with $x$, at least for plausible
values of $x$ \citep[see also][]{vanDokkum08, Conroy09a}.  This
dependency arises because the giant branch dominates the luminosity
for all plausible values of $x$, and so the IMF determines the rate at
which the giant branch is fed by turnoff stars.  Steeper IMFs imply
that the giant branch is more richly populated with time, and
therefore the natural luminosity dimming is reduced.  For sufficiently
steep IMFs (e.g., $x\gtrsim5$), the unevolving dwarfs would dominate
the light, and so the integrated luminosity would be approximately
constant over time.

It is somewhat less-well appreciated that the IMF above $1\,\Msun$
also strongly affects the shape of the SED of composite stellar
populations.  In composite populations the SED is influenced by stars
with a range of masses (see Section \ref{s:csp}), and so the IMF must
play an important role \citep[see e.g.,][]{Pforr12}.

\subsection{Dust}

Interstellar dust is a component of nearly all galaxies, especially
those that are actively star-forming.  Dust plays a dual role in SPS,
both as an obscurer of light in the UV through NIR and as an emitter
of light in the IR.  For both historical and theoretical reasons these
two aspects are often modeled independently of one another. Indeed,
these two aspects are sensitive to different properties of a galaxy
(e.g., dust obscuration is highly sensitive to geometry while dust
emission is more sensitive to the interstellar radiation field), and
so it is not unreasonable to decouple the modeling of these two
components.

\subsubsection{Attenuation}

Dust grains obscure light by both absorbing and scattering starlight.
From observations of individual stars one can infer the total
extinction along the line of sight by comparing an observed spectrum
to the expected spectrum of the star in the absence of dust (the
latter is typically obtained from models in conjunction with an
estimated $\teff$ and $\logg$ of the source).  Thus, extinction
measures the total loss of light along a single line of sight.
Observations in the Milky Way (MW) and Large and Small Magellanic
Clouds (LMC and SMC, respectively) have been obtained for a number of
sightlines, enabling construction of average extinction curves for
these galaxies \citep{Cardelli89, Pei92, Gordon98}.  In the MW and LMC
the only striking feature in the extinction curve is the broad
absorption at 2175\,\AA.  This feature is absent in three out of four
sightlines to the SMC, and it is weaker in the LMC than in the MW.
The grain population responsible for this feature is not known, but
polycyclic aromatic hydrocarbons (PAHs) are a leading candidate
\citep{Draine03}.  In general, dust extinction is a consequence of the
optical properties of the grains and the grain size and shape
distribution \citep{Weingartner01}.

When modeling SEDs of galaxies, the relevant concept is dust
attenuation, which differs from extinction in two important respects:
(1) light can be scattered both out of and into a given line of sight;
(2) the geometrical distribution of dust with respect to the stars
strongly affects the resulting SED \citep[see][for an extensive
discussion of these issues]{Calzetti01}.  The total dust attenuation
in a galaxy can be estimated by analogy with how one estimates dust
extinction: a spectrum of a galaxy is obtained and compared with the
expected spectrum of the same galaxy in the absence of dust.  For
obvious reasons, estimating dust attenuation is considerably more
complex than estimating dust extinction.

Although the shape of the dust attenuation curve depends on the
star-dust geometry, grain size distribution, etc., in a complex
manner, several general rules of thumb can be stated \citep[see
e.g.,][for a more thorough discussion]{Witt00}.  The simplest dust
geometry, that of a homogenous foreground screen, yields an
attenuation curve whose shape depends only weakly on the total dust
column density (the weak dependence arises from the scattered light
component).  More complex geometries generally yield attenuation
curves that become greyer (i.e., shallower) as the column density
increases.  Clumpy interstellar media also result in greyer
attenuation curves than their homogenous counterparts.  In all cases,
the total attenuation optical depth is less than the amount of
extinction that would be produced by the same column density because
some scattered light is always returned to the line-of-sight.
Finally, because the 2175\,\AA\, dust feature is believed to be due to
pure absorption, the effects of radiative transfer will cause this
feature to respond differently to geometrical effects compared to the
rest of the attenuation curve.  The use of a single dust attenuation
curve for analyzing a wide range of SED types is therefore without
theoretical justification.

In practice, most SPS modelers include dust attenuation by fixing the
shape of the attenuation curve and fitting for the normalization.
Popular attenuation curves include the Calzetti law, the MW, LMC, and
SMC extinction curves, or the time-dependent attenuation model from
\citet{Charlot00}.  The impact of the chosen attenuation curve on
derived properties of galaxies will be discussed in later sections.

\subsubsection{Emission}

Beyond $\lambda\sim10\,\mu m$ the SEDs of normal galaxies are dominated
by emission from dust grains.  \citet{Mathis77} were the first to
postulate that the dust grain population is comprised of both silicate
and carbonaceous grains.  In modern theories, the carbonaceous grains
are assumed to be PAHs when they are small, and graphite when they are
large \citep{Draine03}.  The observed dust emission spectrum results
from exposing these grains to a range of interstellar radiation field
strengths.

A variety of models exist that combine grain size distributions and
grain optical properties with models for starlight (or simply the
radiation field) to predict IR emission \citep{Desert90, Silva98,
  Devriendt99, Popescu00, Dale01, Piovan06, Jonsson06, Draine07b,
  Groves08, Popescu11}.  The modeling of PAH emission features (the
most prominent being at $3.6\,\mu m$, $6.2\,\mu m$, $7.7\,\mu m$,
$8.6\,\mu m$, and $11.3\,\mu m$) has become dramatically more
sophisticated since the early attempts by \citet{Desert90},
culminating in state-of-the-art models by \citet{Draine07b}.  At long
wavelengths ($\lambda>50\,\mu m$) the emission is dominated by grains
at a nearly steady temperature of $\sim15-20$ K and contributes
$\sim2/3$ of the total IR luminosity.  At shorter wavelengths the IR
emission arises from single photon heating of dust grains (including
PAHs) and accounts for the remaining $\sim1/3$ of the total IR
emission \citep[see the review by][for details]{Draine03}.  The IR
emission at the shortest wavelengths ($\lambda<12\,\mu m$) is supplied
almost entirely by PAHs. In detail these relative contributions will
depend on the grain composition and size distribution and the
interstellar radiation field.

The models listed above are not always well-suited for interpreting
large numbers of observed IR SEDs because they contain a large number
of parameters, require knowledge of the star-dust geometry, and/or
require radiative transfer calculations.  Simpler models for dust
emission have therefore been developed in parallel to the more complex
ones.  The IR templates of \citet{Chary01} and \citet{Dale01} are
widely used for estimating bolometric luminosities and
$k-$corrections.  These templates are based on sophisticated models,
but are constrained to match observations of normal star-forming and
starburst (i.e., ULIRG) galaxies.  The resulting templates are
functions of only one variable.  For example, the \citet{Chary01}
templates are a function of the bolometric IR luminosity.
\citet{daCunha08} have developed a simple phenomenological model for
dust emission that consists of a series of modified blackbodies for
the thermal dust emission and for the emission from stochastically
heated dust grains.  In addition, they include an empirical spectrum
of M17 to represent the PAH emission.

\subsubsection{Dust Around AGB Stars}

Stars experience very high mass-loss rates as they climb up the AGB
(as high as $10^{-4}\Msun$ yr$^{-1}$ in the superwind phase). The mass
lost is often observed to be dust rich \citep{Bedijn87}.  These stars
are often heavily dust obscured in the optical and emit copiously in
the IR; they are often so dusty that their SEDs peak at $\sim10\,\mu
m$ \citep{Bedijn87}.  Dust around AGB stars is important from the
standpoint of SPS for two reasons: it will diminish the importance of
AGB light in the optical and NIR, and it will contribute additional
flux in the mid-IR beyond what would be expected from standard dust
models \citep[see e.g.,][]{Kelson10}.

Despite the obvious importance of AGB dust, this aspect is included in
few SPS models.  Notable exceptions include the models of
\citet{Bressan98}, \citet{Silva98}, \citet{Piovan03}, and
\citet{Gonzalez10}, which are all based on theoretical models for AGB
stars, their mass-loss rates, and dust formation and composition.
Even the use of currently available empirical libraries of AGB stellar
spectra require reddening the spectra to account for circumstellar
dust \citep{Lancon02}.  Increasingly sophisticated radiative transfer
models of dusty circumstellar envelopes are being developed
\citep[e.g.,][]{Groenewegen12, Sargent11, Srinivasan11}.  The outputs
from these models should be incorporated into SPS codes as a standard
practice.

\begin{figure}
\centerline{\psfig{figure=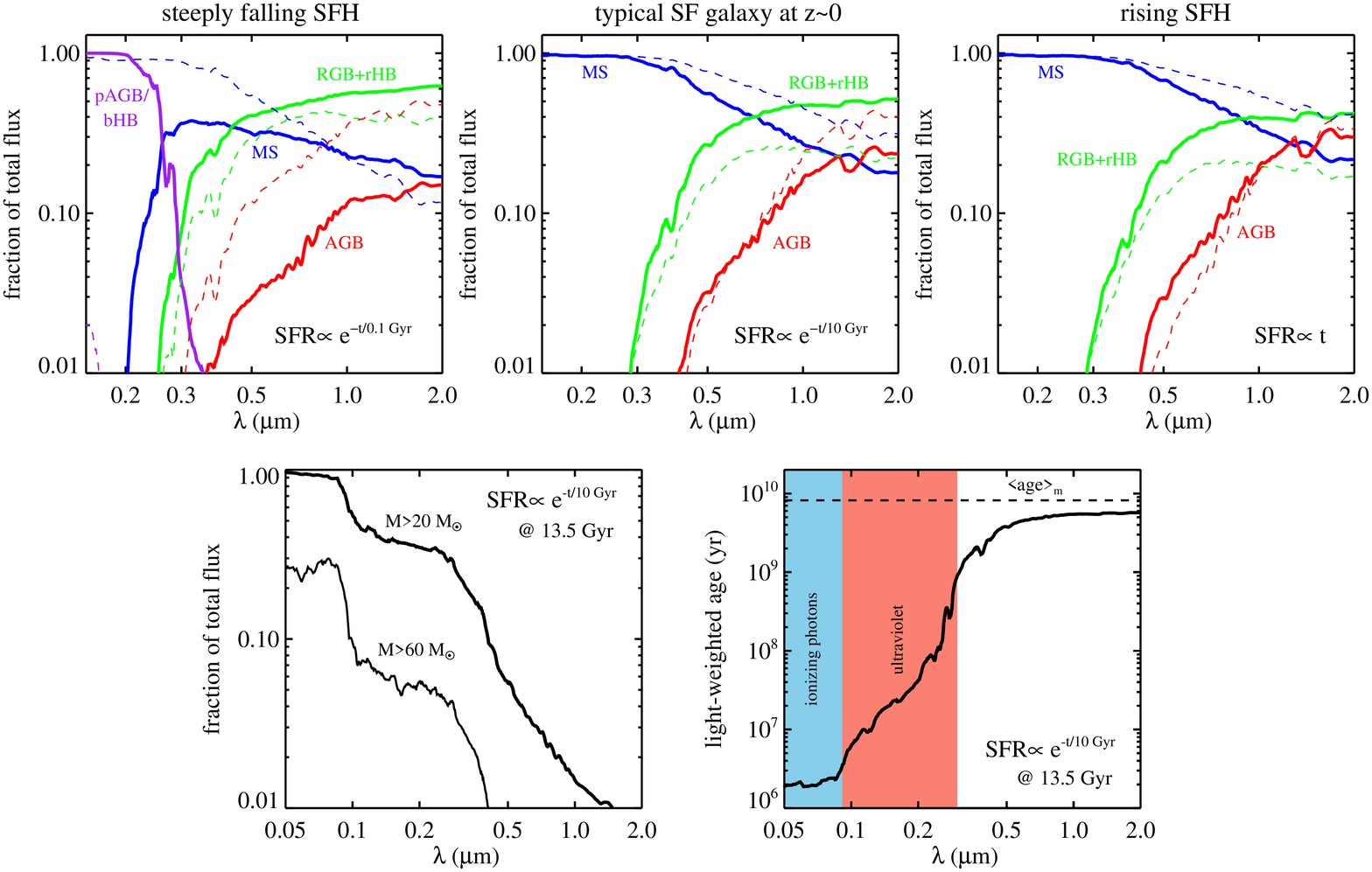,height=4.5in}}
\caption{Top Panels: Fractional contribution to the total flux from
  stars in various evolutionary phases, for three different SFHs.  The
  left panel is representative of a galaxy that formed nearly all of
  its stars very rapidly at early times, the middle panel is
  representative of a typical star-forming galaxy at $z\sim0$, and the
  right panel may be representative of the typical galaxy at high
  redshift.  Flux contributions are at 13 Gyr (solid lines) and 1 Gyr
  (dashed lines) after the commencement of star formation; all models
  are solar metallicity, dust-free, and are from FSPS
  \citep[v2.3;][]{Conroy09a}.  Labeled phases include the main
  sequence (MS), red giant branch (RGB), asymptotic giant branch (AGB,
  including the TP-AGB), post-AGB (pAGB), and the blue and red
  horizontal branch (bHB and rHB).  Bottom Left Panel: Fractional flux
  contributions for stars more massive than $20\,\Msun$ and
  $60\,\Msun$ for the SFH in the middle panel of the top row.  Bottom
  Right Panel: Light-weighted age as a function of wavelength for the
  same SFH.  The dashed line indicates the corresponding mass-weighted
  age.}
\label{f:ev}
\end{figure}

\subsection{Composite Stellar Populations}
\label{s:csp}

The simple stellar populations discussed in Section \ref{s:ssp} are
the building blocks for more complex stellar systems.  Composite
stellar populations (CSPs) differ from simple ones in three respects:
(1) they contain stars with a range of ages given by their SFH; (2)
they contain stars with a range in metallicities as given by their
time-dependent metallicity distribution function, $P(Z,t)$; (3) they
contain dust.  These components are combined in the following way:
\noindent
\be
\label{e:csp}
f_{\rm CSP}(t) = \int_{t'=0}^{t'=t}\int_{Z=0}^{Z_{\rm max}} 
\bigg({\rm SFR}(t-t')\,P(Z,t-t')\,f_{\rm SSP}(t',Z)\,e^{-\tau_d(t')} +
A\,f_{\rm dust}(t',Z)\bigg)\,{\rm
  d}t'\,{\rm d}Z 
\ee
\noindent
where the integration variables are the stellar population age, $t'$,
and metallicity, $Z$.  Time-dependent dust attenuation is modeled via
the dust optical depth, $\tau_d(t')$ and dust emission is incorporated
in the parameter $f_{\rm dust}$.  The normalization constant $A$ is
set by balancing the luminosity absorbed by dust with the total
luminosity re-radiated by dust.

The SFH can in principle be arbitrarily complex, although simple forms
are usually adopted.  By far the most popular is the exponential, or
$\tau-$model, where SFR $\propto e^{-t/\tau}$.  This form arises
naturally in scenarios where the SFR depends linearly on the gas
density in a closed-box model \citep{Schmidt59}.  Recently, rising
SFHs have become popular to explain the SEDs of high-redshift galaxies
\citep{Maraston10, Papovich11}.  Rising SFHs at early times seem to be
a natural consequence of galaxy evolution in a hierarchical universe
\citep{Finlator07, LeeSK10}.  Functional forms that incorporate an
early phase of rising SFRs with late-time decay, such as SFR $\propto
t^\beta\, e^{-t/\tau}$, may therefore become more popular amongst
modelers.

The treatment of metallicity in composite stellar populations is
usually even more simplistic than the treatment of SFRs.  The widely
adopted simplification is to replace $P(Z,t)$ in Equation \ref{e:csp}
with a $\delta-$function.  In other words, a single metallicity is
assumed for the entire composite population.  The impact of this
simplification on the SPS modeling procedure has not been extensively
explored.

The approach to Equation \ref{e:csp} outlined above is standard but
not universal.  A notable exception is the technique of fitting
non-parametric SFHs and metallicity histories in either a pre-defined
or adaptive set of age bins \citep{CidFernandes05, Ocvirk06,
  Tojeiro09}.  The latter approach is computationally expensive since
as many as thirty parameters are simultaneously fit to the data.  Very
high quality data are also a prerequisite for non-parametric
techniques.  Nonetheless, they offer the promise of less biased
reconstruction of the SFH and metallicity history of galaxies based
solely on their SEDs.

Figure \ref{f:ev} presents some basic properties of CSPs.  The top
panels show the fractional flux contribution from stars in different
evolutionary phases for three representative SFHs.  At late times the
RGB and red HB dominate the red and NIR flux of $\tau$-model SFHs, as
is well-known.  However, at young ages and/or for rising SFHs, the red
and NIR are also influenced strongly by AGB stars, suggesting that the
NIR can be susceptible to large uncertainties (due to uncertainties in
the modeling of the AGB, as discussed in detail in later sections).
The lower left panel shows the fractional flux contribution from stars
more massive than $20\,\Msun$ and $60\,\Msun$ for a $\tau=10$ Gyr SFH.
Massive star evolution is uncertain due to the complicating effects of
rotation and binarity, and so this panel provides a rough sense of the
extent to which uncertainties in massive star evolution will affect
SED modeling.  Finally, the lower right panel shows the light-weighted
age as a function of wavelength, again for a $\tau=10$ Gyr SFH.  The
dashed line indicates the mass-weighted mean age.  Already by
$0.5\,\mu m$ (approximately the $V-$band), the light-weighted age
reaches its maximal value, again suggesting that going to redder
restframe wavelengths is not providing substantial new information on
the SFH.  Notice also that the maximum light-weighted age never
reaches the mass-weighted age.  We return to this point in Section
\ref{s:sfr}.

\subsection{Nebular Emission}

Although the effects of nebular emission on SEDs will not be discussed
in detail in this review, a brief overview of this component is given
for completeness.

Nebular emission is comprised of two components: continuum emission
consisting of free-free, free-bound, and two photon emission, and
recombination line emission.  Several photoionization codes exist that
make predictions for the nebular emission as a function of the
physical state of the gas, including CLOUDY \citep{Ferland98} and
MAPPINGSIII \citep{Groves04}.  Other approaches can be taken to the
modeling of nebular emission lines.  For example, \citet{Anders03}
implement non-hydrogen emission lines based on observed line ratios as
a function of metallicity. The nebular emission model must then be
self-consistently coupled to a model for the starlight.  Several
groups have done this, with varying degrees of complexity, and with
some only including nebular continuum, others only line emission, and
others including both \citep[e.g.,][]{Leitherer99, Charlot01,
  Panuzzo03, Groves08, Molla09, Schaerer10}.

The effect of nebular emission on the SED is complex, especially when
line emission is considered.  As a general rule, nebular emission is
more important at low metallicity and at young ages.  In such cases
the contribution of nebular emission to broadband fluxes can be as
high as $20-60\%$ \citep{Anders03}.  Nebular emission will also be
more important at high redshift because a feature with a fixed
restframe EW will occupy a larger fraction of the filter bandpass due
to the redshifting of the spectrum. The effect of nebular emission
therefore cannot be ignored at high redshift, where high SFR, low
metallicity galaxies are common \citep{Schaerer10, Atek11}.

It is noteworthy that amongst the most widely used SPS codes,
including \citet{Bruzual03}, \citet{Maraston05}, PEGASE
\citep{Fioc97}, STARBURST99 \citep{Leitherer99}, and FSPS
\citep{Conroy09a}, only PEGASE and STARBURST99 include nebular
continuum emission, and only PEGASE includes both nebular continuum
and line emission.  Since nebular emission is relatively
straightforward to implement (notwithstanding assumptions about the
physical state of the gas), it should be a standard component of all
SPS codes.

\subsection{Fitting Models to Data}

The SPS models described in this section are most frequently used to
measure physical parameters of stellar populations.  This is achieved
by fitting the models to data, either in the form of broadband SEDs,
moderate-resolution optical/NIR spectra, or spectral indices.  The
SSPs are usually taken as given, and the user then fits for a variety
of parameters including metallicity, dust attenuation, and one or more
parameters for the SFH.  If IR data are available, then additional
variables must be considered.  The fitting techniques vary but are
generally limited to grid-based $\chi^2$ minimization techniques.  As
the number of parameters increases Markov Chain Monte Carlo techniques
become increasingly more efficient \citep{Conroy09a, Acquaviva11}.
One must also be aware of the influence of the chosen priors on the
derived parameters; in cases where parameters are poorly constrained
the prior can have a significant effect on the best-fit values
\citep[see e.g.,][for discussion]{Kauffmann03a, Salim07, Taylor11}.
Moreover, it is highly advisable to derive the `best-fit' parameters
from the marginalized posterior distribution, rather than from the
minimum of $\chi^2$, since the likelihood surface can often be highly
irregular \citep[see e.g.,][]{Bundy05, Taylor11}.

When fitting SEDs it is important to remember that only the shape is
being used to constrain the model parameters. In other words, the SED
shape (or, more crudely, broadband colors) constrains parameters such
as $M/L$, specific SFR (SSFR$\equiv$SFR/$M$), dust attenuation and
metallicity.  To obtain stellar masses one needs to multiply $M/L$ by
the observed luminosity and to obtain SFRs one then multiplies SSFR by
$M$.  This also holds true for $M/L$ ratios and SSFRs inferred from EWs
of emission lines, spectral indices, and full spectral fitting.

The reader is referred to \citet{Walcher11} for further details on SED
fitting techniques.


\section{MASS-TO-LIGHT RATIOS \& STELLAR MASSES}
\label{s:m2l}

\subsection{Techniques \& Uncertainties}

There are three basic techniques for estimating the stellar
mass-to-light ratio, $M/L$, of a galaxy: (1) using tabulated relations
between color and $M/L$; (2) modeling broadband photometry; (3)
modeling moderate resolution spectra.  The first technique is the
simplest to use as it requires photometry in only two bands and no
explicit modeling.  The other techniques require construction of a
library of models and a means to fit those models to the data.  How do
these different techniques compare?

\subsubsection{Color-based $M/L$ Ratios}

The color-based $M/L$ estimators have their origin in the pioneering
work of \citet{Bell01}.  These authors used a preliminary version of
the \citet{Bruzual03} SPS model to chart out the relation between
$M/L$ and color as a function of metallicity and SFH.  Remarkably,
they found that variation in metallicity and SFH (parameterized as a
$\tau$ model) moved galaxies along a well-defined locus in the space
of $M/L_B$ vs. $B-R$, suggesting that the $B-R$ color could be a
useful proxy for $M/L_B$.  Perhaps most importantly, they demonstrated
that the dust reddening vector was approximately parallel to the
inferred color-$M/L$ relation, implying that dust should have only a
second-order effect on derived $M/L$ values.  They also investigated
NIR colors and found much larger variation in $M/L$ at fixed $I-K$,
due mostly to variation in the SFH.  Late bursts of SF complicate the
interpretation of color-$M/L$ relations by driving the $M/L$ ratios
lower at fixed color compared to smoothly declining SFHs.  Their
analysis was for a fixed IMF; allowing for IMF variation will change
the $M/L$ while leaving the color basically unchanged.  These authors
concluded that for a fixed IMF and assuming that large starbursts and
low-metallicity systems are not common properties of galaxies, and
neglecting uncertainties in the SPS models, one could estimate $M/L$
from a single color to an accuracy of $\sim0.1-0.2$ dex.

In subsequent work, \citet{Bell03} analyzed the optical-NIR SEDs of
$\sim12,000$ galaxies with SDSS and 2MASS photometry.  They
constructed a grid of SPS models with varying SFHs, including bursts,
and metallicity. They did not allow for reddening due to dust.  They
derived best-fit $M/L$ values from SED fitting and created new,
observationally-constrained color-$M/L$ relations.  The resulting
$B-R$ vs. $M/L_B$ relation was similar to \citet{Bell01}, but the NIR
relation was considerably more shallow than the earlier work.  At the
bluest colors, the $M/L_K$ ratios differed by $\approx0.3$ dex between
the old and new relations.  The difference arose due to a population
of galaxies with blue observed colors and high inferred $M/L$ ratios,
which \citet{Bell03} interpreted as being due to the lower
metallicities allowed in their fitting procedure compared to the
analytic models of Bell \& de Jong.  These differences are of some
concern because the \citet{Bell03} color-$M/L$ relations are very
widely used to derive `cheap' stellar mass estimates of galaxies.

\citet{Zibetti09} revisited this issue with new SPS models from
Charlot \& Bruzual (in prep).  They followed a somewhat different
approach compared to \citet{Bell03} in that they constructed
color-$M/L$ relations directly from a library of model galaxies.  This
approach is sensitive to how one chooses to populate parameter space,
since the resulting color-$M/L$ relations are simple averages over the
model space.  In contrast, \citet{Bell03} constructed color-$M/L$
relations only from those models that provided good fits to observed
optical-NIR photometry.  Zibetti et al. allowed for dust attenuation,
and they also allowed for larger mass fractions of stars formed in
recent bursts compared to Bell et al.  \citet{Zibetti09} found
substantial differences in the $g-i$ color-$M/L$ relation compared to
Bell et al. ($\sim0.5$ dex at the bluest colors), which they
attributed to their allowance for much stronger bursts of star
formation than allowed in \citet{Bell03}.  \citet{Gallazzi09} also
found a very different color-$M/L$ relation compared to
\citet{Bell03}, which they attributed to the different age
distributions in their model library.  \citet{Taylor11} followed the
philosophy of \citet{Bell03} in fitting models to observed SEDs in
order to derive color-$M/L$ relations.  They found systematic
differences of order $0.1-0.2$ dex compared to the Zibetti et
al. relations, which they attributed mostly to their decision to
select only those models that fit the photometric data, and also the
treatment of dust reddening.

\citet{Zibetti09} found a large difference in $M/L_H$ ratios estimated
from $i-H$ colors between the new Charlot \& Bruzual models and the
previous version \citep{Bruzual03}, with typical offsets of $0.2-0.4$
dex.  This difference is largely due to the different treatment of the
TP-AGB evolutionary phase (see Section \ref{s:mlerr} for details).  In
contrast, optical color-$M/L$ relations showed little difference
between the two SPS models, and also showed slightly less discrepancy
with \citet{Bell03}.  These authors argued that optical color-$M/L$
relations should generally be preferred to NIR-based ones owing to the
larger impact of the uncertain TP-AGB phase in the NIR.  Exceptions to
this rule could be made for galaxies with very large amounts of dust
reddening and/or very strong starbursts, in which cases the optical
SED becomes a poor indicator of the SFH and hence $M/L$.  The upshot
is that NIR color-based $M/L$ estimates should be treated with
caution, since they are subject to large stellar evolution-based
systematics.  However, both optical and NIR-based $M/L$ estimates will
depend on the allowed range of SFHs and metallicities in the model
library, with large systematics appearing for the bluest colors.

\begin{figure}
\centerline{\psfig{figure=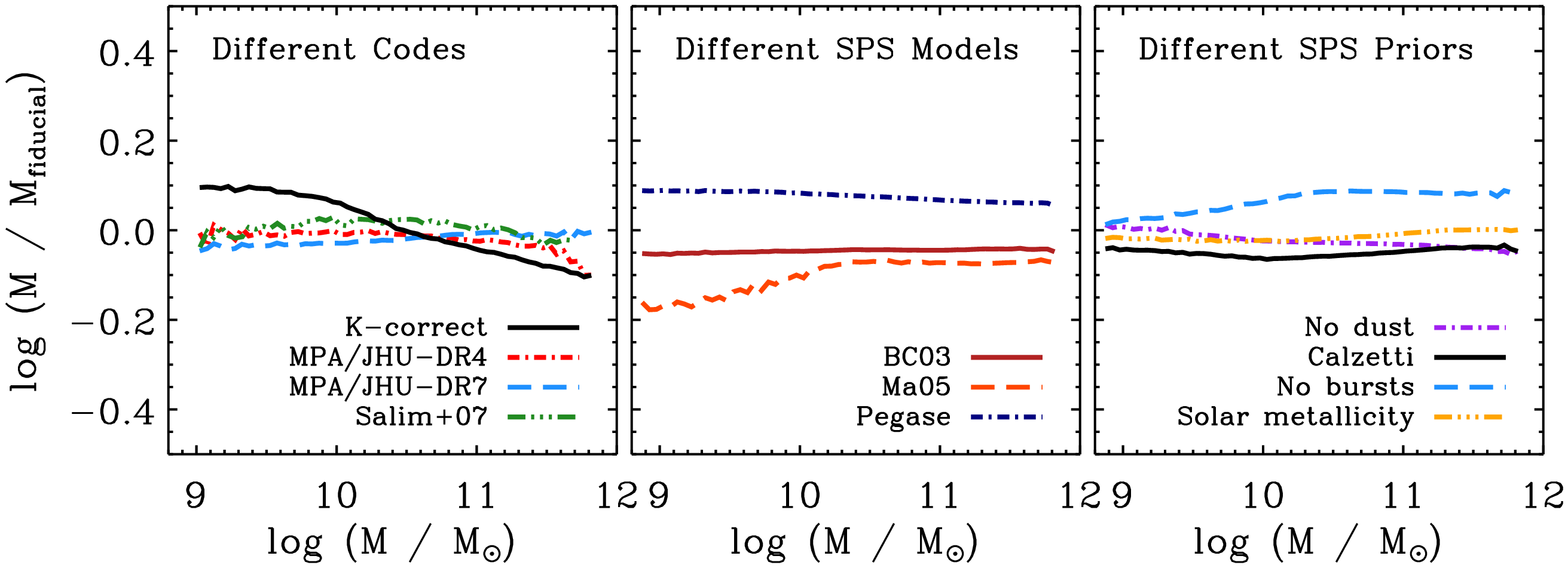,height=2.5in,angle=90}}
\caption{Comparison of different fitting codes, SPS models, and priors
  on derived stellar masses \citep[from][]{Moustakas12}.  The fiducial
  masses are based on fitting SDSS and {\it GALEX} photometry of
  $z\sim0$ galaxies using the \texttt{iSEDfit} code
  \citep{Moustakas12}, with SSPs from FSPS \citep{Conroy09a},
  including dust attenuation, a range in metallicities, and SFHs with
  both smooth and bursty components. The left panel compares stellar
  mass catalogs produced by different groups/codes.  K-correct and
  MPA/JHU-DR7 are based on SDSS photometry; MPA/JHU-DR4 is based on
  SDSS spectral indices, and Salim+07 is based on SDSS and {\it GALEX}
  photometry.  The middle panel shows the effect of different SPS
  models (i.e., different SSPs), and the right panel shows the effect
  of varying the priors on the model library.  The mean systematic
  differences between mass estimators is less than $\pm0.2$ dex.
  Figure courtesy of J. Moustakas.}
\label{f:mass}
\end{figure}

\subsubsection{$M/L$ From Broadband and Spectral Fitting Techniques}

The modern era of SED fitting with SPS models was ushered in by
\citet{Sawicki98} and \citet{Giallongo98}, who analyzed optical-NIR
broadband photometry of high-redshift galaxies.  The model grid of
\citet{Sawicki98} consisted of a $\tau-$model SFH with varying $\tau$
and start time for the SFH, and variation in metallicity and
reddening.  Subsequent modeling of broadband data has largely followed
this approach \citep[e.g.,][]{Brinchmann00, Papovich01, Shapley01,
  Salim07}.  Generically, when fitting broadband SEDs one finds that
stellar masses for `normal' galaxies (i.e., not including pathological
SFHs) can be recovered at the $\approx0.3$ dex level ($1\sigma$
uncertainty).  This uncertainty does not include potential systematics
in the underlying SPS models; see Section \ref{s:mlerr}.  Stellar
masses appear to be the most robust parameter estimated from SED
fitting \citep[e.g.,][]{Papovich01, Shapley01, Wuyts09, Muzzin09,
  LeeSK09}.  The reason for this appears to not be fully understood,
although it probably is at least partly due to the fact that the dust
reddening vector is approximately parallel to SFH and metallicity
variations in color-$M/L$ diagrams \citep{Bell01}.  The choice of the
SFH, in particular whether it is rising, declining, or bursty, can
significantly change the best-fit stellar mass, by perhaps as much as
$0.6$ dex in extreme cases \citep{Pforr12}.  A general rule of thumb
is that $M/L$ ratios estimated via simple SFHs (or single-age models)
will be lower limits to the true $M/L$ ratios \citep{Papovich01,
  Shapley01, Trager08, Graves10a, Pforr12}.  This is a consequence of
the fact that young stars outshine older ones, making it relatively
easy to `hide' old stellar populations in galaxies with a large number
of young stars.  This also explains why color-$M/L$ relations are so
uncertain for very blue colors.  This general rule is not universal:
the modeling of certain galaxy types may lead to systematic biases in
the opposite direction \citep{Gallazzi09}.  Partly for these reasons,
stellar masses estimated for quiescent systems should be more reliable
than for star-forming ones.  Relatedly, \citet{Zibetti09} demonstrated
that stellar masses estimated for individual nearby galaxies based on
a pixel-by-pixel analysis of colors are generally larger than those
estimated from integrated colors.  The differences are typically
$\sim0.05-0.15$ dex, depending on galaxy type.  This should not be
surprising, especially for systems with star-forming disks and
quiescent bulges, since the $M/L$ ratio estimated from the integrated
colors will tend to be biased low by the more luminous young stars in
the star-forming component.  \citet{Wuyts12} performed a similar
analysis on galaxies at $0.5<z<2.5$ with WFC3 {\it Hubble Space
  Telescope (HST)} photometry and found no systematic difference in
stellar masses derived from resolved vs. integrated light, although
the fact that their galaxies were at high redshift means that they
were probing spatial scales an order of magnitude larger than in
Zibetti et al.

There has been some confusion in the literature regarding the
importance of restframe NIR photometry for estimating stellar
masses. First, as indicated in Figure \ref{f:ev}, for smoothly varying
SFHs, photometry at and redward of the $V-$band is sensitive to the
same light-weighted age, and so redder bands do not provide stronger
constraints on the mean stellar age.  \citet{Taylor11} analyzed mock
galaxies and concluded that the addition of NIR data did not yield
more accurate masses.  Taylor et al. also found that different SPS
models produced good agreement in derived properties when NIR data was
excluded from their fits, but poor agreement when NIR was included.
These authors also found much larger residuals in their SED fits when
NIR data were included, suggesting that the models are still poorly
calibrated in this regime.  As discussed in Section \ref{s:pz}, the
NIR is at present probably most useful for constraining metallicities
(within the context of a particular SPS model), and so NIR data may be
useful in cases where there is a degeneracy between $M/L$ and $Z$.  In
general however stellar mass estimates do not appear to be strongly
improved with the addition of NIR data, at least with currently
available models.  Exceptions to this rule may be made for galaxies
with very high dust opacities.

As first emphasized by \citet{Bell01} and \citet{Bell03}, some of the
largest uncertainties in derived $M/L$ ratios stem from uncertainties
in the assumed SFHs, in particular the presence of bursty SF episodes.
The consideration of the Balmer lines with other age and
metallicity-sensitive features, such as the 4000\,\AA\, break
($D_n4000$), can constrain the burstiness of the SFH.  Optical spectra
therefore offers the possibility of providing stronger constraints on
the $M/L$ ratio.  \citet{Kauffmann03a} modeled the H$\delta$ and
$D_n4000$ spectral features measured from SDSS spectra in order to
constrain SFHs and $M/L$ ratios.  They obtained 95\% confidence limits
on stellar masses of $\sim0.2$ dex and $\sim0.3$ dex for quiescent and
star-forming galaxies, respectively.  Again, these are statistical
uncertainties because the underlying SPS model and other aspects such
as the adopted parameterization of the SFH were held fixed.
\citet{Chen12} employed principle component analysis (PCA) to model
the optical spectra of massive galaxies in SDSS.  These authors found
systematic uncertainties in the recovered stellar masses of order
$\sim0.1$ dex depending on the assumed metallicity, dust model, and
SFH.  They also demonstrated that their PCA technique was capable of
measuring parameters at much lower S/N than direct fitting to selected
spectral indices.  They determined statistical uncertainties on their
mass measurements to be $\sim0.2$ dex, which they showed to be
comparable to the formal errors estimated from modeling broadband
photometry.  This result is in agreement with \citet{Gallazzi09}, who
argued that for galaxies with simple SFHs and lacking dust, the
uncertainty on the derived masses are not much larger when using
color-based estimators compared to spectroscopic-based estimators.
With regards to estimating $M/L$, the real value of spectra appears to
be restricted to galaxies with unusual SFHs.

The comparison between spectroscopically-based and
photometrically-based stellar masses is informative because the
approaches suffer from different, though by no means orthogonal
systematics.  The most obvious difference is with regards to dust ---
spectroscopic masses are much less sensitive to dust attenuation than
photometric masses.  \citet{Drory04} compared their own
photometrically-derived masses to the spectroscopic masses from
\citet{Kauffmann03a}.  They found an rms scatter of $\sim0.2$ dex
between the two estimates, with a modest systematic trend that
correlated with H$\alpha$ EW.  \citet{Blanton07} estimated photometric
stellar masses via a technique that is similar to PCA except that the
templates are constrained to be non-negative and are based on an SPS
model.  They compared their derived stellar masses to those of
Kauffmann et al. and found good agreement, with systematic trends
between the two restricted to $\lesssim0.2$ dex.

A comparison between stellar masses for the same galaxies estimated
with different SPS models, priors, and fitting techniques is shown in
Figure \ref{f:mass} \citep[from][]{Moustakas12}.  The galaxies
included in this figure are predominantly `normal' star-forming and
quiescent $z\sim0$ galaxies.  For these galaxies the mean absolute
differences between various mass estimators is less than $0.2$ dex, in
agreement with other work on the systematic uncertainties in stellar
masses of normal galaxies.

\subsection{Uncertainties in $M/L$ due to Stellar Evolution Uncertainties}
\label{s:mlerr}

\citet{Maraston06} was the first to draw attention to the sensitivity
of derived stellar masses to uncertain stellar evolutionary phases, in
particular the TP-AGB.  The \citet{Maraston05} SPS model predicts much
more luminosity arising from TP-AGB stars than in previous models
\citep[e.g.,][]{Fioc97, Bruzual03}.  At ages where the TP-AGB phase is
most prominent ($\sim3\times 10^8-2\times10^9$ yr), Maraston's model
predicts roughly a factor of two more flux at $>1\,\mu m$ compared to
earlier work.  \citet{Maraston06} found that her SPS model implied a
factor of two smaller stellar mass, on average, compared to masses
derived with the \citet{Bruzual03} SPS model for galaxies at $z\sim2$.
When the fits excluded the possibility of dust reddening, Maraston's
model provided a better fit to the optical-NIR SEDs than previous
models, whereas when dust was allowed, the quality of the fits became
indistinguishable, although the offsets in best-fit masses remained.

Subsequent work has largely confirmed the sensitivity of estimated
stellar masses to the adopted SPS model \citep{Wuyts07, Kannappan07,
  Cimatti08, Muzzin09, Longhetti09, Conroy09a}.  Essentially all work
on this topic has focused on comparing Maraston's model to Bruzual \&
Charlot's, where the difference in TP-AGB treatment it probably the
most significant, though not the only difference (other differences
include different RGB temperatures and different treatments for core
convective overshooting).  Many authors have found a maximum factor of
$\sim2-3$ difference in derived $M/L$ ratios between the two models,
especially when NIR data were included.  \citet{Kannappan07} showed
that the differences between Maraston's and Bruzual \& Charlot's
models are relatively modest ($\lesssim1.3$) when NIR data are
excluded.  \citet{Conroy09a} constructed a new SPS model in which the
luminosity contribution from the TP-AGB phase could be arbitrarily
varied.  These authors isolated the importance of the TP-AGB phase and
confirmed previous work indicating that the adopted weight given to
this phase in the models can have a large modulating effect on the
stellar mass.

The contribution of TP-AGB stars to the integrated light peaks at
$\sim3\times10^8-2\times10^9$ yr, depending on metallicity, and so the
importance of this phase to modeling SEDs will depend on the SFH of
the galaxy.  It was for this reason that \citet{Maraston06} focused
their efforts on quiescent galaxies at $z\sim2$; at this epoch even a
quiescent galaxy will have a typical stellar age not older than
several Gyr.  In addition to high-redshift quiescent galaxies,
\citet{Lancon99} suggested that post-starburst galaxies should harbor
large numbers of TP-AGB stars because their optical spectra show
strong Balmer lines, indicative of large numbers of $10^8-10^9$ yr old
stars.  This fact was exploited by \citet{Conroy10c}, \citet{Kriek10},
and \citet{Zibetti12} to constrain the TP-AGB contribution to the
integrated light.  These authors all find evidence for a low
contribution from TP-AGB stars, and in particular they argue that
Maraston's models predict significantly too much flux in the NIR for
these objects.

The situation with TP-AGB stars is extremely complex owing to the fact
that this phase is so sensitive to age and metallicity.  Moreover,
using galaxy SEDs to constrain the importance of this phase is
difficult because many parameters must be simultaneously constrained
(metallicity, SFH, dust, etc.).  A case in point is the analysis of
high-redshift galaxies, where \citet{Maraston06} found large
differences in the quality of the fit of Bruzual \& Charlot models
depending on whether or not dust was included in the fits (as an
aside, if these galaxies do have copious numbers of TP-AGB stars then
one may expect them to also contain dust, since these stars are
believed to be efficient dust factories).  It is also worth stressing
that the conclusion to the TP-AGB controversy may not be an
`either-or' situation in the sense that some models may perform
better for some ranges in age and metallicity while other models may
perform better in different regions of parameter space.  Ultimately of
course we desire models that perform equally well over the full range
of parameter space, and this requires a continual evolution and
improvement of SPS models.

Of course, there are other aspects of stellar evolution that are
poorly constrained, including convective overshooting, blue stragglers
and the HB, and even the temperature of the RGB.  The propagation of
these uncertainties into derived properties such as $M/L$ ratios is
only just beginning \citep[e.g.,][]{Conroy09a}.  \citet{Melbourne12a}
has for example demonstrated that the latest Padova isochrones fail to
capture the observed flux originating from massive core He burning
stars as observed in nearby galaxies.  These stars are very luminous
and can dominate the NIR flux of young stellar populations
($\lesssim300$ Myr).

Finally, it is worth stressing that the effect of TP-AGB stars is
limited to galaxies of a particular type, in particular those that are
dominated by stars with ages in the range $\sim3\times10^8-
2\times10^9$ yr.  Examples include post-starburst galaxies, which are
rare at most epochs, or quiescent galaxies at $z\gtrsim2$.  For
typical galaxies at $z\sim0$, the treatment of this phase seems to be
of little relevance for estimating stellar masses, at least on
average, as is evident in Figure \ref{f:mass}.

\subsection{Stellar Masses at High Redshift}

As first emphasized by \citet{Papovich01}, galaxies with high SFRs can
in principle contain a large population of `hidden' older stars, since
these stars have high $M/L$ ratios.  Papovich et al. found that the
data allowed for significantly larger stellar masses when
two-component SFH models (young+old) were compared to their fiducial
single component models (a median difference of a factor of $\approx3$
with extreme cases differing by an order of magnitude).  This issue
becomes more severe at higher redshifts because the typical SFRs are
higher and because of the increasing possibility of rising SFHs (to be
discussed in detail in Section \ref{s:sfr}).  This makes the analysis
of high redshift galaxy SEDs much more complicated.

However, at the highest redshifts ($z\gtrsim6$), the analysis of SEDs
may actually become simpler.  At $z=8$ the age of the universe is only
$\approx640$ Myr, and if one presumes that galaxy formation commences
after $z\approx20$, then even the oldest stars at $z=8$ will be no
more than $\approx460$ Myr old.  The oldest possible main sequence
turnoff stars will thus be A type stars.  Put another way, the $M/L$
ratio in the blue (U through V bands) at 13 Gyr is $4.5-15$ times
higher for an instantaneous burst of SF compared to a constant SFH,
whereas the $M/L$ ratio for these two SFHs differs by the more modest
$2.5-5$ at 500 Myr.  So there is some expectation that `hidden mass'
will be less hidden when modeling SEDs at the highest redshifts.  This
expectation appears to be borne out by detailed modeling.
\citet{Finkelstein10} analyzed the SEDs of $z\sim7-8$ galaxies and
concluded that even allowing for the possibility of extreme amounts of
old stars (90\% by mass), the best-fit stellar masses increased by no
more than a factor of two compared to a fiducial single-component SFH
\citep[see also][]{Curtis-Lake12}.

At even higher redshifts ($z\gtrsim10$), the analysis of galaxy SEDs
may become even simpler, at least with regards to estimating stellar
masses.  At sufficiently high redshifts the age of the oldest possible
stars will eventually become comparable to the SF timescale probed by
the restframe UV ($\sim10^8$ yr at $\sim2500$\,\AA; see Figure
\ref{f:ev}).  The measured SFR will thus be averaged over roughly the
entire age of the galaxy, implying that a reasonable estimate of the
stellar mass can be obtained by multiplying the UV-derived SFR with
the age of the universe.  This may in fact explain the unexpectedly
strong correlation between stellar mass and UV luminosity at $z>4$
reported by \citet{Gonzalez11}.  Of course, at the highest redshifts
other difficulties arise, including accounting for the effects of very
high EW emission lines and the reliability of the underlying models at
very low metallicities.

\subsection{Summary}

$M/L$ ratios for most galaxies with normal SEDs are probably accurate
at the $\sim0.3$ dex level, for a fixed IMF, with the majority of the
uncertainty dominated by systematics (depending on the type of data
used).  Galaxies with light-weighted ages in the range of $\sim0.1-1$
Gyr will have more uncertain $M/L$ ratios, with errors as high as
factors of several, owing to the uncertain effect of TP-AGB stars and
perhaps other stars in advanced evolutionary phases.  Galaxies that
have very young light-weighted ages (e.g., high-redshift and starburst
galaxies) will also have very uncertain $M/L$ ratios because of the
difficulty in constraining their past SFHs.  Uncertainties due to dust
seem to be subdominant to other uncertainties, at least for standard
reddening laws and modest amounts of total attenuation. Essentially
all stellar masses are subject to an overall normalization offset due
both to the IMF and the uncertain contribution from stellar remnants.
Owing to the lingering model uncertainties in the NIR (e.g., from
TP-AGB and cool core He burning stars), it may be advisable to derive
physical parameters by modeling the restframe UV-optical \citep[see
e.g., the discussion in][]{Taylor11}.


\section{STAR FORMATION RATES, HISTORIES, \& STELLAR AGES}
\label{s:sfr}

\subsection{Measuring SFRs}

The practice of measuring SFRs from monochromatic indicators (and
combinations thereof) is well established \citep[see e.g., the reviews
of][]{Kennicutt98b, Kennicutt12}.  A wide variety of relations have
been established between SFRs and the UV, H$\alpha$, P$\alpha$,
24$\,\mu m$ luminosity, total IR luminosity, radio continuum
luminosity, and even the $X$-ray flux.  As they are not the focus of
the present review, only a few brief remarks will be given regarding
monochromatic SFR indicators.  Each indicator is sensitive to a
different SFR timescale, with ionizing radiation (as probed by e.g.,
the recombination lines) being sensitive to the shortest timescale,
$\sim10^6$ yr, and the UV and total IR being sensitive to longer
timescales, e.g., $\sim10^8$ yr.  In detail, the UV and IR are
sensitive to a range of SF timescales depending on wavelength and
SFH. In the IR this sensitivity arises because shorter wavelengths are
probing dust heated mostly by \ionn{H}{ii} regions, while FIR photons
are sensitive to dust heated both by old and young stars.  Amongst the
recombination lines, `redder is better' since dust attenuation will be
less severe at longer wavelengths, and so P$\alpha$ should be
preferred over H$\alpha$.  For reference, an attenuation of 1 mag in
the $V-$band implies an attenuation of only 0.15 mag at P$\alpha$, for
standard attenuation curves.  The optical and UV SFR indicators
require correction for dust when used on their own.  For this reason,
mixed SFR indicators have become popular, usually via the combination
of UV and IR or H$\alpha$ and IR indicators \citep{Calzetti07,
  Kennicutt07}.

\begin{figure}
\centerline{\psfig{figure=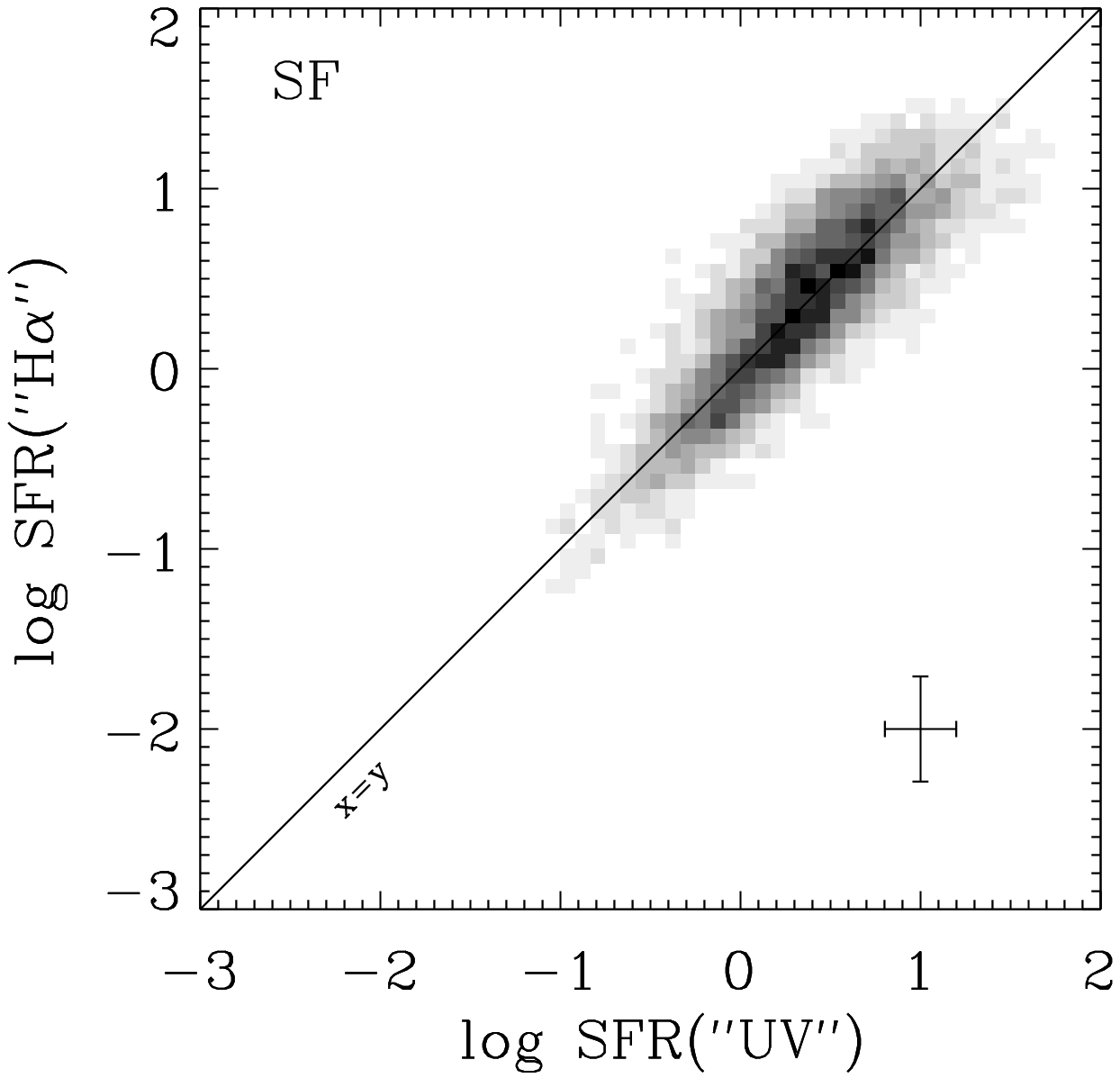,height=2.5in}
\psfig{figure=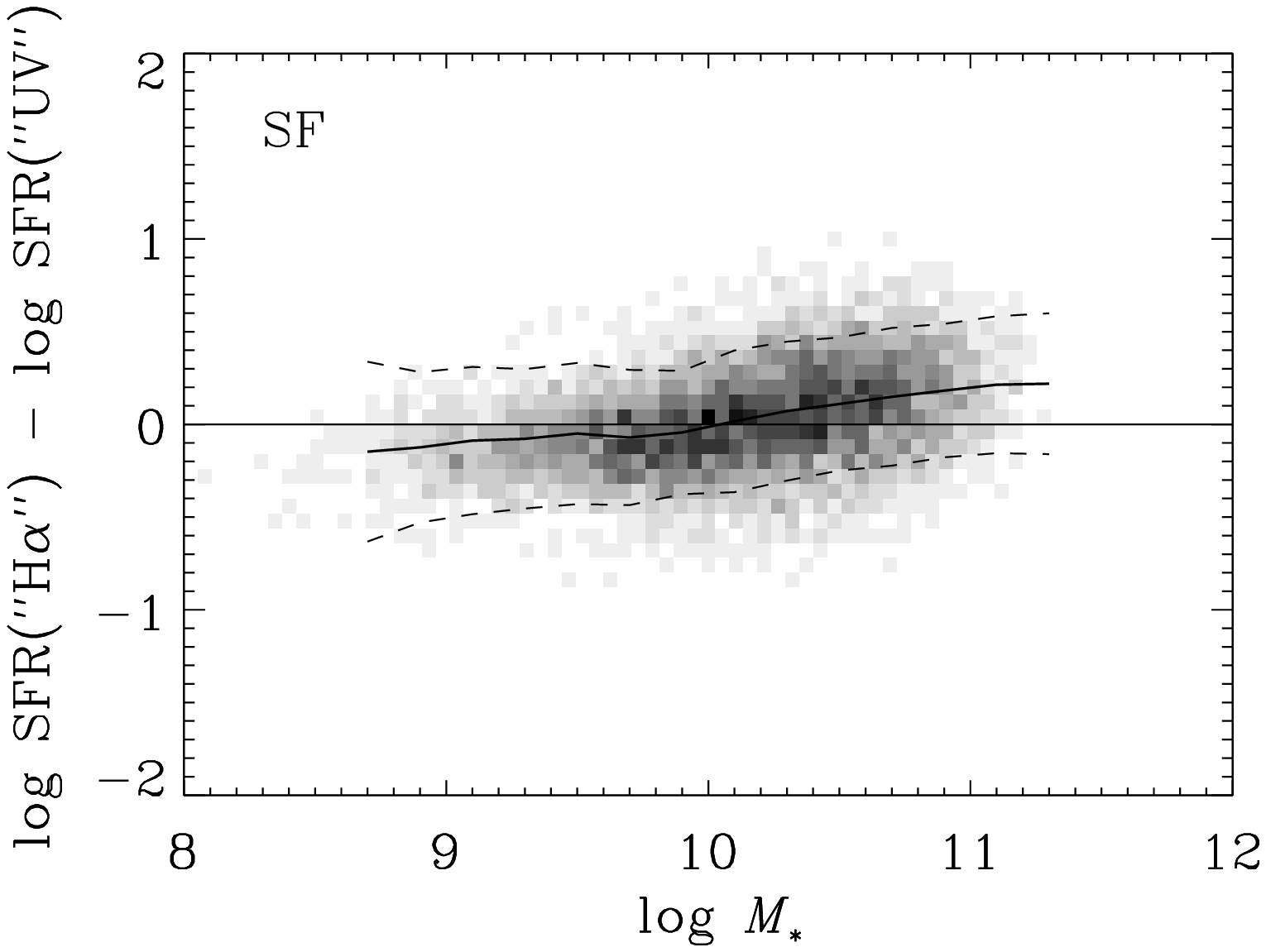,height=2.5in}}
\caption{Comparison between SFRs estimated from emission lines
  \citep[``H$\alpha$";][]{Brinchmann04} and from modeling the
  UV-optical SEDs \citep[``UV";][]{Salim07} of star-forming galaxies
  at $z\sim0$.  The right panel shows the difference between SFR
  estimators as a function of stellar mass.  There is a systematic
  trend between the two estimators, which may be due to modeling
  differences and/or the uncertainty in extrapolating the fiber-based
  H$\alpha$ SFRs to the total galaxy.  From \citet{Salim07}.}
\label{f:uvha}
\end{figure}

Estimating SFRs from SED fitting is very challenging for several
reasons: (1) the age-dust-metallicity degeneracy makes it difficult to
reliably measure ages and hence SFRs unless high quality data are
available (see Section \ref{s:dust}); (2) the choice of model priors
on the dust model and SFH library imposes often severe biases on the
resulting ages and SFRs.  These difficulties were recognized early on
in the study of high-redshift galaxies \citep{Papovich01} and are
still relevant today.

\citet{Brinchmann04} measured SFRs for SDSS galaxies by modeling a
suite of optical emission lines.  Their model included variation in
the total metallicity, dust attenuation, ionization parameter, and
dust-to-metals ratio.  As the SDSS is a fiber-based survey, these
authors had to make a correction for the fact that the spectra sample
only a fraction of the galaxy light.  In addition, they modeled the
stellar continuum in order to subtract off any absorption features
under the emission lines.  \citet{Salim07} analyzed a comparably sized
sample of $z\sim0$ galaxies, focusing on UV photometry from the {\it
  Galaxy Evolution Explorer (GALEX)} and optical photometry from SDSS.
These authors fit the data to SPS models with a range of SFHs
(including exponentially declining SFHs and superimposed starbursts),
metallicities, and dust attenuation.  They presented a direct
comparison between their SED-based SFRs and the emission line-based
SFRs from Brinchmann et al.  The result is shown in Figure
\ref{f:uvha}.  While the overall agreement is good there is a
systematic trend with stellar mass resulting in offsets of $\sim0.2$
dex at the low and high mass ends.  Salim et al. explored the origin
of this offset and concluded that the different derived dust
attenuation optical depths between the two methods were the source of
the discrepancy.  They further speculated that in most cases it was
the emission line-based dust attenuation values that were more
accurate, and therefore also the SFRs.  Salim et al. also demonstrated
that for galaxies with low SFRs and/or galaxies with emission line
ratios indicative of AGN activity, the UV-based SFRs tended to be more
reliable than those estimated by Brinchmann et al.  These results
suggest that SFRs based on modeling UV-optical SEDs carry systematic
uncertainties at the $\lesssim0.3$ dex level.

\citet{Wuyts11} compared the SFRs of galaxies at $z\sim3$ computed
with three different techniques: H$\alpha$-based SFRs, SED-fitting
based SFRs, and a mixed SFR indicator based on the UV and IR emission.
For SFRs $\lesssim100\,\Msun\, {\rm yr}^{-1}$, these authors found
that the three indicators agreed reasonably well (no systematic
offsets were detected), provided that extra attenuation was included
toward \ionn{H}{ii} regions when modeling the H$\alpha$-based SFRs.
They also found that at the highest SFRs $\gtrsim100\,\Msun\, {\rm
  yr}^{-1}$ the SED-based values tended to underpredict the SFR,
compared to the mixed indicator \cite[see also][]{Santini09}.  They
argued that this was due to their assumption of a uniform dust screen,
and at high column density this becomes an increasingly poor
representation of reality.

\begin{figure}
\centerline{\psfig{figure=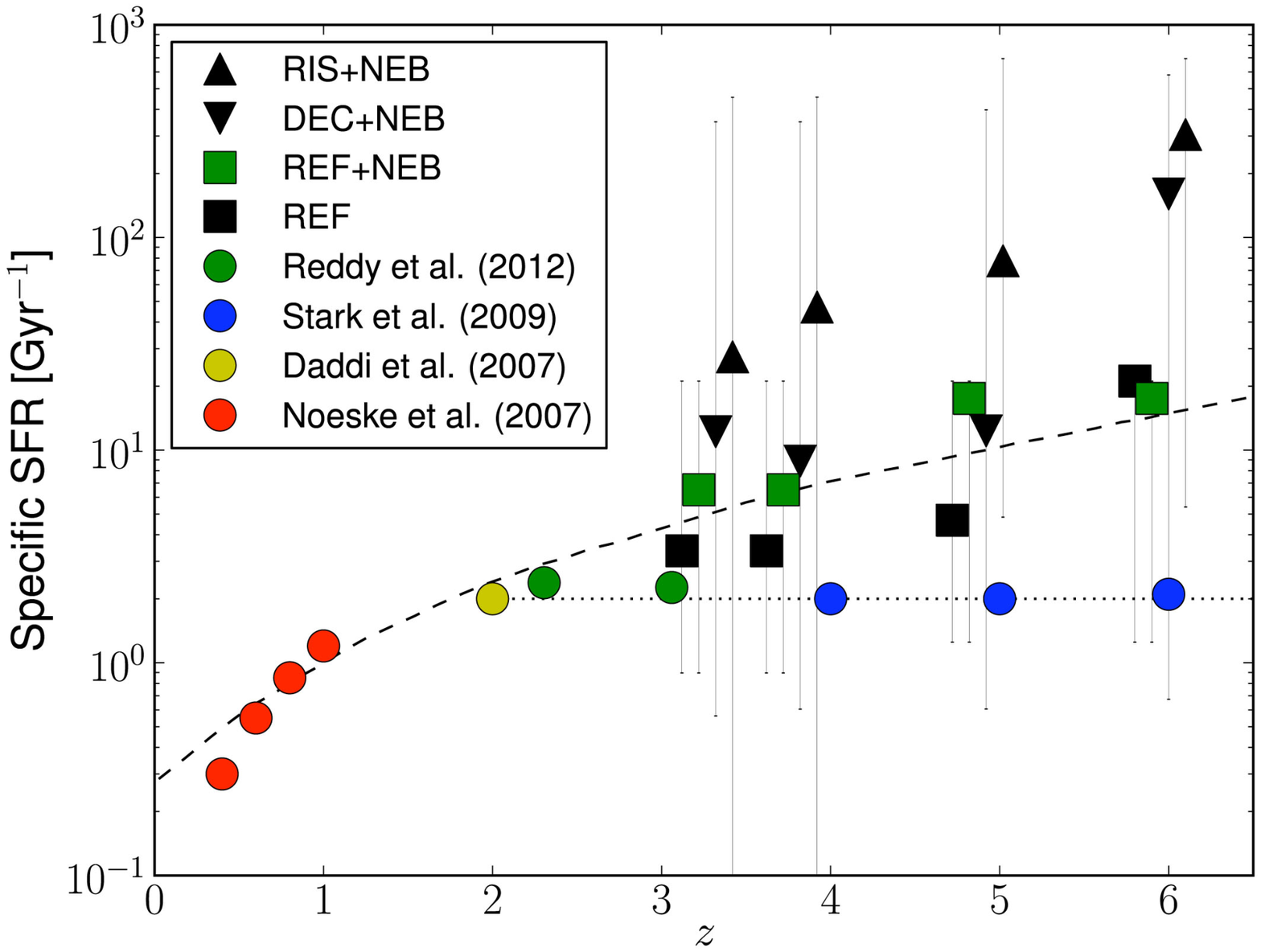,height=4in}}
\caption{Specific SFRs for galaxies with stellar masses of
  $10^{9.5}\,\Msun$ as function of redshift.  The black and green
  symbols reflect different assumptions regarding the SFH (`REF',
  `RIS', `DEC' for constant, rising, and declining) and the correction
  for emission lines (`NEB').  Error bars reflect 68\% confidence
  limits.  From \citet{deBarros12}}
\label{f:ssfr}
\end{figure}

At high redshift the effect of nebular emission lines becomes an
additional source of systematic uncertainty in the modeling, and can
significantly affect the best-fit SFRs.  The impact of emission lines
is larger at higher redshift because of three effects: (1) specific
SFRs are higher; (2) metallicities are lower; and (3) the redshifting
of the spectrum implies that a line of given EW comprises a larger
fraction of the flux through a filter.  In Figure \ref{f:ssfr} the
specific SFR for galaxies with a mass of $10^{9.5}\,\Msun$ is shown as a
function of redshift for various choices regarding the contribution of
emission lines to the observed broadband fluxes.  Evidently the
inclusion or not of emission lines in the modeling can change the
derived specific SFRs in a redshift-dependent manner, by a factor of
roughly two in the most extreme cases \citep[see also][]{Pacifici12,
  Labbe12}.

\subsection{Light-Weighted Mean Ages}

The mean stellar ages of galaxies are another important probe of the
SFH.  Of course, mass-weighted ages offer a direct probe of the
integrated SFH, while light-weighted ages are more directly measurable
from SEDs.  Mass-weighted ages will always be larger than
light-weighted ones, owing to the fact that young stars outshine older
stars.  As discussed further in the next section, the mass-weighted
ages of actively star-forming galaxies turn out to be highly sensitive
to the assumed model SFH used in the SED fitting.  Due to the fact
that light-weighting is highly non-linear, light-weighted ages
computed from spatially resolved photometry will tend to be older than
ages computed from integrated photometry, as demonstrated by
\citet{Wuyts11}.

The mean ages of quiescent galaxies have been estimated largely via
the modeling of spectral features, including $D_n4000$ and the Lick
indices.  A widely adopted procedure is to fit single-age SSPs to the
Lick indices in order to estimate ages.  In this approach it is the
hydrogen Balmer lines that provide the primary constraint on the age,
as these lines are strongest in A type stars and thus for populations
with ages of $\approx10^8-10^9$ yr.  The derived ages are sometimes
referred to as `SSP-equivalent ages' in order to highlight the fact
that no attempt is made to model composite stellar populations.  A
number of authors have demonstrated that these derived ages are lower
limits to the true mass-weighted and light-weighted ages
\citep{Trager00, Serra07, Trager09a, MacArthur09}.  As emphasized by
\citet{Trager09a}, this fact implies that any observed trend of
SSP-equivalent ages with other parameters such as velocity dispersion
will be stronger than the true underlying mass-weighted trend.  The
SSP-equivalent ages are younger even than light-weighted ages largely
because of the sensitivity of the Balmer lines to $\sim1$ Gyr old
stars.  \citet{Trager09a} concluded that SSP-equivalent ages are
mostly measuring residual star formation over $\sim1$ Gyr timescales.

These conclusions are not restricted to the analysis of spectral
indices.  \citet{Ferreras09} estimated mean ages of early-type
galaxies at $z\sim1$ by fitting models to {\it HST} grism spectra.
The ages estimated from single-age models showed almost no correlation
with ages estimated with composite models.  Likewise,
\citet{MacArthur09} performed full spectrum fitting on a sample of
nearby galaxies and found only a modest correlation between
ages derived from single-age models and light-weighted ages derived
from composite models.

It is widely believed that the H$\beta$ index is the ideal age
indicator because it is largely insensitive to metallicity and
abundance patterns \citep{Worthey94, Tripicco95, Korn05}.  However, as
noted by \citet{Worthey94} and later by \citet{deFreitasPacheco95},
\citet{Maraston00}, and \citet{LeeHC00}, the Balmer lines are also
sensitive to the presence of blue HB stars, and there will therefore
be some ambiguity in determining ages from H$\beta$ without additional
constraints on the presence of such stars.  However, \citet{Trager05}
demonstrated that, while the inferred ages of the oldest stellar
populations ($\gtrsim10$ Gyr) are susceptible to uncertainties of
$2-5$ Gyr due to the presence of blue HB stars, the ages of
intermediate-age populations ($\sim1-10$ Gyr) are rather less affected
by blue HB stars.  The reason for this is that the Balmer lines
increase so strongly with decreasing age that very large quantities of
blue HB stars would be required to be present in old stellar
populations to reproduce the high observed H$\beta$ EWs.  As an
example, \citet{Thomas05} found that they could reproduce the highest
H$\beta$ EWs in their sample with a model in which 50\% of the old
metal-rich stars had blue HB stars.  In this two-component model, the
component with blue HB stars needed to have a metallicity $0.2-0.4$
dex higher than the other component.  \citet{Percival11} also found
that models with very extreme HB populations and old ages could
masquerade as a relatively young population with ages as young as
$\sim3$ Gyr, based on H$\beta$ EWs.  There is no compelling evidence
supporting the existence of a population of blue HB stars in
metal-rich quiescent galaxies large enough to explain the full range
of H$\beta$ EWs.  However, it is reasonable to assume that there will
be a population of blue HB stars in such systems, and they will affect
the derived stellar ages.

The higher order Balmer lines are promising because they will be even
more sensitive to warm/hot stars than H$\beta$.  In principle then,
the consideration of multiple Balmer lines should provide constraints
on multiple age components, and/or allow for the separation of age and
blue HB effects \citep{Schiavon04b, Schiavon07, Serra07}.  However,
the potential power of the higher-order Balmer lines is tempered by
the fact that these features are more sensitive to metallicity and
abundance ratios than H$\beta$ \citep{Thomas04, Korn05}.
\citet{Schiavon07} analyzed the H$\beta$, H$\gamma$, and H$\delta$
indices measured from stacked SDSS early-type galaxy spectra with a
variety of composite models including varying blue HB stars, blue
stragglers, and multi-age populations.  He concluded that small
fractions of young/intermediate-age stellar populations, perhaps in
the form of extended SFHs, was the most likely scenario to explain the
observed trends.  This proposal is similar, though not identical to
the `frosting' of young stars proposed by \citet{Trager00} in order to
explain the observed correlations between velocity dispersion, age,
metallicity, and abundance pattern.

There exist other proposed methods for constraining multi-age
components based on spectral indices.  \citet{Rose84} proposed a
\ionn{Ca}{ii} index that is sensitive to the H$\epsilon$ Balmer line.
\citet{Leonardi96} demonstrated that this index in conjunction with
H$\delta$ can break the `age-strength degeneracy', referring to the
age and strength of a recent burst of star formation.
\citet{Smith09b} used this diagnostic to argue against a frosting of
small fractions of young stars ($<1$ Gyr) to explain the young
apparent ages for quiescent galaxies in the Shapley supercluster.
\citet{Kauffmann03a} considered both the H$\delta$ and $D_n4000$
spectral features in an attempt to measure two-component SFHs (in
particular, a burst of SF and an underlying $\tau$ model SFH).
Galaxies with a sizable mass fraction formed in a recent burst of SF
are significant outliers in the H$\delta$-$D_n4000$ plane, and so
strong constraints on the burst fraction can be obtained in this way.
\citet{Percival11} argued that the \ionn{Ca}{ii} Rose index and the
\ionn{Mg}{ii} feature near 2800\,\AA\, could also be used to separate
age from blue HB effects.

\citet{Renzini86} suggested that the onset of the AGB in $\sim0.1-1$
Gyr old populations could be used to age-date galaxies in the NIR.
Subsequent modeling efforts by \citet{Maraston05} confirmed that NIR
spectral features attributable to oxygen-rich and carbon-rich AGB
stars vary strongly over the age range $\sim0.1-1$ Gyr, and may be
used as an independent constraint on the presence of intermediate-age
populations in galaxies.  In principle, the very strong sensitivity of
AGB evolution to mass (age) and metallicity suggests that they should
be good clocks.  However, their sensitivity to mass and metallicity is
due to poorly understood physics, including convection and mass-loss,
and so the calibration of an AGB-based clock is quite uncertain.
\citet{Miner11} attempted to use NIR spectral features to place
constraints on the ages of M32 and NGC 5102 based on Maraston's
models.  They found broad qualitative agreement between NIR-based ages
and CMD-based ages.  Future quantitative modeling of the optical-NIR
spectra will be required to assess the utility of AGB-sensitive
spectral indices as age indicators.

Finally, it is also worth emphasizing that it is fundamentally very
difficult to measure ages for systems older than $\gtrsim9$ Gyr
because evolution in the isochrones is so slow at late times.  For
example, at solar metallicity between $9-13.5$ Gyr, the RGB changes by
only $\approx50$ K while the main sequence turnoff point changes by
$\approx230$ K.  Very accurate models and very high quality data are
required to be able to distinguish such small temperature variations.
In theory the UV should be much more sensitive to age at late times
due to the onset of an extended (blue) HB \citep{Yi99}.  In the
context of a particular model, the UV colors become very sensitive to
age for ages $\gtrsim7$ Gyr.  The problem with this chronometer is
that it is very sensitive to the uncertain physical inputs, such as
the mass-loss prescription and the helium abundance, and also to the
underlying distribution of stellar metallicities \citep{Yi97}.  These
uncertainties will render absolute ages unreliable, although relative
ages (or even simply the rank-ordering of galaxies by age) are
probably more robust.

\subsection{Which Model SFHs to Use?}

\citet{Papovich01} recognized that SED modeling suffered from a
fundamental limitation, namely that the young stars, being so bright,
tend to outshine the older, less luminous stars.  This outshining
effect was known to result in underestimated stellar masses when
single-component SFH models are used \citep[e.g.,][]{Papovich01,
  Daddi04, Shapley05, LeeSK09, Pforr12}, but its effect on recovered
SFRs and ages remained relatively less well-explored until recently.
\citet{LeeSK09} analyzed the SEDs of high-redshift mock galaxies and
concluded that the use of single-component exponentially-decreasing
SFHs ($\tau$ models) resulted in substantially underestimated SFRs and
overestimated mean ages (by factors of two for both parameters).  They
attributed this shortcoming to the fact that the mock galaxies tended
to have rising SFHs in contrast to the declining model SFHs.
Subsequent work by \citet{LeeSK10}, \citet{Maraston10},
\citet{Wuyts11}, and \citet{Pforr12} largely confirmed these findings,
and further concluded that models with rising SFHs tended to provide a
better fit to observed high redshift SEDs and produced SFRs in better
agreement with other indicators.  Relatedly, the choice of the allowed
range of each SFH parameter (i.e., the prior) can significantly affect
the inferred best-fit SFRs and ages \citep[e.g.,][]{LeeSK10, Wuyts11}.

A specific example of the effect of the adopted SFH parameterization
on SFRs is shown in Figure \ref{f:ssfr}.  In this figure the specific
SFR is shown as a function of redshift at a reference stellar mass of
$10^{9.5}\,\Msun$.  At high redshift the scatter between the ``RIS'',
``DEC'', and ``REF'' models is entirely due to the choice of the
assumed SFH (rising, declining, or constant, respectively).  At the
highest redshifts these different choices result in a change in
specific SFRs of more than an order of magnitude.

There has been some difference in opinion regarding which functional
form one should adopt for the rising SFHs.  \citet{Maraston10}, and
\citet{Pforr12} advocated exponentially increasing SFHs while
\citet{LeeSK10} advocated delayed $\tau$ models of the form SFR$\propto
t\,e^{-t/\tau}$.  The data do not presently favor one functional form
over the other.  The basic conclusion to draw from these analyses is
that the model SFH library must be sufficiently diverse to allow for a
wide range in SFH types.  Model SFHs that are too restricted will
inevitably lead to biased results.  Of course, the criterion for
`sufficiently diverse' SFHs will depend on the galaxy type in
question and has not been thoroughly examined in a general sense.  At
present, all that can be stated with confidence is that for certain
SED types, especially highly star-forming galaxies, great care must be
taken when constructing the model library of SFHs.

Another option is to use a library of SFHs drawn from semi-analytic or
hydrodynamic models of galaxy formation.  This technique was
introduced by \citet{Finlator07}, who analyzed SEDs of six high
redshift galaxies with SFHs drawn from their hydrodynamic simulations.
The advantage of this approach is that the model SFH library is more
likely to contain realistic SFHs (with rising and falling SFHs,
multiple bursts, etc.)  than simple analytic models.  Recently,
\citet{Pacifici12} explored a similar technique using SFHs from a
semi-analytic model rather than a hydrodynamic simulation.  The
advantage of using a semi-analytic model is that a much larger range
of SFHs can be computed in a given amount of computational time.  As
with any other SFH library, the results can be sensitive to the choice
of priors.  In this case, the `prior' is in some sense the reliability
of the model in producing a realistic range of SFHs.  A possible
disadvantage of this approach is that it can be much more difficult to
quantify the influence of this prior on the derived physical
properties than when using analytic SFH libraries.

\subsection{Non-Parametric Star Formation Histories}

\begin{figure}
\centerline{\psfig{figure=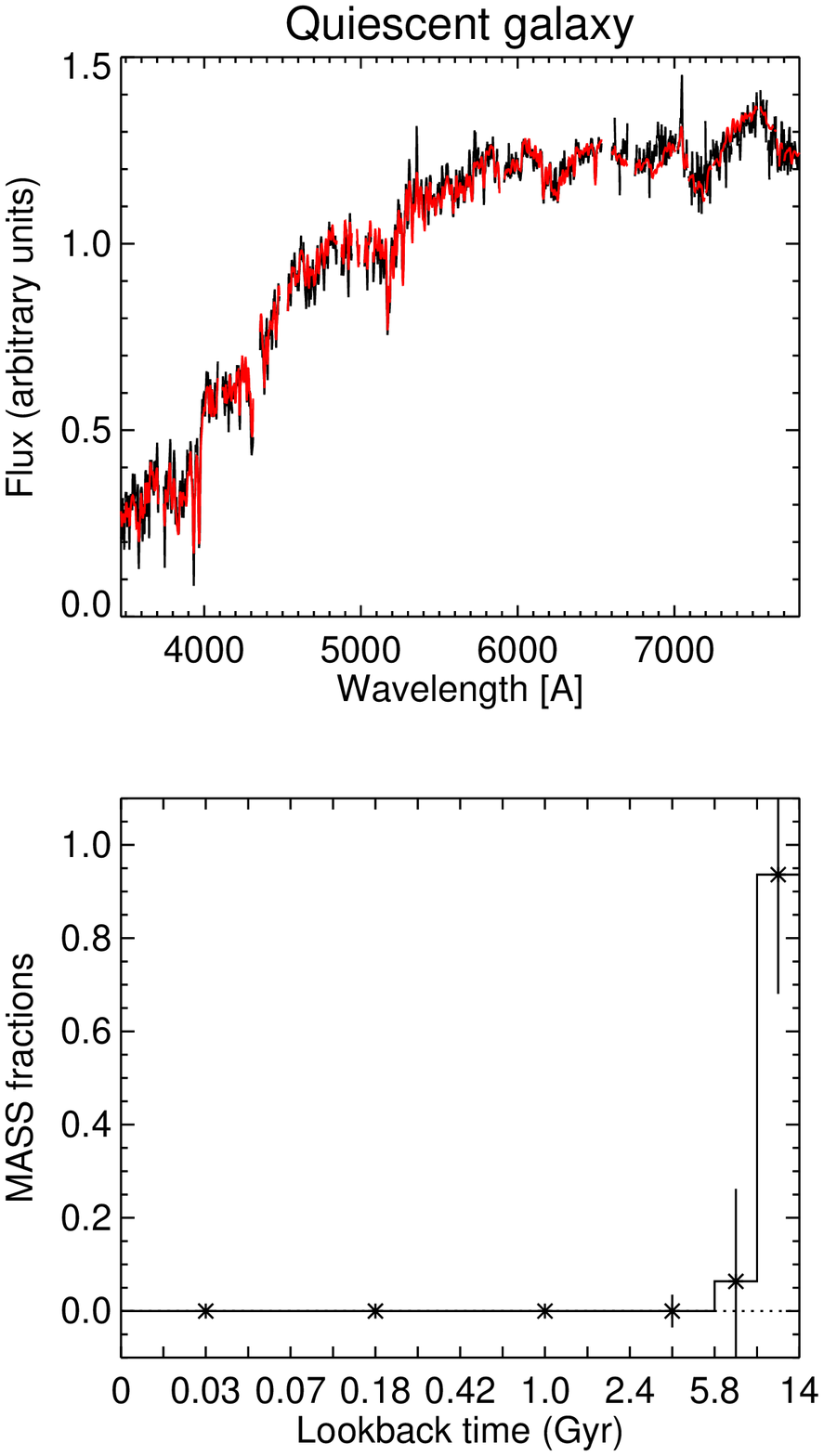,height=4in}
\psfig{figure=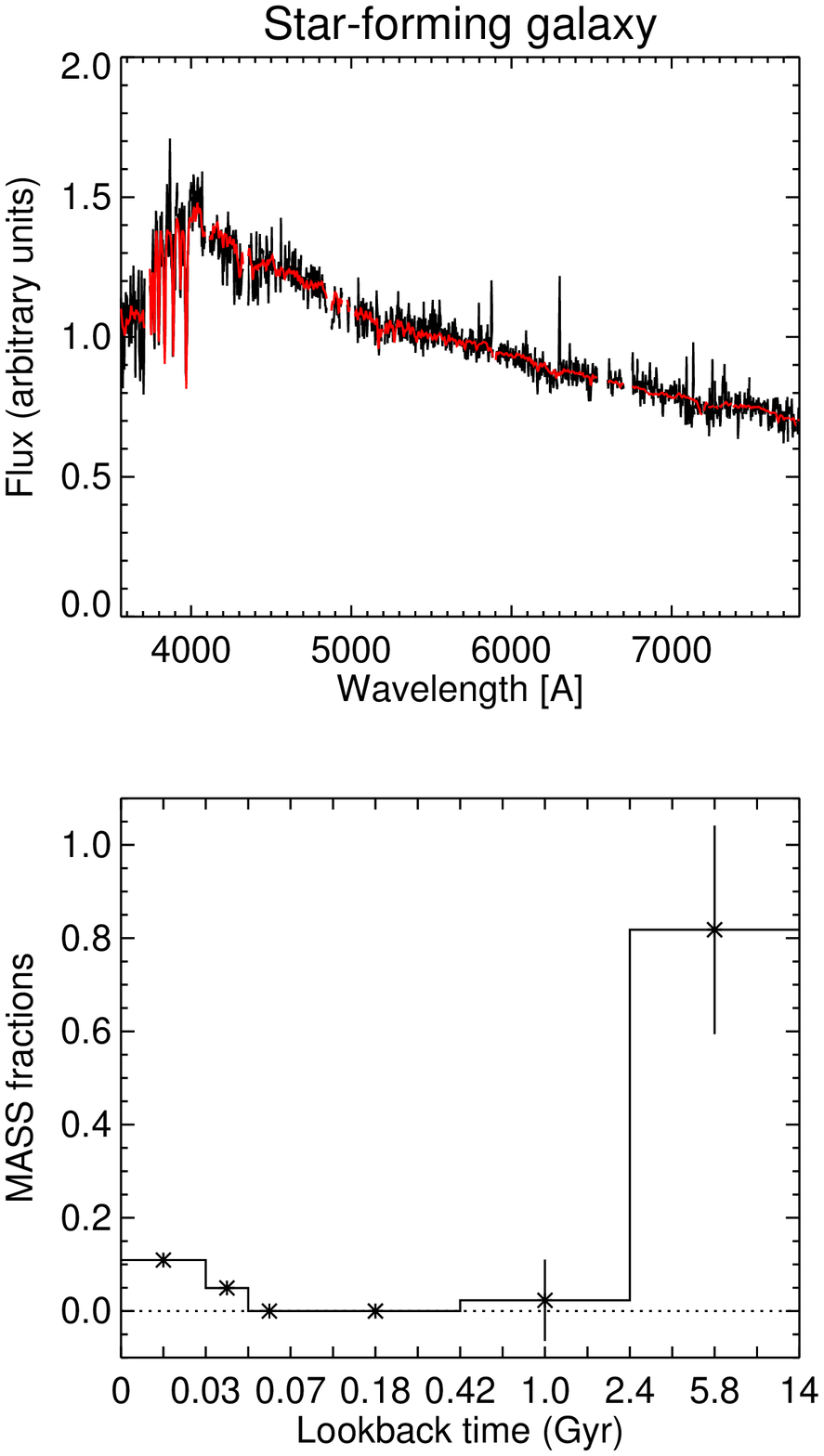,height=4in}}
\caption{Example fits to the optical spectrum of a quiescent and a
  star-forming galaxy with VESPA \citep{Tojeiro09}.  The SFHs are
  estimated non-parametrically with an age binning that is adaptive
  depending on the S/N and information content of the spectrum.  The
  upper panels show the best-fits (red line is the model, black line
  is the data; emission lines have been excluded from the fits); lower
  panels show the recovered SFHs plotted as the mass fraction formed
  within each age bin.  Figure courtesy of R. Tojeiro.}
\label{f:vespa}
\end{figure}

When modeling SEDs one typically must adopt some parameterization of
the SFH, and it is this step that clearly introduces a number of
poorly quantified systematics.  An appealing solution to this problem
is to model SEDs with non-parametric SFHs.  This is precisely the
approach taken by several groups, including MOPED \citep{Heavens00},
STARLIGHT \citep{CidFernandes05}, STECMAP \citep{Ocvirk06}, VESPA
\citep{Tojeiro07}, ULySS \citep{Koleva09}, and \citet{MacArthur09}.
In the simplest implementation one specifies a fixed set of age bins
and then fits for the fraction of mass formed within each bin.  More
sophisticated algorithms utilize adaptive age binning that depends on
the SED type and S/N \citep[e.g.,][]{Tojeiro07}.  In all cases one is
engaging in full spectrum fitting (in the restframe optical; for an
example of the technique see Figure \ref{f:vespa}).  In principle one
could use such codes to fit broadband SEDs, but the fits would likely
be seriously under-constrained.  These methods have been tested in a
variety of ways, and it appears that they are capable of recovering
complex SFHs remarkably well, at least when the mock galaxies are
built with the same SPS models as used in the fitting routines
\citep{Ocvirk06, Tojeiro07, Koleva09}.  Not surprisingly, the
requirements on the data quality are demanding: a wide wavelength
coverage, high spectral resolution, and high S/N (typically $>50$/\AA)
are required in order to robustly recover complex SFHs
\citep{Ocvirk06, Tojeiro07}.  In addition, there are clear systematics
in the recovered SFHs due to different SPS models, particularly for
the SFHs at ages of $\sim0.1-1$ Gyr \citep{Tojeiro09, Tojeiro11}.
This is perhaps related to the different treatment of AGB stars and/or
core convective overshooting amongst SPS models.  Moreover, current
implementations of non-parametric SFH recovery utilize SPS models that
only allow variation in metallicity and age --- the abundance pattern
is fixed to the solar value.  When fitting to spectra this practice
can introduce additional systematics since many of the features that
are being fit are sensitive to the detailed abundance pattern of the
system.

There have been few comparisons between the SFHs derived via
non-parametric methods with more conventional techniques.
\citet{Tojeiro09} compared SFRs estimated with VESPA to emission
line-based SFRs from \citet{Brinchmann04}.  No obvious discrepancies
were found, but further detailed analyses would be desirable.
Moreover, there appears to be no published work comparing the mean
stellar ages (or some other moment of the SFH) between these various
techniques.

\begin{figure}
\centerline{\psfig{figure=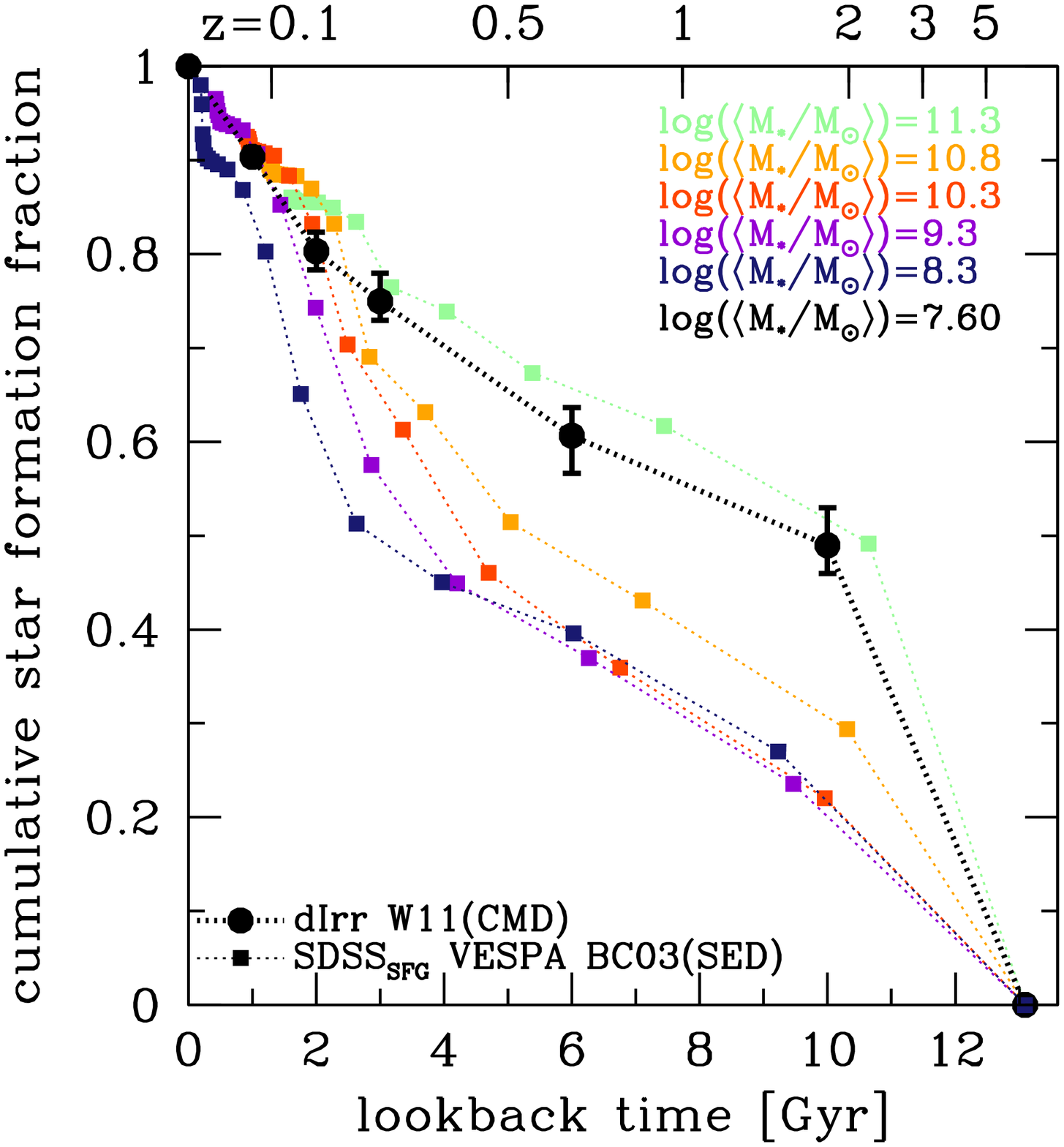,height=2.8in}}
\caption{Comparison between SFHs estimated from integrated light SEDs
  to SFHs estimated from CMDs.  The SED-based SFHs are based on SDSS
  spectra, as derived in \citet{Tojeiro09}, and binned by stellar
  mass. The CMD-based SFHs are from \citet{Weisz11} based on data from
  the ANGST survey and are for low-mass dwarf irregulars.  It is
  striking that the CMD-based SFHs for low-mass dwarfs look closer to
  the SED-based SFHs for the most massive galaxies.  From
  \citet{Leitner12}.}
\label{f:sfh}
\end{figure}

\paragraph{SFHs from CMDs versus SFHs from Integrated Light}
Ideally one would like an independent constraint on the SFH to compare
against the SFH derived from integrated light spectra.  Unfortunately
no direct tests of this sort have been performed to date.  The most
robust method for determing SFHs is via the modeling of resolved color
magnitude diagrams of the stars within galaxies.  This technique is
well-developed, and has recently been applied by \citet{Weisz11} to a
large sample of dwarf galaxies in the local volume compiled by the ACS
Nearby Galaxy Treasury Survey \citep[ANGST;][]{Dalcanton09}.
\citet{Leitner12} compared the SFHs derived by \citet{Weisz11} to the
SFHs derived from integrated light spectra by \citet{Tojeiro09}.  The
result is shown in Figure \ref{f:sfh}.  The dwarf galaxies from the
ANGST Survey have masses slightly lower than the lowest mass bin in
the integrated light sample, and it is worth emphasizing that there is
no overlap between the two samples.  And yet, the dwarfs have SFHs
comparable to the highest mass bin in the integrated light sample.
There are many complications in making this comparison, not the least
of which is the question of whether or not nearby dwarfs are
representative of the cosmological average.  Nonetheless, the
difference in the qualitative trend derived from CMD-based and
integrated light-based SFHs is striking and worthy of further study.

\subsection{Low-Level Star Formation}
\label{s:lowsfr}

Probing low levels of star formation in galaxies is challenging
because at low levels classic SF indicators can become contaminated by
other processes.  At low luminosities, emission lines can be
influenced by low-level AGN activity and they become difficult to
measure on top of the strong absorption features in the continuum,
dust emission can be due to heating from old stars (rather than
tracing the formation of new stars), and ultraviolet light can be
attributed to hot evolved low-mass stars.  The former two indicators
have been effectively abandoned as low-level SF indicators, while the
third, UV light, still holds some promise.

It is beyond the scope of this review to discuss in detail the
complications in interpreting UV (especially FUV) emission from
predominantly quiescent systems, but a few remarks will serve to round
out this section.  The reader is referred to \citet{Oconnell99} for a
thorough review, and \citet{Yi08} for a recent update.  Low-level UV
emission from galaxies can be due either to young massive stars or to
hot evolved low-mass stars.  In the latter category includes extreme
HB stars, post-AGB stars, AGB-Manqu\'e, etc.  Extreme HB stars are
known to exist in essentially all low metallicity and some high
metallicity globular clusters.  Post-AGB stars should exist at all
metallicities, and AGB-Manqu\'e are thought to occur only in special
circumstances.  It is very difficult to predict the specific numbers
of these stars as a function of metallicity and time, and it is for
this reason that low-levels of UV flux are so difficult to interpret.
The recent catalog of evolved hot stars in Galactic globular clusters
by \citet{Schiavon12b} may help to calibrate the models for these
advanced evolutionary phases.

There have been several important developments since the review of
\citet{Oconnell99}, which are mainly due to progress provided by {\it
  HST} and {\it GALEX}.  It is now clear that galaxies on the
optically-defined red sequence (e.g., in $B-V$ or $g-r$ color space)
frequently harbor low levels of star formation, at the rate of
$\sim0.1-1\,\Msun\,{\rm yr}^{-1}$ \citep{Yi05, Kaviraj07a, Salim10,
  Salim12, Fang12}, although for the most massive galaxies SF levels
appear to be even lower.  The implication is that optical SEDs are not
sensitive to specific SF at levels of $\lesssim10^{-11}$ yr$^{-1}$.
It is also now quite clear that optically-defined red sequence
galaxies have a wide range of UV colors, from those with low levels of
star formation to those that are extremely red.  It has become common
practice to define an arbitrary UV-optical color boundary in order to
separate star forming from quiescent galaxies \citep{Yi05, Jeong09}.
Of course, such a practice does not provide an answer to the question
of how low star formation rates actually are in the reddest galaxies.
Instead it signals the difficulty in measuring SFRs for such red
systems because any UV flux becomes easily confused with hot evolved
low-mass stars.  Further progress can be made by directly resolving
luminous young stars and star clusters in nearby galaxies.  This was
precisely the strategy taken by \citet{Ford12}, who observed four
nearby elliptical galaxies with {\it HST}.  These authors argued that
the galaxies in their sample have current specific SFRs in the range
of $10^{-14}-10^{-16}$ yr$^{-1}$, far below what can be reliably
probed via SED modeling.  UV spectroscopic features may also aide in
distinguishing hot evolved low-mass stars from hot upper main sequence
stars \citep{Oconnell99}.

\subsection{Summary}

Measuring the detailed SFH of a galaxy from its observed SED is one of
the holy grails of SPS modeling.  Unfortunately, two unrelated
physical effects make this task very difficult in practice: (1) high
mass stars are so luminous that they dominate the SEDs of star-forming
galaxies, easily outshining the more numerous, older, lower mass
stars; (2) at late times isochrones evolve very little, so that it is
difficult in practice to distinguish an $\sim8$ Gyr population from a
$\sim12$ Gyr one.  The first effect implies that light-weighted ages
will always underestimate the true mass-weighted age.  It also implies
that the priors on the model SFHs can strongly influence the best-fit
parameters.  Non-parametric SFH reconstruction methods offer the best
hope of providing unbiased results, but they require data of very high
quality.  As a general rule of thumb, optical SEDs are sensitive to
specific SFRs as low as $~10^{-11}$ yr$^{-1}$, UV fluxes are sensitive
to levels as low as $~10^{-12}$ yr$^{-1}$, while resolved UV
photometry is in principle sensitive to specific SFRs down to at least
$\lesssim10^{-14}$ yr$^{-1}$.  SED-based SFRs are probably accurate at
the factor of two level (at a fixed IMF), except for extreme systems
where uncertainties can be much larger.  There appears to be little
consensus on the accuracy of light-weighted and mass-weighted stellar
ages derived from SED fitting.


\section{STELLAR METALLICITIES \& ABUNDANCE PATTERNS}

\subsection{The Age-Metallicity Degeneracy}

The metallicity, $Z$, affects the SED in two distinct ways: (1) an
increase in metallicity results in lower effective temperatures,
including a cooler main sequence and giant branch; (2) at fixed
$\teff$, an increase in metallicity results in stronger spectroscopic
absorption features and generally redder colors.  Both of these
effects contribute to the overall reddening of an SED with increasing
metallicity.  In addition to the overall metal content, the detailed
elemental abundance pattern is imprinted on the SED, especially in the
strength of spectral absorption features.

Age will also tend to redden the SED through the effect of age on the
isochrones.  \citet{Worthey94} studied the degeneracy between age and
metallicity in detail using his SPS models and concluded that at ages
$>5$ Gyr neither broadband colors nor most spectral indices were able
to reliably separate age from metallicity.  Worthey introduced his
``3/2 rule'', whereby an increase/decrease in the population age by a
factor of three is almost perfectly degenerate with an
increase/decrease in metallicity by a factor of two.  He identified a
small number of spectral indices that were unusually sensitive to age,
including the hydrogen Balmer lines, and several indices that were
unusually sensitive to metallicity, including the Fe4668 and Fe5270
indices \citep[see also][for early attempts at separating age and
metallicity]{Oconnell80, Rabin82, Rose85}.  Worthey therefore
demonstrated conclusively that the age-metallicity degeneracy could be
broken by jointly considering a handful of carefully chosen indices.
He also pointed out that the age-metallicity degeneracy is less severe
at low metallicity, where the turnoff point and giant branch have very
different age sensitivity at fixed metallicity.

\begin{figure}
\centerline{\psfig{figure=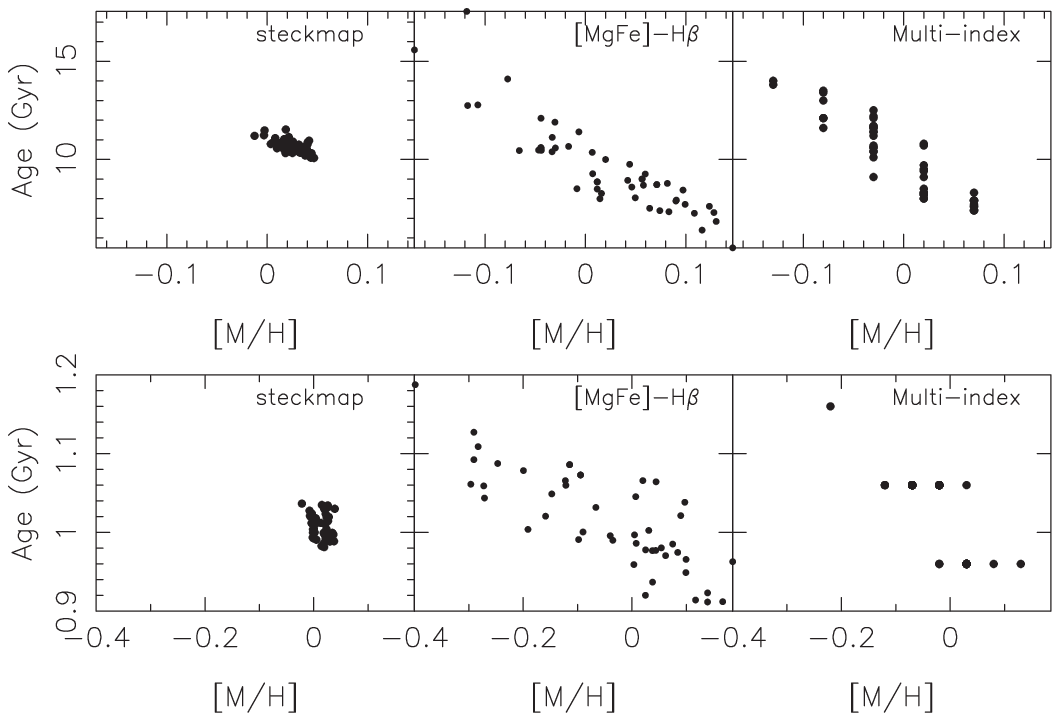,height=4in}}
\caption{Illustration of the age-metallicity degeneracy in spectra of
  modest S/N.  Three techniques are compared, the full spectrum
  fitting technique of \citet[][left panel]{Ocvirk06}, classic
  two-index fitting (middle panel), and multi-index fitting (right
  panel).  Results are shown for a mock galaxy with S/N$=50/$\AA,
  [M/H]$=0.0$, and an age of 10 Gyr (top panels) and 1 Gyr (bottom
  panels).  Each point represents the best-fit for a single
  realization of the mock galaxy spectrum.  From
  \citet{Sanchez-Blazquez11}. }
\label{f:agez}
\end{figure}

It has long been appreciated that a clean separation of age and
metallicity can be achieved with Lick indices only for data of very
high S/N \citep[$\gtrsim100$/\AA;][]{Trager00a, Kuntschner00}.  Figure
\ref{f:agez} demonstrates that an age-metallicity degeneracy persists
when exploiting Lick indices at lower S/N. In contrast, full spectral
fitting offers a much more robust separation of age and metallicity at
moderate to low S/N \citep[at least within the context of an
underlying set of model SSPs; see][]{MacArthur09, Sanchez-Blazquez11}.
This is simply a consequence of the fact that much more information is
available in the full optical spectrum than can be captured by a
limited number of spectral indices.

\subsection{Photometric Metallicities}
\label{s:pz}

The age-metallicity degeneracy discussed in \citet{Worthey94} strictly
applies to systems that are coeval with ages $>5$ Gyr; models at
younger ages were not considered by Worthey.  In subsequent work it
became clear that the broadband optical-NIR colors of composite
stellar populations could separate age and metallicity effects.  The
physics underlying the idea is that blue colors probe the
age-sensitive main sequence turnoff point while red/NIR colors probe
the metallicity-sensitive giant branches.  \citet{Bell00} and
\citet{MaCarthur04}, amongst others, showed that age and metallicity
vectors were nearly orthogonal in optical-NIR colors (e.g., $B-R$
vs. $R-H$).  These authors therefore concluded that estimates of
stellar metallicities were possible from photometric data alone, for
systems whose light-weighted ages were not too old.  \citet{Eminian08}
showed that galaxies grouped by spectroscopically-derived ages and
metallicities separate relatively well in $g-r$ vs. $Y-K$ color-color
space.  Notice that this statement does not depend on the predicted
model colors, thereby providing strong evidence that separation of age
and metallicity is at least in principle possible based on broadband
optical-NIR data alone.

\citet{LeeHC07} investigated the model uncertainties associated with
separating age and metallicity based on broadband colors.  They echoed
previous work in that the young populations in composite stellar
models dominate the light and therefore allow a separation of age and
metallicity effects when optical-NIR colors are employed.  However,
they also emphasized that different SPS models produce very different
age-metallicity vectors in color-color space, a point also emphasized
by \citet{Eminian08}.  The result is that age and metallicity can be
separately constrained within the context of a particular SPS model,
but the derived ages and metallicities will vary from model to model.
The underlying cause of this problem is that the NIR light is heavily
influenced by AGB stars for populations of moderage age, and so
different treatments of this uncertain evolutionary phase cause
substantial differences in the age-metallicity grids.  Because of the
difficulty in measuring metallicities from photometric data, many SPS
studies simply treat metallicity as a nuisance parameter to be
marginalized over.  Other studies take the even simpler approach of
fixing the metallicity to the solar value.  The danger in making the
latter simplification is that assumptions in the assumed metallicity
will generally affect other parameters of interest, such as the
stellar mass, dust opacity, and star formation rate \citep[see
e.g.,][]{Wuyts09, Marchesini09, Muzzin09, Pforr12}.

\subsection{Spectroscopic Metallicities}

The more robust approach to estimating stellar metallicities is to
employ spectroscopic features.  Optical spectra of galaxies are rich
in atomic and molecular absorption features that are readily apparent
even at low spectral resolution ($R\sim1000$).  For a fixed population
age, the strengths of the absorption features will depend not only on
the overall metallicity, $Z$, but also on the detailed elemental
abundance pattern.  In general, this greatly complicates
interpretation of absorption features and spectral indices.  However,
with the aide of models that allow for variation in both metallicity
and abundance pattern, one can search for combinations of features
that are relatively robust against abundance pattern variations.
\citet{Thomas03} refined a proposal by \citet{Gonzalez93} of an index
comprised of magnesium and iron lines that is almost entirely
insensitive to the level of $\alpha$ enhancement, [$\alpha$/Fe], and
is therefore a relatively robust tracer of $Z$.  The index is defined
as [MgFe]'$\equiv\sqrt{{\rm Mgb}\,(0.72\,{\rm Fe}5270+0.28\,{\rm
    Fe}5335)}$, where Mgb, Fe5270, and Fe5335 are Lick indices defined
in \citet{Worthey94b}.  The power of this index, and others like it
\citep[see e.g.,][]{Bruzual03}, is that variable abundance pattern
models are then not required to measure total metallicities from
galaxy spectra.  However, a significant caveat to this approach is the
assumption that abundance patterns in galaxies can be described by two
parameters: [Fe/H], and [$\alpha$/Fe].  In reality, analyses of high
resolution stellar spectra in the Galaxy consistently demonstrate
that the $\alpha$ elements do not vary in lock-step
\citep{Edvardsson93, Venn04, Fulbright07}.  This means that the true
total metallicity, $Z$, can only be accurately estimated if the
detailed abundance patterns are known.

\citet{Gallazzi05} applied the idea that [MgFe]' is insensitive to
[$\alpha$/Fe] (assuming that the $\alpha$ elements, C and N all vary
in lock-step) to the analysis of 175,000 galaxies from the SDSS with
the aide of the \citet{Bruzual03} SPS models.  Several age-sensitive
and $Z-$sensitive spectral indices were used in their analysis. They
derived light-weighted stellar metallicities for individual galaxies
with statistical errors as low as 0.1 dex for the highest S/N spectra.
These authors also explored a variety of systematic effects and
concluded that these could add $\sim0.1$ dex systematic uncertainty to
the error budget on $Z$.  Probably the dominant systematic in the
results of Gallazzi et al., and others based on similar techniques, is
the assumption that the $\alpha$ elements and C and N all vary in
lock-step.

Another technique is to fit the entire optical spectrum with models
that allow for variation in age and total metallicity, and to simply
ignore the possibility that the observed spectrum may have non-solar
abundance patterns.  This approach has become increasingly common as
automated spectral fitting codes garner wider use.  \citet{Tojeiro11}
fit stacked spectra of luminous red galaxies from SDSS as a function
of luminosity and redshift with the VESPA code.  Remarkably, they
found essentially no variation in $Z$ with galaxy luminosity (though
the sample was restricted to very luminous galaxies), and for the
\citet{Conroy09a} and \citet{Maraston05} SPS models there was no
evolution in $Z$ with redshift over the interval $0.1<z<0.5$.  When
the \citet{Bruzual03} models are employed, VESPA favored higher
metallicities at higher redshift.  Variation in the best-fit
metallicity between SPS models points toward systematic uncertainties
in the models at the $\sim0.2$ dex level.

\citet{Koleva08} investigated the accuracy of full spectrum fitting
codes at recovering metallicities.  They considered both the ability
of one SPS model to recover the known metallicity of a second model
and the ability of an SPS model to recover the known metallicities of
Galactic globular clusters.  They concluded that the high-resolution
version of the \citet{Bruzual03} model carries large systematic
uncertainties due to its use of the STELIB stellar library.  STELIB
does not sample a wide enough range in metallicity, and so
metallicities derived from the Bruzual \& Charlot models will carry
large uncertainties.  The other models tested, including the
Vazdekis/MILES and Pegase-HR models, reproduced the known globular
cluster metallicities to an accuracy of $\approx0.14$ dex.

Of course, the most robust method for estimating stellar metallicities
is with models that self-consistently include abundance pattern
variation.  Such models have existed for the Lick indices since the
late 1990s while models for the full optical-NIR spectrum have only
recently become available \citep{Coelho07, Walcher09, Conroy12a}.

\citet{Trager00a, Trager00} modeled three spectral indices with SPS
models that allowed for variation in abundance patterns.  Trager and
colleagues focused on fitting for the age, total metallicity, $Z$, and
enhancement ratio, [$\alpha$/Fe].  In practice, they grouped elements
into two types, the `depressed' and the `enhanced' group, with the
latter comprised of $\alpha$ elements except Ca, plus C and N, while
the former comprised the iron-peak elements and Ca.  In subsequent
work, \citet{Thomas05} modeled the same set of spectral indices as in
Trager et al., though with their own SPS models and with an expanded
dataset.  Their conclusions largely echoed those of Trager and
collaborators, with the novel finding that at fixed velocity
dispersion galaxies in denser environments appeared to be more
metal-rich than galaxies in less dense environments.

As discussed in Section \ref{s:sfr}, most analyses of Lick indices fit
single-age models to the data.  \citet{Serra07} and \citet{Trager09a}
considered the effect of multiple age components in the fitting of the
Lick indices of mock galaxies and found that the derived metallicities
agreed well with the light-weighted mean metallicities when the
youngest component was older than $\sim3$ Gyr.  For young components
with ages $1-3$ Gyr there was a slight bias in the sense that the
derived metallicities were $\sim0.1$ dex higher than the true
light-weighted mean metallicities.  These authors did not consider
young components younger than 1 Gyr.

The estimation of total metallicity with models that include abundance
pattern variation are the most robust but also the most restricted.
Models such as those by Thomas et al., Conroy \& van Dokkum, Worthey,
or Schiavon are limited to ages $>1$ Gyr, and therefore are routinely
applied only to early-type galaxies.  A major area for future growth
in this field is the extension of these models to younger ages in
order to model galaxies of all types.

\subsection{Abundance Patterns}

\citet{Wallerstein62} was the first to show that non-solar abundance
patterns were common amongst metal-poor stars in the Galaxy.  It was
therefore natural to wonder if other galaxies harbored non-solar
abundance patterns.  Early data suggested that magnesium may vary more
than iron amongst the early-type galaxies \citep{Oconnell76,
  Peterson76, Peletier89}, but it was with the availability of new,
relatively high-quality spectral synthesis models that
\citet{Worthey92} were able to firmly establish that the [Mg/Fe] ratio
varied within the early-type galaxy population.  The discovery of
non-solar abundance patterns has been one of the most significant
discoveries afforded by population synthesis models.  The ratio
between the abundance of $\alpha-$elements and iron-peak elements,
[$\alpha$/Fe], is considered to be particularly valuable, as it
provides insight into the SFH of galaxies \citep{Tinsley79}.
Following the development of models that allowed for arbitrary
variation in abundance patterns, the measurement of abundance ratios
became routine.  Particular attention has been paid to [Mg/Fe], in
large part because it is the most readily measurable abundance ratio
in moderate resolution spectra, thanks to the numerous strong
\ionn{Fe}{i} features and the Mg{\it b} feature at $5200$\,\AA\, (see
Figure \ref{f:specabund}).

As discussed above, \citet{Trager00a, Trager00} presented the first
systematic investigation of abundance patterns in early-type galaxies.
They found enhancement ratios, [$\alpha$/Fe] generally in excess of
zero, with a strong correlation with velocity dispersion.  As Trager
and collaborators only fit H$\beta$, Mgb, and two iron indices, their
[$\alpha$/Fe] is in reality a statement about [Mg/Fe].  Since then,
numerous authors, using different SPS models, have all reached the
same conclusion that [Mg/Fe] exceeds zero and increases with galaxy
velocity dispersion \citep{Thomas05, Schiavon07, Smith09, Graves09a,
  Johansson12}.

As noted in the previous section, early studies grouped elements into
two classes, `enhanced' and `depressed', and so only the total
metallicity and the enhancement factor could be measured from the data
\citep{Trager00, Thomas05}. More recent work has allowed for the
abundance of a larger number of elements to vary, including C, N, Ca,
and Ti \citep[see][for a complementary review to the one presented
below]{Schiavon10}.

The calcium abundance is most commonly measured from the \ionn{Ca}{i}
line at 4227\,\AA, as probed by the Ca4227 Lick index (see Figure
\ref{f:specabund}).  Other regions of the spectrum with strong
sensitivity to calcium abundance include the \ionn{Ca}{ii} H\&K lines
and the triplet of \ionn{Ca}{ii} lines at 8600\,\AA\, (CaT).  The
calcium abundance in early-type galaxies has been described as a
`puzzle' because its abundance does not seem to track magnesium,
despite the fact that they are both $\alpha$ elements.  This was noted
qualitatively in the early work of \citet{Vazdekis97} and
\citet{Trager98} and quantified with variable abundance models by
\citet{Thomas03b}.  The CaT feature is weaker than even solar-scaled
models, for reasons that will be discussed in Section \ref{s:imf}.
\citet{Schiavon07} and \citet{Johansson12} fit the Ca4227 index (along
with others) of early-type galaxies from SDSS and find [Ca/Fe] values
close to zero.

Carbon and nitrogen abundances can be measured from the NH, CH, C$_2$,
and CN molecular absorption features in the blue spectral region (see
Figure \ref{f:specabund}).  \citet{Kelson06} were the first to probe
the carbon and nitrogen content of early-type galaxies based on Lick
indices.  \citet{Schiavon07} measured [C/Fe] and [N/Fe] from Lick
indices as a function of early-type galaxy luminosity based on stacked
spectra from SDSS.  He found that the abundance of these elements
increased significantly with increasing luminosity.
\citet{Johansson12} reached similar conclusions based on analysis of
individual early-type galaxies in SDSS.  The NH feature at
$\approx3360$\,\AA\, is in principle a powerful probe of [N/Fe]
because it is sensitive only to N (unlike CN).  To date this feature
has not been extensively utilized, both because it is near the
atmospheric transmission cutoff and thus difficult to observe and
because variable abundance models are not well-developed in this
spectral region.  Nonetheless, several authors have investigated
empirical trends of this feature with other properties of early-type
galaxies, with mixed results \citep{Davidge94, Ponder98, Serven11}.

Moderate resolution spectra of early-type galaxies respond strongly to
changes in the abundances of Fe, Mg, C, N, Ca, and Ti, and thus these
elements are the most easily measured.  However, the optical and NIR
spectrum of an old stellar population contains information on many
more elements.  In a forward-looking study, \citet{Serven05}
considered the spectral response due to variation in the abundance of
23 separate elements and concluded that for S/N$=100$ spectra and a
galaxy velocity dispersion of $\sigma=200\kms$, one could in principle
measure the abundance of C, N, O, Na, Mg, Al, Si, Ca, Sc, Ti, V, Cr,
Mn, Fe, Co, Ni, Sr, and Ba.  Of course, as the effects are often quite
subtle, all relevant systematics must be well-controlled.  As a first
step toward extracting these more subtle features, \citet{Conroy13a}
reported measurements of the abundances of the neutron-capture
elements Sr and Ba from analysis of moderate-resolution spectra of
nearby early-type galaxies.  Studies such as these should provide
important new clues to the formation histories of old stellar systems.

Attention here has focused exclusively on early-type galaxies with
little or no ongoing star formation, and for good reason.  The hot
massive stars that dominate the SEDs of actively star-forming galaxies
have generally weak metal absorption lines, especially in the
optical-NIR, and so their presence acts primarily to dilute the
strength of the metal lines of the older, cooler stars.  More
fundamentally, models simply do not currently exist to study the
abundance patterns of actively star-forming galaxies.

\paragraph{What About Oxygen?}
Oxygen is the most abundant element in the universe after hydrogen and
helium.  Its abundance relative to iron, [O/Fe], is therefore a
critical variable in both stellar interior models \citep{VandenBerg01}
and stellar atmospheres.  There are no transitions of atomic oxygen
visible in moderate resolution spectra, but molecules involving oxygen
(including TiO, H$_2$O, CO, and OH) do create significant features in
the optical-NIR spectra of cool stars.  TiO is the only molecule with
transitions in the optical (beginning at $\approx4500$\,\AA), and
unfortunately the work on response functions of the Lick indices to
abundance changes have neglected TiO line lists from their
computations \citep{Tripicco95, Korn05}.  The situation with oxygen is
quite complex because CO has the highest dissociation potential of any
molecule, and thus increasing the abundance of oxygen has a cascading
effect on other species.  For example, increasing oxygen causes more
carbon to be consumed by the formation of CO, which in turn causes a
decrease in the concentration of CN and C$_2$.  The oxygen abundance
must therefore be known, or assumed, before one can interpret other
spectral features.  The OH lines in the $H$-band probably afford the
most direct constraint on the oxygen abundance, but these features
have not been exploited because of a lack of available models.  This
limitation has now been overcome with the variable abundance ratio
models of \citet{Conroy12a} that extend to the $K$-band.  The upshot
is that direct constraints on the photospheric abundance of oxygen in
other galaxies currently do not exist, but they are possible with
newly available models.  This presents a significant limitation to
current models because of the strong effect of oxygen on the effective
temperatures, etc., of stars.  More troubling is the observation that
in the Galactic bulge the abundances of oxygen and magnesium do not
track each other \citep{Fulbright07}.  In fact, bulge stars with
[Fe/H]$>0$, have [O/Fe]$<0.2$, [Si/Fe]$<0.2$, [Ca/Fe]$<0.2$, and
[Ti/Fe]$<0.2$, while [Mg/Fe]$\sim0.3$ for the same stars.  The lesson
to draw from the Galactic bulge is that one should be very cautious in
using [Mg/Fe] as a tracer of [$\alpha$/Fe].  Because of these
uncertainties surrounding oxygen, \citet{Thomas05} investigated its
effect on their analysis of Lick indices.  They concluded that if
oxygen is in reality decoupled from the other $\alpha-$elements (e.g.,
[O/Mg] is non-zero) then the ages, [Fe/H], and [Mg/Fe] abundances are
largely unchanged with respect to their standard model where oxygen
tracks the other $\alpha-$elements.  However, these authors only
considered the effect of oxygen on the stellar spectra; they did not
consider the non-trivial effect of oxygen on the isochrones
\citep{Dotter07}.  Further work is clearly needed regarding the
important role of oxygen in SPS models.

\paragraph{Z vs. [Fe/H]}
The iron abundance, [Fe/H], is often regarded as a proxy for the total
metallicity, $Z$, but, as discussed above, the correspondence between
these two quantities breaks down for non-solar abundance patterns.  As
shown in \citet{Trager00a}, the relation between [Z/H] and [Fe/H] can
be parameterized by [Z/H]$=$[Fe/H]$+A$[$\alpha$/Fe].  The constant $A$
depends on the detailed abundance pattern; for models in which all
$\alpha$ elements and C and N are grouped together, $A=0.93$, while
for models with [O/Fe]$=$[C/Fe]$=0$, $A=0.77$.  It needs to be
emphasized here that formulae such as the one presented above only
apply in the simple case in which all $\alpha$ elements vary in
lock-step and C and N either are solar-scaled or vary with $\alpha$.
This is probably not a good approximation to reality.  From the
modeling perspective, some have chosen to enhance or depress element
groups at fixed $Z$ \citep[e.g.,][]{Trager00a, Thomas03}, while others
have chosen to fix [Fe/H] as other elements vary
\citep[e.g.,][]{Schiavon07, Conroy12a}.  The choice would be rather
arbitrary and inconsequential were it not for the fact that most
models adopt a fixed set of isochrones as the abundances are varied.
The important question is then whether isochrones vary more at fixed
$Z$ or fixed [Fe/H].  At least for variation in [O/Fe],
\citet{Dotter07} have demonstrated that the isochrones vary in $\teff$
by approximately the same absolute amount (though with a different
sign) when the variation is computed at fixed $Z$ or fixed [Fe/H].
The disadvantage, at least conceptually, of computing models at fixed
$Z$ is that a variation in one element forces variation in another so
that $Z$ is held constant.

\subsection{The Low-Metallicity Tail}

Stars within the Galaxy have a wide range of metallicities, and so one
might expect other massive galaxies to harbor stars of varying
metallicity.  This expectation was confirmed with resolved color
magnitude diagrams of nearby early-type galaxies
\citep[e.g.,][]{Grillmair96, Harris00, Monachesi11}.  The difference
between a monometallic metal-rich stellar population and one that
includes even a small fraction of metal-poor stars will be most
apparent in the UV.  This occurs for two reasons: at low metallicity
the main sequence turn-off point is considerably hotter than at solar
metallicity, and at low metallicity stellar populations are observed
to contain significant numbers of blue HB stars.  The closed box model
of galactic chemical evolution predicts that $\approx10$\% of the
stars in an evolved galaxy will have [Fe/H]$<-1$.  Old stellar
populations with [Fe/H]$\lesssim-1$ can be expected to harbor blue HB
stars (by analogy with the Galactic globular clusters), and thus there
will be a large difference in the UV between a closed box metallicity
distribution function and one that is monometallic.

\citet{Bressan94} constructed spectrophotometric evolutionary models
for elliptical galaxies and found that closed box models produced an
excess of UV flux compared to observations.  They argued that the
failure of the model was due to the excess of metal-poor stars and
concluded that real early-type galaxies contain a more narrowly peaked
metallicity distribution function.  This conclusion mirrors the
observation in the solar neighborhood, where fewer low-metallicity
stars are observed than predicted by the closed box model \citep[this
is known as the `G-dwarf problem';][]{Tinsley80}.  \citet{Worthey96}
arrived at a similar conclusion from analysis of a UV-optical spectrum
of M31 and spectral indices for a sample of early-type and S0
galaxies.  These authors argued for a metal-poor component a factor of
$2-3$ lower than predicted by the closed box model.  From the analysis
of Lick indices of a sample of early-type galaxies, \citet{Greggio97}
also concluded that a closed box model for the metallicity
distribution failed to reproduce the data.

While it is reasonably clear that early-type galaxies contain fewer
low-metallicity stars than predicted by the closed box model, the more
general task of measuring the detailed metallicity distribution
function, or even the fraction of low-metallicity stars from
integrated light measurements is much more complex.  As discussed in
Section \ref{s:lowsfr}, the UV spectral region is sensitive to
minority populations of hot stars of many types, including metal-poor
main sequence stars, young stars, and exotic and uncertain
evolutionary phases such as blue stragglers, extreme HB stars, and
post-AGB stars.  The latter two are very hot, and so are more
influential at $\lambda<2000$\,\AA, where the UV upturn phenomenon in
early-type galaxies exists.  From consideration of the mid-UV
(2000\,\AA$<\lambda<$4000\,\AA) spectra of M31 and other nearby
galaxies, \citet{Worthey96} argued for metal-poor fractions $\leq5$\%.
\citet{Maraston00} modeled the mid-UV spectra of four nearby galaxies
and found metal-poor fractions $\leq6$\%, and \citet{Lotz00} argued in
favor of small, but non-zero fractions of metal-poor stars.
\citet{Maraston00} also highlight the point, first raised by
\citet{Greggio97}, that known metallicity gradients within galaxies,
when combined with projection effects, will `contaminate' nuclear
spectra of galaxies with metal-poor stars.  It is notable that
\citet{Maraston00} find evidence for a spread in metal-poor fractions,
ranging from $0-6$\%.  This suggests that one should include the
metal-poor fraction as an additional free parameter when fitting the
mid-UV of early-type galaxies.  Care must be taken when interpreting
these fractions because the definition of `metal-poor' varies,
although it generally refers to metallicities at least as low as
[Fe/H]$\lesssim-1.5$.  The problem boils down to the fact that the hot
star fraction can be measured from UV (and perhaps blue spectral)
data, but it is very challenging to convert this hot star population
into a fraction of metal-poor stars owing to uncertainties in advanced
phases of stellar evolution.

For actively star-forming galaxies essentially nothing is known
regarding their internal distribution of metallicities, as the young
stars overwhelm the subtle UV features signaling the presence or
absence of low-metallicity stars.

\subsection{Summary}

The technique of measuring metallicities and abundance patterns from
old stellar systems is well-established.  The age-metallicity
degeneracy can be robustly broken by considering the hydrogen Balmer
lines in conjunction with iron features in moderate resolution optical
spectra. Abundance ratios including [Mg/Fe], [C/Fe], [N/Fe], and
[Ca/Fe] are now routinely measured from optical spectra.  Other
elements including O, Na, Si, Sc, Ti, V, Cr, Mn, Co, Ni, Sr, and Ba
should also be measurable from high S/N optical-NIR spectra of old
stellar systems, and models are currently being developed to extract
this information.  It is possible to measure metallicities based on
photometric data, but the derived values depend on the underlying SPS
model and on various assumptions like the form of the SFH.
Photometric metallicities are therefore probably accurate only in a
relative sense.  Spectroscopic metallicities seem to be accurate at
the $\sim0.2$ dex level, with model systematics again being the
dominant source of error.  There is clearly information on the
distribution of stellar metallicities within galaxies, and previous
work suggests that it is possible to measure at least the low
metallicity tail of the distribution.  Further modeling is required to
understand precisely how much information on the metallicity
distribution function is available in the SEDs of quiescent galaxies.


\section{DUST}
\label{s:dust}

\begin{figure}
\centerline{\psfig{figure=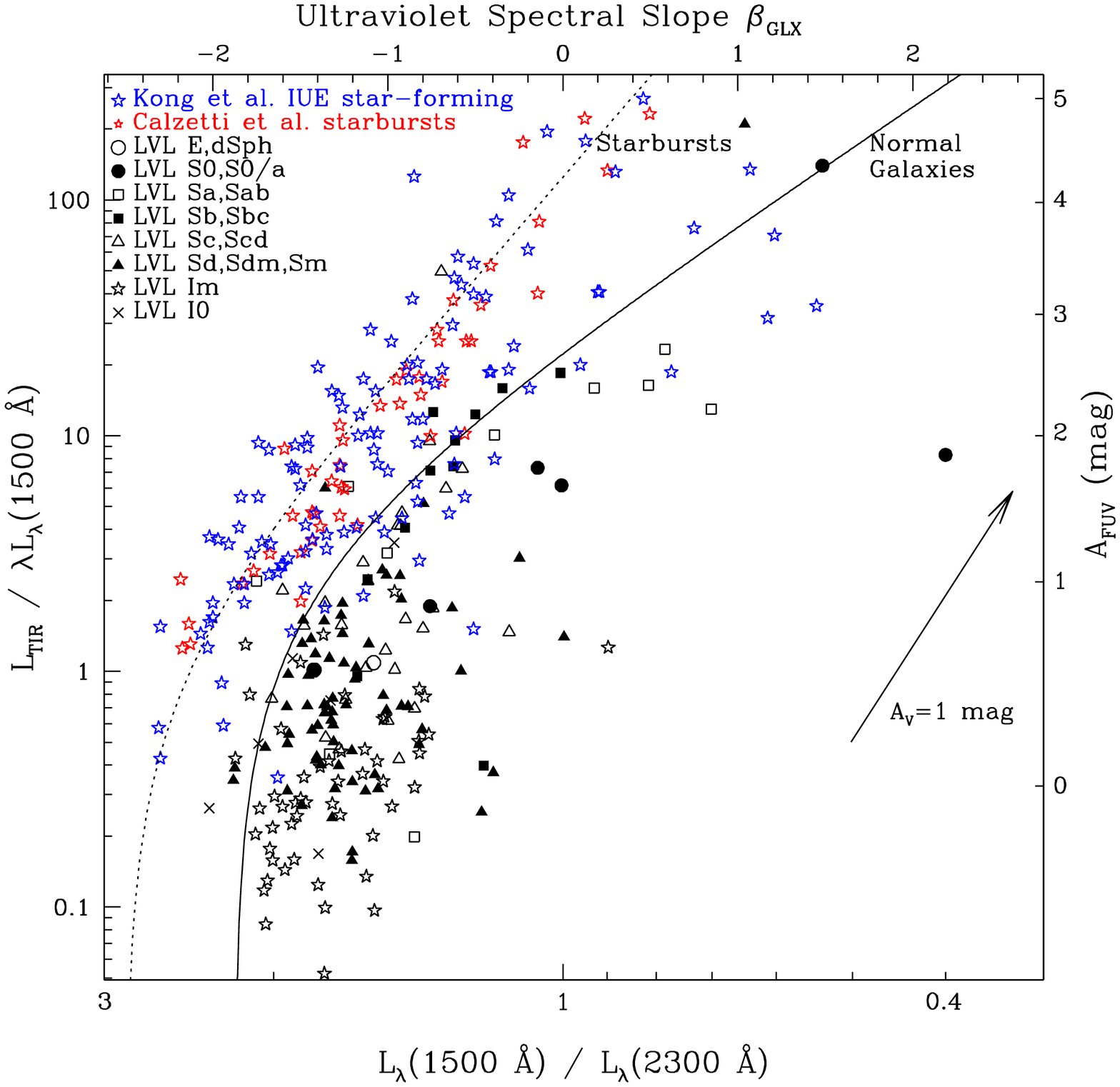,height=2.6in}\psfig{figure=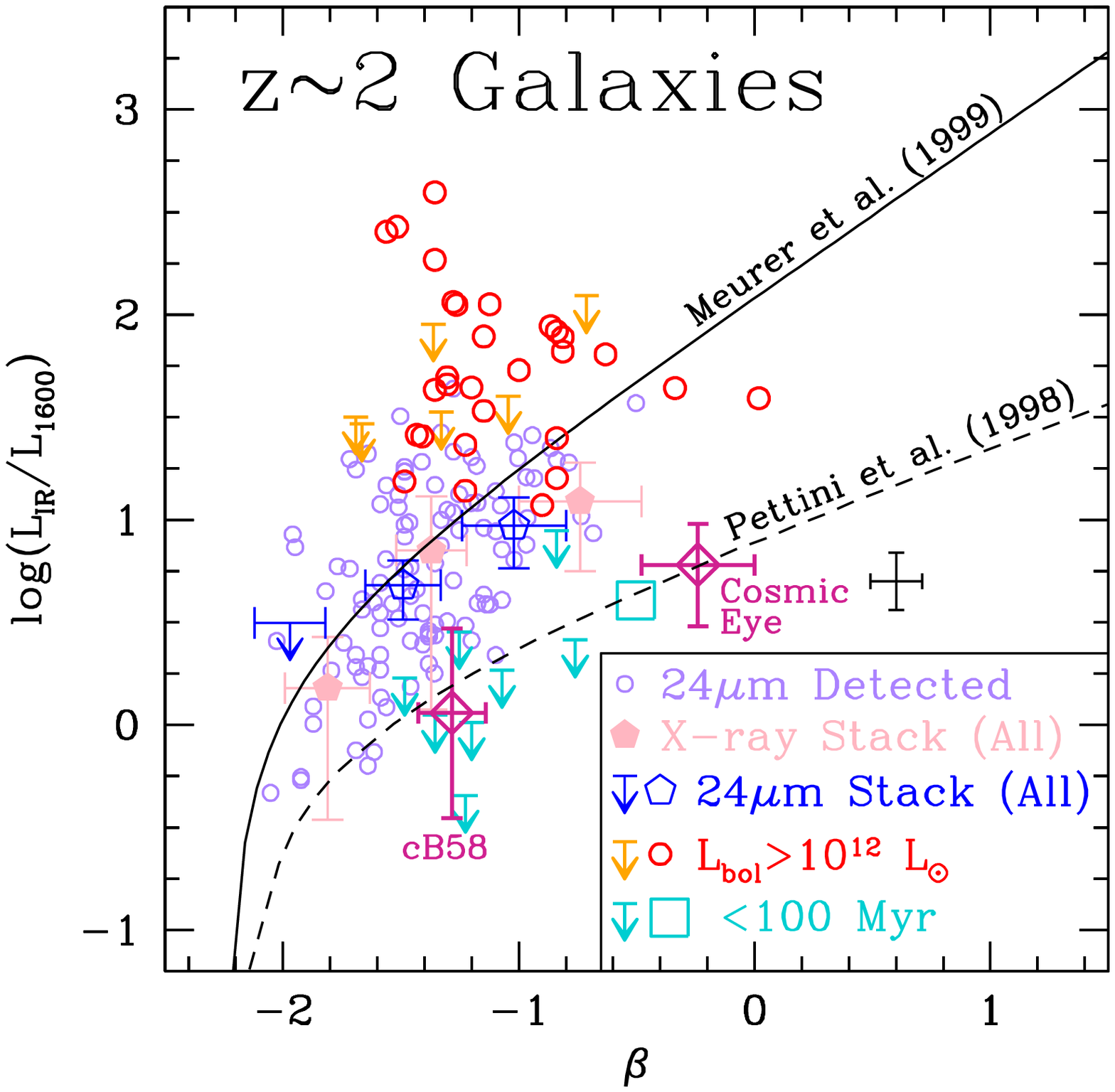,height=2.4in}}
\caption{The IRX-$\beta$ relation at $z=0$ from the Local Volume
  Legacy Survey \citep[LVL;][left panel]{Dale09} and at $z\sim2$ from
  \citet[][right panel]{Reddy10}.  The starburst (Meurer) relation is
  shown as a dotted (solid) line in the left (right) panel.  In the
  left panel, symbols indicate starburst and star-forming galaxies,
  along with galaxies split by morphological type.  In the right
  panel, symbols indicate galaxies selected in several different ways,
  either as having a young light-weighted age, high bolometric
  luminosity, or on the basis of x-ray or 24$\,\mu m$ fluxes.  There
  is an overall trend that redder UV slopes correspond to higher IRX
  values. However, the more striking impression is the tremendous
  scatter in IRX at fixed $\beta$, especially in the region of the
  diagram where most galaxies reside, at $-2<\beta<-1$.}
\label{f:irx}
\end{figure}

\subsection{Total Dust Attenuation}

There exist four primary techniques for measuring dust attenuation in
galaxies.  (1) As the obscuring effects of dust increase with
decreasing wavelength, multi-wavelength analysis (including
moderate-resolution spectra) of the UV-NIR SED can yield constraints
on the total dust attenuation.  (2) One can exploit the fact that the
Balmer line ratio H$\alpha$/H$\beta$ (sometimes referred to as the
Balmer decrement) in the absence of dust can be computed from first
principles and depends only weakly on the temperature of the gas
\citep{Osterbrock89}.  Comparison of the observed to the intrinsic
ratio provides a measurement of the attenuation toward \ionn{H}{ii}
regions. (3) Energy conservation demands that whatever photons are
absorbed in the UV/optical must be re-radiated in the IR.  The ratio
of IR to UV luminosity (often referred to as `IRX') therefore provides
a measure of the total dust absorption.  (4) A sufficiently luminous
background source with a known intrinsic spectrum can provide a
constraint on the attenuation in the foreground galaxy.  This last
technique will not be discussed herein, as it falls outside the scope
of what one can learn from the SEDs of galaxies.

Techniques (1)-(3) provide complementary constraints on the dust
within galaxies.  This was most clearly demonstrated in the work of
\citet{Calzetti94}, who analyzed the reddening in a sample of 39
starburst and blue compact galaxies from both the Balmer line ratio
H$\alpha$/H$\beta$ and the slope of the UV continuum.  Calzetti and
collaborators found that the dust optical depth measured from the
Balmer lines was nearly twice that estimated from the continuum
(sometimes referred to as the `1:2' rule).  They speculated that this
could occur if young hot stars are embedded in dusty natal clouds
while older stars are subject to attenuation only from the diffuse
ISM.  \citet{Charlot00} later provided a detailed model for dust
absorption which adopted this physical picture wherein young stars are
embedded in their natal dusty clouds and, in addition, all stars
experience dust attenuation due to the diffuse ISM.  In this
two-component model one must specify a transition time, after which
the natal cloud is dispersed (typically taken to be $10^7$ yr), and
attenuation curves for both the natal cloud and the diffuse ISM.  In
the standard implementation, both curves are taken to be power-laws
with the same exponent ($\tau_d\propto\lambda^{-0.7}$) and the
normalization of the natal cloud attenuation curve is taken to be
twice that of the diffuse ISM.  The Charlot \& Fall dust model is
widely employed when modeling dusty SEDs.

In subsequent work, \citet{Meurer99} compared the UV continuum slope,
$\beta$, to the ratio between FIR and UV flux,
log$(L_{IR}/L_{UV})\equiv$\,IRX, and concluded that the starburst
galaxies formed a well-defined sequence in the IRX-$\beta$ plane (see
Figure \ref{f:irx}).  With additional assumptions about the underlying
stellar populations and dust attenuation curve, these authors were
able to relate IRX (and hence $\beta$) to the total FUV dust
attenuation.  The resulting `Meurer relation' has experienced
widespread use for estimating the dust attenuation in distant galaxies
for which only measurements of $\beta$ are available.  However, many
subsequent studies have cast doubt on the practice of using $\beta$ to
estimate UV dust attenuation.  \citet{Bell02} noted that the
relationship between $\beta$ and UV attenuation measured in
\ionn{H}{ii} regions in the LMC differed from the Meurer relation.
\citet{Kong04} demonstrated that, while dust content was the main
variable responsible for the Meurer relation, other stellar population
parameters, particularly the recent SFH, drive galaxies to other
locations in the IRX-$\beta$ plane \citep[see also][]{Boquien12}.  The
underlying extinction curve and star-dust geometry also play an
important role in determining where galaxies lie in this plane
\citep{Charlot00, Kong04, Panuzzo07, Conroy10d}.

Based on UV-IR photometry of 1000 galaxies, \citet{Johnson07a} cast
further doubt on the utility of the IRX-$\beta$ relation for
estimating dust content.  Their conclusion rested partly on the
observation that the relation defined by normal star-forming galaxies
is so steep that even small errors in $\beta$ will translate into
large errors on IRX.  Figure \ref{f:irx} shows the IRX-$\beta$
relation for galaxies from the Local Volume Legacy Survey
\citep{Dale09}, and the IRX-$\beta$ relation for $z\sim2$ galaxies
from \citet{Reddy10}.  It is clear that, while galaxies of a
particular type define their own IRX-$\beta$ relations, the overall
galaxy population presents a complicated picture.  Fortunately, as
multi-wavelength datasets become increasingly common, the estimation
of dust content from crude measures such as the IRX-$\beta$ relation
will give way to more robust techniques.

SED fitting of UV-NIR data has become remarkably common in the past
decade.  In nearly all cases the dust content is a parameter included
in the fit, although the derived dust opacities are rarely discussed.
The reason for this seems to be the fact that the dust opacity is a
relatively poorly constrained quantity when data only reach the NIR.
As an example of the difficulty in constraining the dust opacity from
broadband data, \citet{Salim07} fit UV-NIR photometry to models and
were able to recover $V-$band dust opacities with an uncertainty of
$\sim0.4$ mag.  Furthermore, there was only a weak correlation between
the dust opacity derived from broadband data and the opacity derived
from emission lines via \citet{Brinchmann04}.  Some of the difference
can be attributed to aperture effects and real differences in the
opacity measured from emission lines and stellar continua, but overall
the weak correlation confirms that dust content measured from
broadband data is quite uncertain.  Poorly constrained dust opacities
appears to be a generic result of the modeling of UV-NIR broadband
SEDs \citep[e.g.,][]{Papovich01, Shapley06, Kriek08, Taylor11}.

Broadband UV-NIR data do not provide strong constraints on dust
attenuation because of the well-known degeneracy between age and dust
\citep[e.g.,][]{Papovich01}.  This degeneracy can be broken, and
stronger constraints on the dust opacity can be obtained, when either
spectroscopic or FIR data are available.  Spectroscopic data offer an
estimate of the mean stellar age and metallicity independent of dust
content because narrow spectroscopic features are relatively immune to
overall changes in the continuum shape.  \citet{Kauffmann03a}
exploited this technique to measure mean ages of $\sim10^5$ SDSS
galaxies from H$\delta$ and the 4000\,\AA\, break.  With the ages thus
constrained, an estimate of the dust attenuation could be obtained by
comparing the model colors to the observed ones.  This study was
limited primarily by the fact that the SDSS spectra sample a modest
fraction of the total galaxy, and so a correction must be made for
this aperture effect (ongoing and future planned IFU surveys such as
CALIFA and MaNGA will remedy this limitation).  Narrowband photometric
surveys also offer the promise of separately constraining age and dust
attenuation, primarily because of the strongly age-sensitive $D_n4000$
spectral feature, which is so strong that it can be probed with low
resolution data \citep[e.g.,][]{Whitaker10, PerezGonzalez12}.  FIR
data can also provide a robust measurement of the total dust
attenuation because they significantly reduce the degeneracy between
age and dust \citep[e.g.,][]{Burgarella05, Noll09b}.

\citet{Wuyts07} and \citet{Williams09} pioneered the use of restframe
optical-NIR color-color diagrams (specifically $U-V$ vs. $V-J$;
commonly referred to as the UVJ diagram) to efficiently separate
quiescent and star-forming galaxies.  The UVJ diagram has since become
a popular tool for separating these two galaxy types, especially at
high-redshift where data are limited.  In a broad sense, the UVJ
diagram allows one to break the age-dust degeneracy, but only in the
sense of separating quiescent and dusty star-forming galaxies.  Within
the star-forming sequence, the age-dust degeneracy is still manifest
when only broadband information is available.

Extremely blue UV continua can place strong constraints on the amount
of dust in the system.  \citet{Bouwens12b} and \citet{Finkelstein12}
measured the UV continuum slopes of high-redshift galaxies and found
that $\beta$ systematically decreases toward higher redshift and
fainter galaxy luminosities.  At the highest redshifts and faintest
luminosities they find $-2.0<\beta<-2.5$, which not only suggests that
these galaxies are dust-free but also begins to place interesting
constraints on their stellar populations.  The models of
\citet{Schaerer03} reach such blue slopes only at very young ages
($\sim10^6$ yr) and/or very low metallicities ($<10^{-3}\,Z_\odot$),
and the inclusion of nebular continuum emission, which presumably must
be present whenever stars of these young ages are still alive, sets a
lower limit to the UV continuum slope of $-2.5<\beta<-3.0$, depending
on the details of the SFH.  Clearly more data probing the restframe UV
at very high redshifts would be valuable.

\subsection{Constraints on the Attenuation Curve}

\citet{Calzetti94} estimated a mean attenuation curve from 39
starburst and blue compact galaxies based on UV-optical spectra.  By
assuming that their sample was comprised of galaxies with the same
underlying stellar populations, they were able to construct average
attenuation curves by comparing galaxies with high and low Balmer line
ratios, i.e., by comparing dusty to dust-free galaxies.  The resulting
average attenuation curve, measured over
$\approx1200$\,\AA$-8000$\,\AA, was greyer (shallower) than either the
MW or LMC extinction curve and lacked the broad 2175\,\AA\, dust
feature visible in the MW and LMC.  The derived curve is now referred
to as the ``Calzetti attenuation law'' and is widely used when
modeling the broadband SEDs of galaxies.

The Calzetti law was estimated from a relatively small sample of
unusual (i.e., starburst and blue compact) galaxies at $z=0$ and so
one may wonder to what extent it is applicable to other systems.  As
noted in the previous section, the IRX$-\beta$ relation is sensitive
to the attenuation curve.  The large range in $\beta$ at a fixed IRX,
evident in Figure \ref{f:irx}, suggests considerable variation in the
attenuation curve, but interpretation of the IRX$-\beta$ plane is
complicated by the additional variables affecting IRX and $\beta$.

Progress can be made by estimating stellar population parameters such
as the SFH and metallicity independently of the dust content (e.g.,
via narrow spectral features).  With such quantities in hand, one can
exploit the technique pioneered by Calzetti and collaborators by
computing ratios of SEDs of more to less dusty galaxies with the same
underlying stellar populations.  \citet{Johnson07a} estimated mean
attenuation curves in this way for galaxies as a function of the
4000\,\AA\, break strength and stellar mass.  The curves were
constrained from broadband FUV-NIR photometry and were consistent both
with a simple power-law, $\tau_d\propto\lambda^{-0.7}$, and with the
Calzetti law.  \citet{Wild11} exploited a similar technique using
pairs of galaxies with similar gas-phase metallicities, specific star
formation rates, and inclinations, and significantly different Balmer
decrements to construct attenuation curves from broadband FUV-NIR
data.  These authors found significant variation in the mean
attenuation curve as a function of specific SFR, inclination, and
stellar mass surface density, based on a sample of 20,000 $z\sim0$
star-forming galaxies.  They also found variation in the ratio between
dust opacity in the stellar continuum and in the nebular line
emission, implying that the `1/2' rule of \citet{Calzetti94} --- one
unit of continuum opacity for ever two units of line opacity --- is
not universal.  \citet{Buat12} modeled the UV-FIR SEDs of galaxies at
$1<z<2$, allowing for variation in the dust attenuation curve.  They
found evidence for a steeper attenuation curve (i.e., faster rise in
the UV) than the Calzetti law in $20-40$\% of their sample.

\subsubsection{The 2175\,\AA\, dust feature}

The absence of the 2175\,\AA\, dust feature from the Calzetti
attenuation law is striking given its ubiquity in the MW and LMC.
\citet{Witt00} were able to reproduce the Calzetti law in their
radiative transfer models only by assuming SMC-type dust, i.e., by
adopting an underlying extinction curve that lacked the 2175\,\AA\,
dust feature.  In contrast, \citet{Granato00} were able to reproduce
the Calzetti law in their radiative transfer models with MW-type dust.
In their model the star-dust geometry plays a central role in shaping
the starburst attenuation curve.  Their molecular clouds are optically
thick and so the attenuation curve is determined primarily by the
fraction of stars inside molecular clouds as a function of age
\cite[see also][]{Panuzzo07}.  \citet{Fischera11} were able to
reproduce a Calzetti-like attenuation curve if the carrier responsible
for the 2175\,\AA\, feature (e.g., PAHs) was destroyed at high column
density.

In all radiative transfer dust models the expectation is that normal
star-forming galaxies should show evidence for the 2175\,\AA\, dust
feature provided that the underlying grain population is similar to
the MW or LMC.  Owing to a paucity of restframe UV spectra of
star-forming galaxies, it has proven difficult to verify this
expectation.  \citet{Burgarella05} analyzed the UV-FIR SEDs of 180
star-forming galaxies and concluded that the data required an
attenuation curve with a 2175\,\AA\, dust feature that was on average
weaker than observed in the MW extinction curve.  \citet{Noll09}
presented stacked restframe UV spectra of $z\sim2$ star-forming
galaxies and found strong evidence for the presence of the 2175\,\AA\,
dust feature in a subsample of their objects.  \citet{Conroy10d}
analyzed UV-NIR photometry of $z\sim0$ star-forming galaxies as a
function of inclination and found evidence for a 2175\,\AA\, dust
feature with a strength slightly weaker than observed in the MW
extinction curve.  \citet{Hoversten11} analyzed medium band UV
photometry from the {\it Swift} UV/Optical Telescope, and found
evidence for a prominent 2175\,\AA\, bump in M81.  \citet{Wild11} found
evidence in their derived attenuation curves for the presence of the
2175\,\AA\, dust feature, with a slight tendency for galaxies with
higher specific SFRs to have lower bump strengths.
\citet{Wijesinghe11} compared dust-corrected SFRs estimated from
H$\alpha$ and the UV and found agreement only when the 2175\,\AA\,
feature was removed from the attenuation curve, but these authors did
not consider variation in the shape of the attenuation curve.
\citet{Buat12} found evidence that the 2175\,\AA\, dust feature is
present in 20\% of their sample, with an amplitude that is half the
strength found in the MW extinction curve.  Similar to Wild et al.,
they found an anti-correlation between bump strength and specific SFR.
The emerging consensus appears to be that the 2175\,\AA\, dust feature
is present in the typical star-forming galaxy over the redshift range
$0<z<2$, with a strength that is generally lower than in the MW
extinction curve, and that increases with increasing inclination and
decreasing specific SFR.  The absence of the dust feature in the
Calzetti law may then simply be a reflection of the fact that the
galaxies in the Calzetti sample have unusually high specific SFRs.  An
outstanding question is whether these trends are due to varying
star-dust geometries or changes in the underlying grain population.
As emphasized in \citet{Conroy10d} and elsewhere, these results have
consequences for interpreting the IRX$-\beta$ relation, since $\beta$
is frequently measured over a wavelength range that includes the
2175\,\AA\, feature.

\subsection{Physical Dust Properties}

The IR emission by dust grains provides strong constraints on a
variety of physical properties of a galaxy including the bolometric
luminosity, dust temperature(s), dust mass, and relative abundance of
PAH molecules (or, more generally, the grain size distribution).  A
constraint on the bolometric luminosity helps to break the age-dust
degeneracy in the UV-NIR, as discussed in the previous section.  The
other parameters provide clues to the nature of dust in galaxies and
its spatial relation to the stars and gas.

It has long been recognized that constraints on the dust mass and dust
temperatures required data beyond the peak in the thermal dust
emission spectrum at $\sim100\,\mu m$.  The SCUBA bolometer array at
the James Clerk Maxwell Telescope, operating at $450\,\mu m$ and
$850\,\mu m$, has been extremely influential in this field, and now
the SPIRE instrument onboard the {\it Herschel Space Observatory},
operating at $250-500\,\mu m$, promises to revolutionize the field by
dramatically increasing sample sizes over a wide range in redshifts.
With data obtained for the SCUBA Local Universe Galaxy Survey,
\citet{Dunne00} and \citet{Dunne01} constrained dust masses and
temperatures for 104 nearby galaxies.  The data favored models that
contained both cold and warm dust components (at $\sim20$K and
$\sim40$K, respectively).  They also found that dust masses estimated
with single-temperature models were on average lower by a factor of
$\sim2$ compared to two-temperature models, and the two-temperature
dust masses yielded dust-to-gas ratios in agreement with measurements
in the Galaxy.

\citet{Draine07} presented a thorough analysis of the IR SEDs of
galaxies from the SINGS survey of 65 nearby galaxies.  These authors
analyzed data from {\it Spitzer}, {\it IRAS}, and SCUBA with the
physical dust models of \citet{Draine07b}.  The model considers grain
populations exposed to a variety of starlight intensities and includes
both thermal emission and single photon heating of dust particles.  In
addition, the fraction of dust mass contained in PAH molecules is a
free variable.  Draine et al. confirmed the results of previous work
that the sub-mm data, provided by SCUBA, were essential for deriving
dust masses with an uncertainty of $<50$\%.  They also derived strong
constraints on the fraction of dust mass in PAHs, the interstellar
radiation field strength in the diffuse ISM, and the flux-weighted
mean radiation field strength averaged over the galaxy.  The
well-sampled spectral coverage from the mid-IR through the sub-mm was
essential in order to separately constrain each of these components.

\citet{Dale12} analyzed FIR and sub-mm photometry for SINGS galaxies
from {\it Herschel} in conjunction with previously available 2MASS,
{\it Spitzer}, {\it IRAS}, ISO, and SCUBA data.  These authors modeled
the SEDs with the \citet{Draine07b} dust model, both with and without
the {\it Herschel} data.  They found modest differences (factors of
$\approx1.6$ on average) in the best-fit dust masses, PAH mass
fractions, and fraction of dust emission arising from the diffuse ISM
when the FIR and sub-mm data were either included or excluded.  They
also demonstrated that dust masses estimated via simple modified
blackbody fits significantly underestimated the dust masses (by
factors of $\approx3$ in the worst cases), with the underestimation
becoming more severe as bluer wavelengths were included in the fits.
The {\it Herschel} data are of such high quality that simplistic,
modified blackbody dust models are no longer capable of providing
adequate fits to the data.

\begin{figure}
\centerline{\psfig{figure=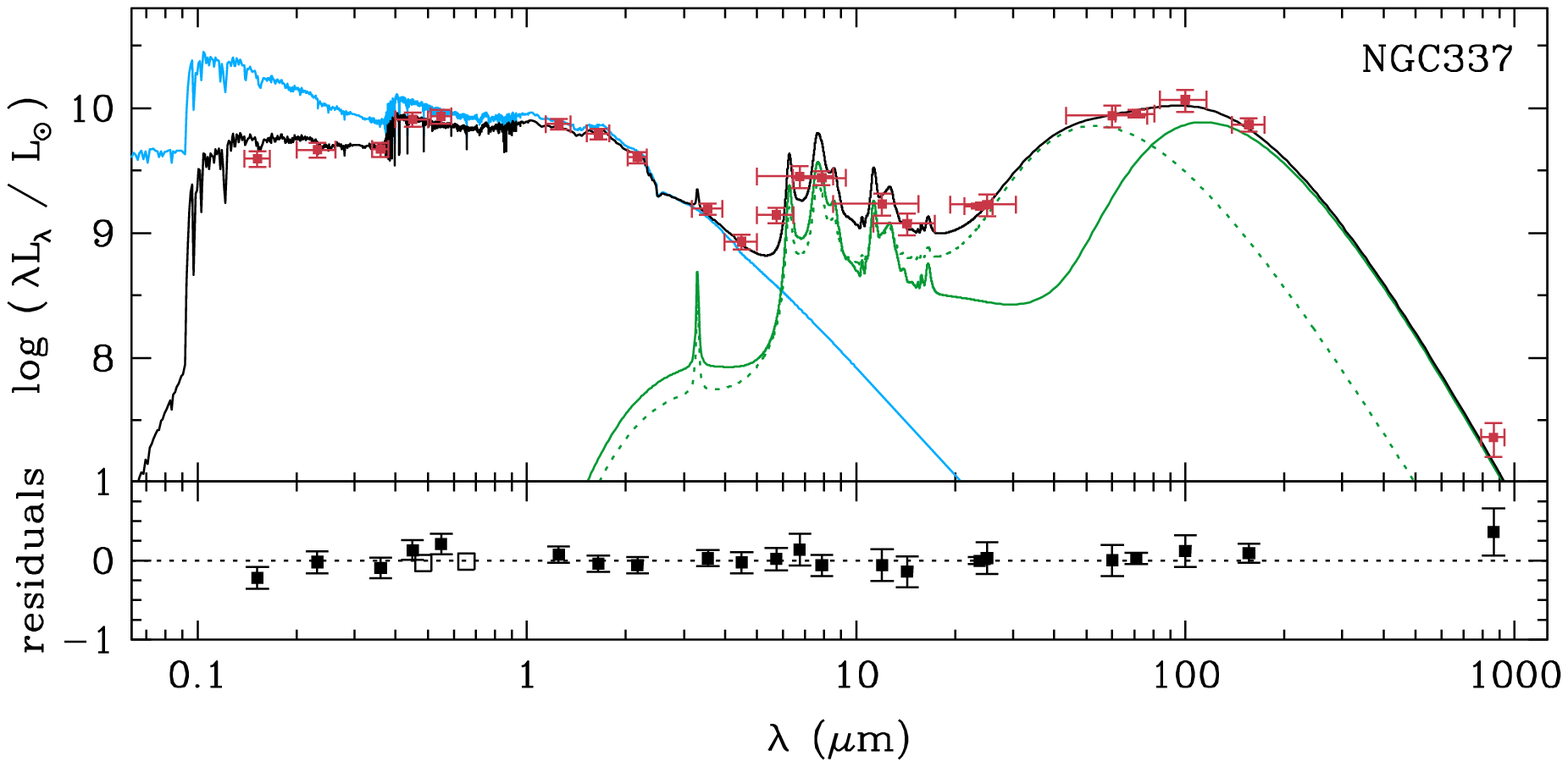,height=3in}}
\caption{Model fit to the FUV through FIR SED of NGC 337, from
  \citet{daCunha08}.  The black line is the best-fit model.  Also
  shown is the unattenuated model starlight (blue line) and the dust
  emission separated into dust in the diffuse ISM (solid green line)
  and dust in birth clouds (dotted green line).  Figure courtesy of
  E. da Cunha.}
\label{f:dacunha}
\end{figure}

\citet{daCunha08} developed a phenomenological model for the UV-FIR
SEDs of galaxies that is based on the \citet{Bruzual03} SPS models for
the starlight, the \citet{Charlot00} dust attenuation model, and dust
emission comprised of multiple components including an empirical
spectrum for the PAH emission, warm and cold thermal dust emission,
and emission from stochastically heated grains (see Figure
\ref{f:dacunha}).  Based on tests with mock galaxies, they concluded
that a variety of parameters could be constrained within the context
of their model, including the temperature of the cold dust component,
the fractional contribution of the cold dust to the total FIR
emission, and the total dust mass.  They analyzed the same SINGS
galaxies as in Draine et al. and derived dust masses that agreed to
within $50$\%, demonstrating that dust masses can be reliably measured
with reasonably well-sampled mid-IR, FIR, and sub-mm data.

\citet{Galliano08} employed the dust model of \citet{Zubko04} to
constrain a variety of parameters including the total dust mass,
fraction of dust mass in PAHs, fraction of dust luminosity originating
from photodissociation regions (PDRs), and the gas-phase metallicity.
These authors measured these and other parameters for 35 galaxies with
optical, IR, and sub-mm photometry and mid-IR spectroscopy.  Although
no direct comparison was made with the results from \citet{Draine07},
there appears to be a quantitative difference in the derived PAH mass
fractions.  As discussed extensively in \citet{Draine07b}, the optical
properties of the PAH molecules are uncertain and have in many cases
been tuned to fit extragalactic data.  It would therefore be valuable
to compare the results obtained from dust models that assume different
treatments of the uncertain PAH properties.

\subsection{Cosmic Evolution of IR SEDs}

Measuring the cosmic evolution of IR SEDs is important not only for
constraining models of galaxy evolution but also because local IR SED
templates are frequently employed to interpret high-redshift data.
For example, the local templates are often used to convert $24\,\mu m$
observations of high-redshift galaxies into SFRs.  If IR SEDs at fixed
$\lbol$ evolve with time then it would considerably complicate the
interpretation of higher redshift data.

\citet{Pope06} presented evidence that the SEDs of ULIRGs at $z\sim2$
peak at longer wavelengths than local ULIRGs, implying that the dust
temperature in high-redshift ULIRGs is on average $\sim5$K cooler than
local ULIRGs.  \citet{Muzzin10} analyzed high-quality data for two
$z\sim2$ ULIRGs and found that these galaxies contained colder dust
than local ULIRGs, in agreement with \citet{Pope06}.  Muzzin et
al. showed that the SED shapes of their ULIRGs were well-fit by the
$z=0$ \citet{Chary01} templates that were an order of magnitude less
luminous than their $z\sim2$ galaxies.  In other words, the $z\sim2$
ULIRGs had SED shapes that were similar to local LIRGs.  Exploiting
the power of {\it Herschel} data, \citet{Hwang10} estimated dust
temperatures by fitting modified blackbodies to the data, and found
evidence for slightly cooler dust temperatures at $z\sim1$, by $2-5$ K
on average, for galaxies with $L_{\rm IR}>10^{11}L_\odot$.
\citet{Daddi07a} compared restframe $8\,\mu m$ luminosities of
$z\sim2$ galaxies to local templates and found evidence for excess
mid-IR emission for galaxies with $L_{8\,\mu m}>10^{11}L_\odot$.  This
mid-IR excess problem was confirmed by \citet{Papovich07} and later by
\citet{Elbaz10} with {\it Herschel} data. It is noteworthy that the
problem only appears at $z>1.5$.  Bolometrically luminous
high-redshift galaxies thus have colder dust and more flux in the
mid-IR compared to galaxies at the same $\lbol$ at $z=0$.  The mid-IR
excess problem has important implications for the common practice of
estimating SFRs for galaxies at $z>2$ with {\it Spitzer} $24\,\mu m$
observations (i.e., restframe $\sim8\,\mu m$).  The sign of the excess
is such that SFRs estimated by extrapolating observed $24\,\mu m$ flux
with local templates will tend to overestimate the true SFRs, by
factors of several in the most extreme cases
\citep[e.g.,][]{Papovich07}.

\citet{Elbaz11} investigated the origin of the differences between
high and low redshift IR SEDs with very deep {\it Herschel}
observations.  They argued that it is the distribution of SED types
that is changing at high luminosity as a function of redshift.  At low
redshift, high luminosity systems (i.e., ULIRGs) are predominantly
star-bursting and compact, while at higher redshift ULIRGs are mostly
the high-luminosity extension of normal star-forming galaxies.  Local
ULIRGs therefore have hotter dust temperatures and a depressed
emission component from PAHs in the mid-IR compared to distant ULIRGs
because the former are compact starbursts while the latter are normal,
extended star-forming galaxies \citep[see also][who reached similar
conclusions]{Rujopakarn12}.  The physical origin of this result is not
yet fully understood, but a plausible explanation is that compact
starbursts have harder and more intense radiation fields.  The
radiation field obviously has a direct influence on the dust
temperature, and it may also modulate the dust mass fraction in PAHs
and therefore the mid-IR luminosity \citep{Voit92, Madden06,
  Draine07}.

\subsection{Summary}

The Calzetti attenuation curve works remarkably well at describing the
mean attenuation properties of star-forming galaxies over most of
cosmic history.  However, closer scrutiny of the data reveals
variation in the attenuation curve of galaxies as a function of galaxy
properties, as expected on theoretical grounds.  In contrast to the
Calzetti attenuation curve for starburst galaxies, the 2175\,\AA\,
dust feature is apparently common in normal star-forming galaxies,
with a strength that is weaker than observed in the MW extinction
curve.  The IRX$-\beta$ relation should be used with caution, as there
is evidently no universal relation for all galaxies.  In addition, the
relation is so steep over the range in $\beta$ where most galaxies
reside that estimating dust opacity in this way is highly unreliable.
Dust masses can be measured to an accuracy of $\sim50$\% when FIR and
sub-mm data are available.  Physical dust models, such as the model of
\citet{Draine07b}, are capable of extracting a variety of parameters
from the global SEDs of galaxies, including the fraction of dust mass
in PAHs, the fraction of dust emission due to PDRs vs. the diffuse
ISM, and the typical interstellar radiation field strength heating the
dust grains (or, alternatively, the typical dust temperature).
Further work appears to be needed to sort out which derived parameters
are robust, and which are dependent on uncertain components of the
model.  Below $\lbol\sim10^{11}L_\odot$, the IR SEDs of galaxies
evolve little over the interval $0<z<2$.  At higher luminosities there
is a marked shift such that high redshift ULIRGs have more mid-IR flux
and a colder dust component that peaks at longer wavelengths.  A
plausible explanation for this trend is that ULIRGs at high-redshift
are at the massive end of the normal star-forming galaxy sequence,
while ULIRGs in the local universe are unusual, compact, star-bursting
systems.


\section{THE INITIAL MASS FUNCTION}
\label{s:imf}

The stellar IMF is of fundamental importance for many areas of
astrophysics. The form of the IMF is well-constrained in the Galactic
disk \citep{Salpeter55, Miller79, Kroupa01, Chabrier03}. However, it
is not clear whether the IMF has had the same form over all of cosmic
time and in all environments. The low mass end is of particular
importance, as $\sim 60-80$\,\% of the stellar mass density in the
Universe is in the form of stars with masses $<0.5$\,M$_{\odot}$.  Low
mass stars are very faint and so contribute only a few percent to the
integrated light of an old population.  For this reason, when fitting
the SEDs of galaxies, an IMF is typically assumed, and so the IMF is
one of the largest sources of systematic uncertainty in studies of
extragalactic stellar populations.

It has been known since at least the 1960s that dwarf-sensitive and,
to a lesser degree, giant-sensitive spectral features could be used to
count the number of low-mass stars in the integrated light of old
stellar populations \citep[e.g.,][]{Spinrad62, Spinrad71, Whitford77,
  Frogel78, Frogel80, Vazdekis96, Schiavon97a, Schiavon97b,
  Schiavon00}.  This idea is limited to quiescent systems because the
massive stars associated with ongoing star formation would
dramatically outshine the faint low mass stars.  The fundamental
difficulty then, as now, is the separation of abundance effects from
giant-to-dwarf ratio effects.  Both effects can change the strength of
the gravity-sensitive lines.

\citet{Spinrad71} concluded that the nuclei of the nearby galaxies
M31, M32, and M81 are dominated by dwarfs, with an implied $M/L=44$.
These conclusions were based on photoelectric scanner observations
reaching to $1\,\mu m$.  Subsequently, \citet{Cohen78} measured the
IMF-sensitive \ionn{Na}{i}, CaT, FeH and TiO features in the nucleus
of M31 and M32 and concluded that an IMF comparable to that of the
solar neighborhood was favored, along with metallicity enhancement in
M31.  \citet{Faber80} also measured the strength of the \ionn{Na}{i}
doublet in M31 and M32 and instead argued for a dwarf-rich IMF, along
the lines of \citet{Spinrad71}, with $M/L_B=28$.  \citet{Frogel78}
analyzed optical-NIR colors and the CO bandhead at $2.3\,\mu m$ for 51
early-type galaxies and concluded that dwarf-rich IMFs were
inconsistent with the data and favored models with $M/L_V<10$.  With
updated detector technology and more sophisticated models,
\citet{Carter86} argued against dwarf-rich IMFs on the basis of the
\ionn{Na}{i}, CaT, TiO and FeH features in 14 elliptical and
lenticular galaxies.  Their results implied mass-to-light ratios of
$M/L\sim6$.  In a series of papers, Hardy, Couture and collaborators
\citep{Hardy88, Delisle92, Couture93} analyzed the IMF features
\ionn{Na}{i} and FeH for of order a dozen galaxies and concluded that
dwarf-rich IMFs were strongly disfavored.  Measurements of the FeH
band at $1\,\mu m$ (also known as the Wing-Ford band) were
particularly important, as this feature becomes very strong for IMF
exponents steeper than Salpeter \citep{Whitford77}.

\begin{figure}
\centerline{\psfig{figure=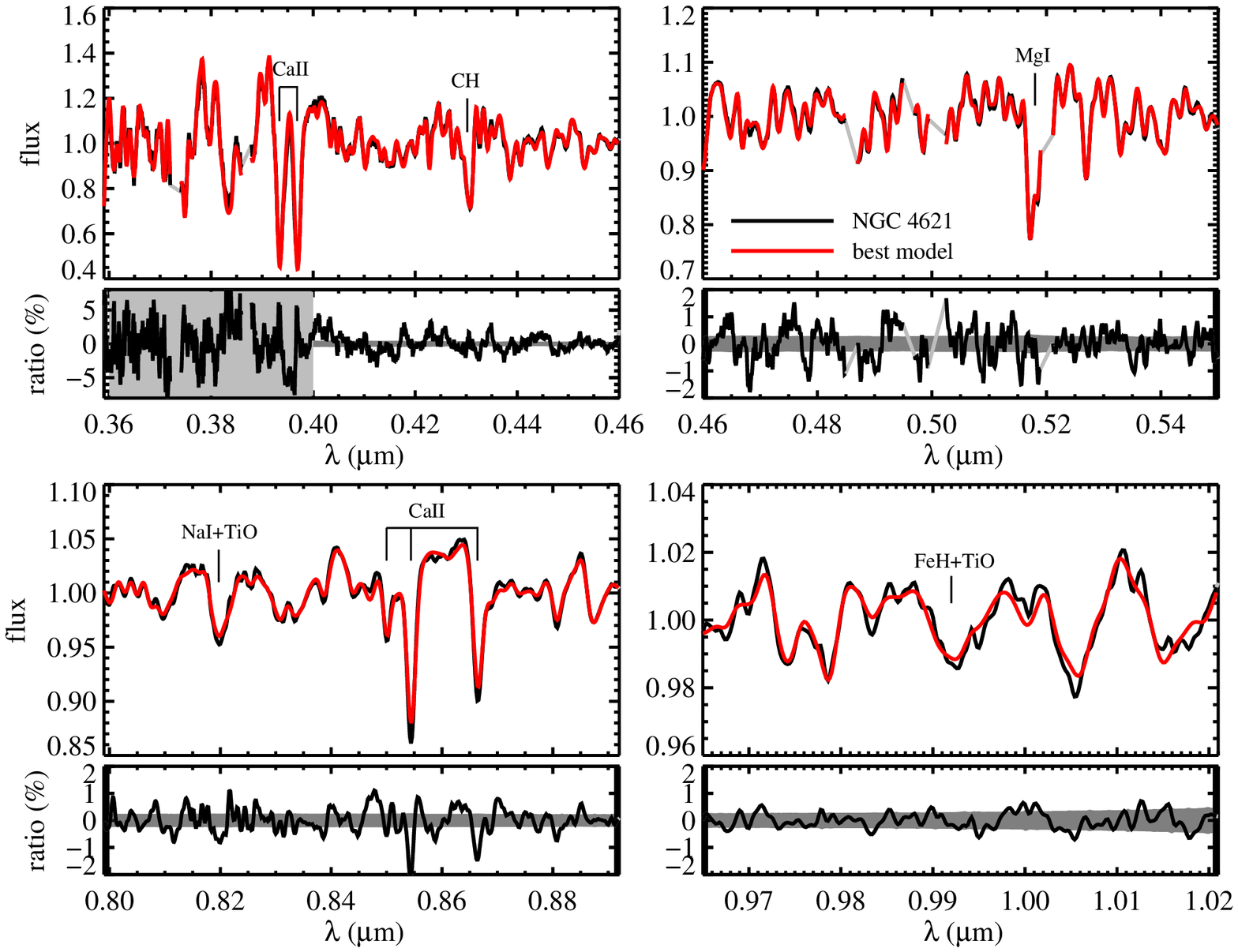,height=5in}}
\caption{Comparison between the best-fit model and observed spectrum
  for NGC 4621, a massive early-type galaxy.  Both data and model have
  been continuum normalized.  The $S/N$ is indicated as a grey shaded
  band; the data at $<4000$\,\AA\, were not included in the fit but
  are included here for completeness.  The model has 21 free
  parameters including the abundance of 11 elements and the shape of
  the low-mass IMF.  The rms difference between the model and data is
  0.8\%.  IMF-sensitive features include the \ionn{Na}{i} doublet at
  $0.82\,\mu $m, the \ionn{Ca}{ii} triplet at $\approx0.86\,\mu m$,
  and the FeH bandhead (i.e., the Wing-Ford band) at $0.99\,\mu m$.
  The best-fit $M/L$ is $\approx2\times$ as large as would be expected
  for a MW IMF, suggesting that the IMF is not universal.  From
  \citet{Conroy12b}.}
\label{f:imf}
\end{figure}

Early work aimed at measuring the IMF from integrated light spectra
suffered from serious limitations, including (1) the lack of accurate
stellar evolution calculations across the main sequence and through
advanced evolutionary phases; (2) the use of empirical stellar spectra
collected from the solar neighborhood, which implied that the effect
of non-solar abundance ratios could not be adequately assessed; (3)
poor NIR detector technology, which made it very difficult to measure
red spectra at the sub-percent level precision necessary to measure
the low-mass IMF in integrated light.  These limitations contributed
to the wide range of reported IMF exponents and mass-to-light ratios
noted above. In the past two decades, each of these limitations has
been significantly reduced, though not eliminated.  Stellar evolution
calculations have improved dramatically since the 1970s, the response
of stellar spectra to elemental abundance variations has been studied
in detail \citep[e.g.,][]{Tripicco95, Korn05, Serven05, Conroy12a},
and NIR detector technology has steadily improved.

The debate over IMF variation was renewed by \citet{Saglia02} and
\citet{Cenarro03}, who independently reported an anti-correlation
between the strength of the IMF-sensitive CaT index and velocity
dispersion for elliptical galaxies.  Comparison to SPS models with
solar-scaled abundance ratios required dwarf-rich IMFs, with an
exponent of $\alpha\approx4$ at the highest dispersions (compared to
the Salpeter value of 2.35).  \citet{Saglia02} argued against this
interpretation, as such a steep IMF would imply $M/L\approx40$, well
in excess of newly available dynamical constraints.  \citet{Worthey11}
investigated several calcium-sensitive spectral features, and
concluded that a modest decrease in [Ca/Fe] with velocity dispersion
could also explain the observed trends.  \citet{Worthey11} also noted
that such dwarf-rich IMFs would yield optical-NIR colors (e.g., $V-K$)
much redder than even the reddest observed early-type galaxies.

An IMF different from the Galactic disk (often referred to as a
non-universal IMF) was also reported by \citet{vanDokkum10} based on
the analysis of high S/N spectra extending to $1\,\mu m$ for eight
massive galaxies in the Coma and Virgo galaxies.  These data were of
extremely high quality thanks to new detector technology installed on
the Low Resolution Imaging Spectrometer (LRIS) instrument at the Keck
Telescope.  The data were interpreted with new models limited to
solar-scaled abundance patterns.  Dwarf-rich IMFs were favored, with
an IMF exponent in the range $2.3<x<3$.  It is worth pausing here to
note that the meaning of `dwarf-rich' has evolved over the years.  In
the 1970s dwarf-rich IMFs indicated mass-to-light ratios exceeding 20,
often reaching as high as 40.  With the IMF in the solar neighborhood
now well-established \citep{Kroupa01, Chabrier03}, the reference point
has shifted.  Now, `dwarf-rich' IMFs are those that contain relatively
more low-mass stars than found in the solar neighborhood and would
indicate, amongst other implications, a non-universal IMF.  In this
sense, the Salpeter IMF is `dwarf-rich', as it contains more low-mass
stars than the \citet{Kroupa01} IMF for the solar neighborhood.

\citet{Conroy12b} were the first to analyze high-quality absorption
line spectra of 34 early-type galaxies and the nuclear bulge of M31
with a model that allowed for both IMF variation and abundance pattern
variation (in addition to variation in age and several nuisance
parameters).  The model, presented in \citet{Conroy12a}, allows for
variation in the abundance pattern of 11 elements, and is based on the
latest isochrone tables and the MILES and IRTF stellar libraries.
\citet{Conroy12b} simultaneously fit the full blue and red galaxy
spectra to their models, including the classic IMF-sensitive features
\ionn{Na}{i}, CaT, and FeH, as well as IMF-sensitive features in the
blue spectral region.  An example fit to a massive early-type galaxy
is shown in Figure \ref{f:imf}.  They argued that the models favor
IMFs that become progressively more dwarf-rich (i.e., more bottom
heavy) for more massive and more $\alpha-$enhanced galaxies.  Over the
full sample, the mass-to-light ratios predicted by their best-fit IMFs
vary by only a factor of 3 at fixed age and metallicity.  The IMF
variation advocated by \citet{Conroy12b} is thus much subtler than
previous claims for dwarf-rich IMFs.  The best-fit mass-to-light
ratios do not violate dynamical constraints.  Similar conclusions were
reached by \citet{Smith12} based on the analysis of $1\,\mu m$
spectroscopy of 92 galaxies in the Coma cluster, and by
\citet{Spiniello12}, who analyzed indices sensitive to \ionn{Na}{i}
and TiO for a sample of early-type galaxies.

Meanwhile, there have been several less-direct arguments favoring
top-heavy (or bottom-light) IMFs in some types of galaxies and in some
phases of a galaxy's evolution.  \citet{Tinsley80} suggested that a
constraint on the IMF near the turn-off mass could be obtained by
comparing the evolution of the $M/L$ ratio and color of passively
evolving galaxies.  \citet{vanDokkum08} applied this idea to the
evolution of galaxies on the fundamental plane, and found evidence
that the IMF is flatter than the MW IMF at $\sim1\,\Msun$ in
early-type galaxies.  However, \citet{vanDokkum12} demonstrated that
when galaxies are compared at fixed velocity dispersion, as opposed to
fixed stellar mass, the evolution in the fundamental plane is
consistent with the standard Salpeter slope at $\sim1\,\Msun$
\citep[see also][]{Holden10}.  Numerous authors have demonstrated that
the evolution of the cosmic star formation rate and cosmic stellar
mass density are mutually inconsistent but can be brought into
agreement by adopting a top-heavy IMF \citep{Hopkins06b, Dave08,
  Wilkins08}.  However, recent improvement in the modeling of the
data, including luminosity-dependent dust corrections, rising SFHs,
and revised values of the faint end slope of the mass function have
reconciled the cosmic star formation and mass densities derived with
standard IMFs \citep{Reddy09, Papovich11, Behroozi12}.  A third
argument in favor of top-heavy IMFs is the apparent inability of
galaxy formation models to reproduce the observed number of luminous
submillimeter galaxies \citep[][]{Baugh05}.  Here too recent
observations and more sophisticated models have significantly reduced
the tension between models and data \citep{Hayward12}.  It thus
appears that much of the evidence originally favoring a top-heavy IMF
at certain epochs and under certain conditions can now be explained by
more mundane effects.

\subsection{Summary}

Extremely dwarf-rich IMFs are now routinely ruled out both by
sophisticated SPS models and dynamical constraints, and earlier
evidence for top-heavy IMFs has not held up to further scrutiny.
However, more modest IMF variations appear to be supported by the
data, at the level of a factor of $2-3$ in $M/L$.  It is intriguing
that constraints on the IMF based solely on dynamical modeling also
favor a mild steepening (i.e., relatively more low-mass stars) with
increasing velocity dispersion \citep{Treu10, Auger10, Graves10a,
  Spiniello12, ThomasJ11, Sonnenfeld12, Cappellari12, Dutton12a,
  Dutton12b, Dutton12c}.  As both dynamical and stellar
population-based techniques suffer from non-negligible but largely
orthogonal systematics, the most promising direction for future
constraints will come from joint analyses of the same systems with SPS
and dynamical techniques.


\section{CONCLUDING REMARKS}

The goal of this review has been to summarize what we have learned
about galaxies from their panchromatic SEDs using the tools of stellar
population synthesis.  By way of concluding, I would like to turn
toward the future and highlight areas where additional work is needed
to make SED modeling both a more firmly quantitative science and also
an effective engine for new discoveries.  In short, and not
surprisingly, the future requires better data, better models, better
comparison of models to data, and a better understanding of what is in
principle knowable from the analysis of galaxy SEDs.

A clear theme of this review has been the power of combining broadband
data with moderate resolution spectra.  The SDSS has been a
revolutionizing force in this regard, as it has provided high-quality
photometry and optical spectra for over one million objects.  The
great drawback of the SDSS is that it is fiber-based and thus the
spectra only sample the central 3'' (in diameter) of galaxies.  This
drawback will be alleviated with IFU spectroscopic surveys of nearby
galaxies, including the recently completed SAURON survey
\citep{Bacon01}, the ongoing ATLAS3D survey \citep{Cappellari11} of
260 early-type galaxies, the CALIFA survey of $\sim600$ galaxies
\citep{Sanchez12}, and the proposed MaNGA survey of a mass-limited
sample of $\sim10,000$ galaxies.  Such surveys will provide
high-quality spectra that are well-matched to the broadband data.
Restframe NIR spectra will be a new frontier for SPS studies as next
generation NIR facilities become operational, including ground-based
spectrographs (FIRE, FMOS, KMOS, MOSFIRE, FLAMINGOS-2, etc.), and the
eventual launch of the {\it James Webb Space Telescope}.  High-quality
models are only now being developed to interpret such data.  In
addition to spectra, narrow-band photometry will also be a valuable
addition to the landscape, as pioneered by the COMBO-17 survey
\citep{Wolf04} and now utilized by surveys such as ALHAMBRA
\citep{Moles08}, the NEWFIRM Medium-Band Survey \citep{Whitaker11},
and SHARDS \citep{PerezGonzalez12}.  Grism data will also help bridge
the gap between broadband photometry and moderate resolution spectra,
as demonstrated by the 3D-{\it HST} survey \citep{Brammer12}.  In
addition to these object-by-object surveys, the construction of
composite SEDs from galaxies spanning a range of redshifts allows for
the creation of very high-quality and well-sampled SEDs that will be
invaluable for SPS studies \citep{Assef08, Kriek11}.

Another theme of this review has been the growing realization that
uncertainties in the SPS models are becoming a critical limiting
factor to the interpretation of galaxy SEDs.  The challenge here is
not simply to enumerate the uncertainties but rather to identify areas
where clear progress can be made.  As a first step, all SPS models
should include contributions from nebular emission and dust around AGB
stars, as these processes are known to occur and the incorporation of
such effects into the models is reasonably straightforward, even if
the details are uncertain.  Panchromatic models (i.e., FUV-FIR
coverage) should also become standard both because IR data are now
widely available and because sophisticated dust emission models are
well-developed \citep[e.g.,][]{Draine07b}.  The stellar atmospheric
and synthetic spectral models will benefit from new asteroseismology
measurements from the {\it Kepler} mission, interferometric
observations (e.g., by CHARA), and new very high resolution UV-NIR
spectral atlases of nearby stars across the HR diagram
\citep{Bagnulo03, Lebzelter12}.  Perhaps the most vexing issues lie
with the stellar evolution uncertainties, as obvious calibrating data
are lacking.  The well-known problem is that, while globular clusters
are the canonical testing ground for stellar evolution, metal-rich
clusters are rare, and so the models tend to be poorly constrained
precisely in the metallicity range most relevant for modeling
galaxies.  Moreover, the largest uncertainties are associated with
fast evolutionary phases and so stars in such phases will be rare in
all but the most massive star clusters.  Efforts to constrain
uncertain stellar evolutionary phases from the SEDs of galaxies is
promising because the right metallicity ranges are probed and there
are sufficient numbers of stars to overcome Poisson noise, but the
obvious complexity of dealing with composite stellar populations makes
this approach challenging \citep[e.g.,][]{MacArthur10, Kriek10,
  Zibetti12}.  The Panchromatic Hubble Andromeda Treasury survey
\citep{Dalcanton12} is an {\it HST} program covering $\sim1/3$ of
M31's star-forming disk in six filters.  It promises to provide new
and powerful constraints on luminous and advanced stellar evolutionary
phases at moderately high metallicities.

More accurate models and higher-quality data will necessitate a more
sophisticated approach to comparing the two.  Presently, model fitting
is something of an art, owing to the fact that large regions of
parameter space are often severely under-constrained, which implies
that the choice of priors on the model parameters can have a large
impact on the derived results.  A few basic guidelines should be
followed to ensure that results are robust.  For example, one should
not simply fix a parameter to a particular value because it is
under-constrained.  Moreover, because the likelihood surface often
contains multiple peaks and valleys and is frequently computed on a
coarse grid, the best-fit parameters ought not be chosen based on the
minimum of $\chi^2$ \citep[see e.g.,][]{Taylor11}.  Rather, the full
posterior distributions should be used to derive best-fit values and
associated uncertainties.  The choice of priors needs to be considered
carefully, and in fact the model space should probably depend on the
type of data being fit, the redshift of the object, and even its
spectral type (quiescent vs. star-forming vs. peculiar).  As the
number of parameters increases, Markov Chain Monte Carlo techniques
will see more widespread use owing to their efficient exploration of
parameter space.  With regards to the modeling of moderate resolution
spectra, the analysis of spectral indices should eventually give way
to full spectral fitting as the latter not only allows for the
extraction of more information but also allows the modeler to visually
inspect the fits in a way that is not possible when only EWs are
extracted from the data.  This is important for identifying model
systematics and areas for future improvement.

Finally, further work is needed to understand what is knowable, in
principle, from the modeling of galaxy SEDs.  Questions such as `how
many discrete SF episodes can be measured from high quality optical
spectra?', or `how many moments of the metallicity distribution
function can be extracted from SEDs?', or `can the detailed dust
attenuation curve be measured with sufficiently high quality data on
an object-by-object basis?' have yet to be thoroughly explored.
Addressing these questions will be difficult because they depend
sensitively on the quality of the data, the SED type, and the
reliability of the models.  Fitting routines such as STARLIGHT,
STECMAP, and VESPA that attempt a non-parametric recovery of the SFH
and metallicity distribution function offer probably the most reliable
tools to explore these questions.  Theoretical studies aimed at
understanding `what is knowable' will help guide the next generation
of surveys aimed at studying the detailed physical properties of
galaxies.


\section{DISCLOSURE STATEMENT}

The author is not aware of any affiliations, memberships, funding, or
financial holdings that might be perceived as affecting the
objectivity of this review.


\section{ACKNOWLEDGMENTS}

I would like to thank my collaborators for the continuous lively
conversations that have helped form my views on this topic.  I would
also like to thank the authors who generously shared their figures for
this review, and especially Elisabete da Cunha, John Moustakas, Naveen
Reddy and Rita Tojeiro for providing new or modified figures.
St\'ephane Charlot, Daniel Dale, Sandy Faber, Jerome Fang, Ricardo
Schiavon, Rita Tojeiro and Scott Trager are thanked for very useful
comments on an earlier version of this manuscript.



\end{document}